\documentclass[useAMS,usenatbib,a4paper]{mn2e}

\usepackage{graphicx}
\usepackage{amssymb}

\newcommand{\ha}{H\,$\alpha$}
\newcommand{\hb}{H\,$\beta$}

\newcommand{\oiii}{\mbox{[O\,{\sc iii}]}}
\newcommand{\oi}{\mbox{O\,{\sc i}}}
\newcommand{\oii}{\mbox{O\,{\sc ii}}}
\newcommand{\nii}{\mbox{N\,{\sc ii}}}
\newcommand{\nv}{\mbox{N\,{\sc v}}}
\newcommand{\sii}{\mbox{S\,{\sc ii}}}
\newcommand{\feii}{\mbox{Fe\,{\sc ii}}}
\newcommand{\mgii}{\mbox{Mg\,{\sc ii}}}
\newcommand{\ciii}{\mbox{C\,{\sc iii}]}}
\newcommand{\civ}{\mbox{C\,{\sc iv}}}
\newcommand{\degree}{{}^{\circ}}

\newcommand{\kms}{km~s$^{-1}${}}
\newcommand{\ud}{\mathrm{d}}

\newcommand{\rten}{$R_{\mathrm{10 GHz}}$}

\newcommand{\loglum}{$\log L_{178 \mathrm{MHz}}$}

\newcommand{\pd}{\mathrm{D}}
\newcommand{\pmm}{\mathrm{M}}

\newcommand{\ps}{\mathrm{S}}
\newcommand{\pprob}{\mathrm{Prob}}
\newcommand{\rg}{R_\mathrm{G}}

\newcommand{\swmc}{$10^{-9}\, \mathrm{W} \mathrm{m}^{-3}$}

\title[Modelling the orientation of accretion disks in quasars using \ha{} emission]{Modelling the orientation of accretion disks in quasars using \ha{} emission}
\author[E. J. Down, S. Rawlings, D. S. Sivia and J. C. Baker]{E. J. Down$^{1,2}$\thanks{E-mail:
emily.down@nrc-cnrc.gc.ca}, S. Rawlings$^{1}$, D. S. Sivia$^{3}$ and J. C. Baker$^{4}$\\
$^{1}$Oxford Astrophysics, Keble Road, Oxford, OX1 3RH, UK\\
$^{2}$NRC Herzberg Institute of Astrophysics, 5071 West Saanich Road, Victoria BC, V9E 2E7, Canada\\
$^{3}$St John's College, Oxford, OX1 3JP, UK\\
$^{4}$Nature, The Macmillan Building, 4 Crinan Street, London, N1 9XW, UK}
\begin{document}

\date{Accepted 2009. Received 2009; in original form 2008 December 30}

\pagerange{\pageref{firstpage}--\pageref{lastpage}} \pubyear{2008}

\maketitle

\label{firstpage}

\begin{abstract}
  Infrared spectroscopy of the \ha{} emission lines of a sub-sample of
  19 high-redshift (0.8 $< z <$ 2.3) Molonglo quasars, selected at 408
  MHz, is presented. These emission lines are fitted with composite
  models of broad and narrow emission, which include combinations of
  classical broad-line regions of fast-moving gas clouds lying outside
  the quasar nucleus, and/or a theoretical model of emission from an
  optically-thick, flattened, rotating accretion disk, with velocity
  shifts allowed between the components. All bar one of the nineteen
  sources are found to have emission consistent with the presence of
  an optically-emitting accretion disk, with the exception appearing
  to display complex emission including at least three broad
  components. Ten of the quasars have strong Bayesian evidence for
  broad-line emission arising from an accretion disk together with a
  standard broad-line region, selected in preference to a model with
  two simple broad lines. Thus the best explanation for the complexity
  required to fit the broad \ha{} lines in this sample is optical
  emission from an accretion disk in addition a region of fast-moving
  clouds. We derive estimates of the angle between the rotation axis
  of the accretion disk and the line of sight. Deprojecting radio
  sources on the assumption of jets emerging perpendicular to the
  accretion disk gives rough agreement with expectations of radio
  source models. The distribution in disk angles is broadly consistent
  with models in which a Doppler boosted core contributes to the
  chances of observing a source at low inclination to the line of
  sight, and in which the radio jets expand at constant speed up to a
  size of $\sim 1$ Mpc. A weak correlation is found between the
  accretion disk angle and the logarithm of the low-frequency radio
  luminosity. This is direct, albeit tenuous, evidence for the
  receding torus model first suggested by \citet{lawrence91} in which
  the opening angle of the torus widens with increasing radio
  luminosity. The highest accretion disk angle measured is
  48$\degree$, consistent with the opening angle predicted for
  radio-luminous sources. Velocity shifts of the broad \ha{}
  components are analysed and the results found to be consistent with
  a two-component model comprising one single-peaked broad line
  emitted at the same redshift as the narrow lines, and emission from
  an accretion disk which appears to be preferentially
  redshifted with respect to the narrow lines for high-redshift
  sources and blueshifted relative to the narrow lines for
  low-redshift sources. An additional analysis is performed in which
  the disk emission is fixed at the redshift of the narrow-line
  region; although only two quasars show a robust change in fitted
  angle, the radio luminosity -- disk angle correlation falls
    sharply in probability, and so is strongly model dependent in this
    sample.
\end{abstract}

\begin{keywords}
galaxies: active -- galaxies: jets -- quasars: emission lines -- quasars: general.
\end{keywords}

\section{Introduction}

\subsection{Orientation effects}
\label{sec:orientation}

Optical emission spectra of Active Galactic Nuclei (AGN) matched in
radio and optical luminosity are now firmly believed to be strongly influenced by
orientation effects, with only small underlying differences in the
sources themselves. There are two separate orientation-dependent
effects which alter the optical spectra of AGN.

The Type 1/Type 2 classification of AGN is made
according to the presence of broad emission lines. Type 2 AGN
possess only narrow emission lines, $\lesssim 2000$ \kms{}
\citep[e.g.][]{peterson97} and weak non-stellar continuum
emission. Type 1 AGN have broad emission lines of $\sim 2000$
-- $20000$ \kms, and strong non-stellar continuum emission, in
addition to narrow lines similar to the Type 2 sources. An
explanation for this disparity grew from the discoveries of broad
emission lines seen in polarised light from Type 2 Seyfert
sources \citep[e.g.][]{antonucci85}, which suggested
orientation-dependent obscuration caused by an intervening screen of
matter, such as a dusty molecular torus \citep{krolik86}. It has
recently become clear that obscuration by dust in starbursting
galaxies can also be responsible for concealing Type 2 AGN
\citep[e.g.][]{martinez05}.

The second orientation-dependent effect arises from the relativistic
motion of the plasma in the radio jets. \citet{scheuer79} first
suggested that viewing a radio source with opposing relativistic jets
would cause a large contrast in the luminosities of the approaching
and receding jets, and that objects with jet axes close to the line of
sight would be seen more often due to Doppler boosting of the core.
\citet{orr82} made the connection between Doppler boosting of the
core, and a measure of quasar orientation from the core-to-lobe radio
flux density ratio; this allowed them to unify the ``core-dominated''
quasars with flat optical spectra, viewed at angles close to the line
of sight, with the ``steep-spectrum'' or radio-lobe-dominated quasars
viewed at larger angles. \citet{wills86} linked the radio properties
to the optical properties by discovering an anticorrelation between
the core-to-lobe radio flux ratio and the width of the \hb{} line,
interpreting this connection as the result of beaming of radio
emission from a jet emerging along the rotation axis of an accretion
disk; the broad \hb{} lines arise from the accretion disk with a width
correlated with the angle between the line of sight and the disk axis.

The two optical schemes were reconciled by \citet{barthel89}, who
gave a consistent picture in which FRII narrow line radio galaxies,
steep-spectrum quasars and flat-spectrum quasars are all drawn from
the same parent population, but viewed at decreasing angles to the
line of sight. A review of these so-called unified schemes for AGN can
be found in \citet{urry95} or \citet{antonucci93}.

A solid understanding of how quasar emission lines arise, and how they
are affected by the AGN environment along different sight lines, is
not only an interesting study in terms of quasar structure, but is
also vital in order to disentangle orientation effects from
large-scale AGN surveys which enable the study of cosmic evolution.

\subsection{Accretion disks}
\label{sec:disks}

There is a growing body of evidence that AGN are powered by accretion
of gas and dust onto supermassive black holes, and this is now the
accepted paradigm. As the black hole feeds on the surrounding
material, it is expected that this will form an accretion disk of
infalling matter \citep{shakura73}. The current theory is that
accretion disks have two parts: a puffed-up, X-ray-emitting inner
disk, and a flattened, outer disk which gives rise to broad optical
emission lines.

\citet{collin80} first suggested that a thickened inner accretion
disk, shielding and reprocessing the hard X-ray emission from the
black hole, gives rise to the optical low-ionisation \feii{} emission
seen in Type 1 Seyferts, from the atmosphere above the
geometrically-thin outer part of the disk. \citet{rees82} postulated
that a geometrically- and optically-thick ion-supported torus
surrounds the supermassive black hole at the centre of a radio galaxy,
collimating the emerging radio jets. \citet{tanaka95} observed a
broad, asymmetric iron K$\alpha$ line consistent with emission from a
disk of this description, situated between $\sim 3$ -- $10 \rg$ (where
$\rg = G M / \mathrm{c}^2$ is the gravitational radius) from the
nucleus of the AGN.

\citet{filippenko88} reviewed the different lines of evidence from
optical and UV data that flattened, extended accretion disks fuel the
central black holes of galaxies. For example, \citet{baldwin77}
recorded the anticorrelation of the UV continuum luminosity with the
equivalent widths of broad \civ{} emission (the ``Baldwin
Effect''). Both this observation and the ``big blue bump'' of excess
UV continuum emission found by \citet{malkan82} may be explained by
emission from an optically-thick, geometrically-thin accretion disk
\citep{netzer85}. The strongest direct evidence for this scenario is
the presence of double-peaked, low-ionisation optical emission lines
seen in some AGN, which arise from the Doppler effect acting on the
emission from rotating material in the outer accretion disk
(e.g. \citet{chen89}, \citet{perez88}).

The thin, optically-emitting disk must be illuminated by some
mechanism. The photoionising flux may originate either from the
central non-thermal source, or from the inner, X-ray-emitting disk
\citep{collin80}; and the radiation may illuminate the outer disk
directly, or be scattered from a highly-ionised diffuse medium above
the outer disk \citep{chen89b}.

Optical double-peaked lines have to date only been found in a
relatively low percentage of radio-loud AGN ($\sim$ 10$\%$, see
\citet{eracleous94}). \citet{strateva03} discovered that radio-quiet
quasars also emit double-peaked lines, although these appear to be
rarer still: double-peaked emission was seen in $\sim$ 3$\%$ of the
SDSS AGN, including both radio-loud and radio-quiet
sources. Double-peaked profiles are not unique to Balmer lines:
\citet{strateva03} found double-peaked \mgii{} emission lines in some
SDSS AGN.

It is not clear why the double-peaked line profiles should be rare,
although there are several possibilities: the outer accretion disk may
simply be obscured by broad-line-emitting clouds surrounding it; or if
the outer accretion disk causes a wind, then the broad lines which
arise from this wind are predicted to be single-peaked
\citep{murray97}. In any case, it should not be expected that the
low-ionisation broad lines seen in an AGN originate solely from the
rotating disk; single-peaked emission may be seen in addition to
double-peaked profiles.

The accretion disk model used in this paper is taken from
\citet{chen89}, and consists of a thick, hot torus, whose outer edge
may reach up to $100\, \rg$ from the black hole. Inverse
Compton-scattered X-rays from this inner disk illuminate an
optically-thick, flattened outer disk \citep{halpern89}. The thin,
circular disk, which produces the double-peaked emission lines, may
extend up to $\sim 10^5\, \rg$ from the central engine. The
distinctive line profiles are caused by the rotation of the disk,
splitting the emission into redshifted receding material and
blueshifted approaching material; the blueshifted peak is of higher
intensity than the redshifted peak, as a result of Doppler boosting.

The chosen model was necessarily simple, to limit the number of free
parameters.  More complex disks, such as an elliptical disk, which may
arise when a single star is disrupted near the black hole
\citep{gurzadyan79}, or a warped disk, thought to occur around
rotating black holes \citep{bachev99}, give rise to a wider range of
line profiles, including double-peaked profiles with a redward peak of
higher intensity than the blueward peak. \citet{strateva03} found that
assuming all low-ionisation broad-line emission comes from a disk,
non-axisymmetric disks would be required in $\sim 60\%$ of their SDSS
sample, while \citet{eracleous03} found that in their sample of 106
radio-loud AGN, 20\% have double-peaked \ha{} profiles visible to the
eye, of which $\sim 40\%$ require a model more complex than the
circular Keplerian disk.

In this paper, a small, but close to complete, sample of radio-loud
quasars are analysed to determine if their spectra include emission
from circular, planar accretion disks, either as the sole component of
broad optical emission, or in combination with a single-peaked broad
line.

\subsection{Velocity shifts}
\label{sec:velshifts}

It has been generally accepted that the narrow-line region (NLR) of an
AGN falls near the systemic redshift. The NLR is
extended, and appears to be reasonably independent of viewing angle
effects, and so the narrowness of the lines constrains the velocity of
this gas to be small. \citet{heckman81} showed, for a handful of
low-$z${} Seyferts and radio galaxies, that the narrow lines had
small blueshifts of between $\sim$ 0 -- 300 \kms{} relative to the
neutral hydrogen emission of the host galaxies. \citet{vandenberk01}
discovered that, for composite spectra created with several thousand
Sloan Digital Sky Survey (SDSS) quasar spectra covering a wide
redshift range ($0.04 \lesssim z \lesssim 4.8$), the narrow lines are
blueshifted by small velocities ($\lesssim 100$ \kms) which correlate
with ionisation potential.

The broad-line region (BLR) has a complex structure, with emission
line redshifts which vary according to species, implying separate
regions of gas (see Figure 6 of \citet{collin80}). It is thought that
the high-ionisation lines (HILs) arise from a compact, spherical
region close to the central black hole, while the low-ionisation lines
(LILs) are formed in a flattened structure further from the AGN
centre, possibly an accretion disk \citep{krolik91}.  \citet{collin88}
suggested that the HILs might be produced by shocks in an outflowing
wind.

\citet{gaskell82} demonstrated for a sample of flat-spectrum quasars
with z $\sim$ 0.2 -- 2.3 that the HILs, such as \ciii, \civ{} and
\nv{}, are blueshifted by $\sim 600$ \kms{} with respect to the LILs,
which include \mgii{}, \oi{} and the Balmer lines. \citet{wilkes84}
confirmed this shift for high-$z${} quasars ($2 \lesssim z \lesssim 3$),
finding a slightly higher range of shifts, up to $\sim$ 1400
\kms. Blueshifted HIL zones have also been observed by
\citet{espey89}, who found shifts of $\sim$ 1000 \kms{} in a small
sample of $1.3 \lesssim z \lesssim 2.4$ sources, and \citet{corbin90},
who found shifts exceeding 4000 \kms{} for luminous,
optically-selected sources with $z \gtrsim 1$. \citet{richards02} also
confirm this trend from studies of $\sim 800$ quasars with $1.5
\lesssim z \lesssim 2.2$ from the SDSS, though they find a wide
distribution of velocity shifts of the high-ionisation \civ{} with
respect to the low-ionisation \mgii, ranging from redshifts of $\sim$
500 \kms{} to blueshifts of over 2000 \kms, and they take pains to
point out that they do not believe it is a line shift so much as a
lack of flux in the red wing of the line.

\citet{mcintosh99} found that the HIL zone is at the same redshift as
the narrow lines in low-$z${} sources, but for a sample of quasars
with $2 \lesssim z \lesssim 2.5$, broad \hb{} had a redshift of $\sim$
500 \kms{} relative to \oiii.  Although the LILs are typically
considered to have low velocity shifts, there have been recorded
instances of Balmer lines with very high redshifts, e.g. $\sim$ 2100
\kms{} for 3C277 \citep{osterbrock76} and $\sim$ 2600 \kms{} for OQ208
\citep{osterbrock79}.

\section[]{Sample Selection}
\label{sec:selection}

A sub-sample of 19 quasars was defined from the Molonglo Quasar Sample
\citep{molonglo3}. The Molonglo sample of radio sources was selected
from the Molonglo Reference Catalog (MRC) \citep{large81}, a 408 MHz
survey conducted with the Molonglo Synthesis Telescope which is 99.9\%
complete at 1 Jy. The radio sources were then identified from VLA
1-arcsecond resolution radio images, optical imaging and spectroscopy,
the quasars \citep{molonglo3} being distinguished from the radio
galaxies \citep{molonglo1} by the presence of broad optical emission
lines.

The Molonglo quasar sample includes all quasars with flux densities
\mbox{$S_{408 \mathrm{MHz}} > 0.95$ Jy} in a $10\degree$-wide strip in
the Southern sky, \mbox{$-20\degree > \delta > -30\degree$}, excluding
sources with low Galactic latitude \mbox{($\mid b \mid > 20\degree$)}
and also a strip in Galactic R.A. (details in \citet{molonglo3}) to
define a sample of manageable size. It should be noted that as the
Molonglo sample was selected at the mid-range frequency of 408 MHz,
there are likely to be some sources included in the sample by virtue
of their strong radio core emission, and are therefore inclined at
small angles to the line of sight; this builds an orientation bias
into this survey.

The quasar sub-sample was selected on the basis of four criteria:
observability during the relevant time period; redshift such that
\ha{} and \hb{} emission falls in wavelength windows
corresponding to regions of high atmospheric transparency ($0.8 < z <
1.0$, $1.5 < z < 1.65$, $2.2 < z < 2.3$); sufficient J-band brightness
to be observable in a reasonable integration time; and exclusion of
the RA range 14h -- 03h in accordance with scheduling constraints.
The J-band magnitude limit is the only factor which adds a significant
bias in the sub-sample selection. The objects were selected to be
brighter than \mbox{J $\sim 18.5$}, and this excluded one source from
the sample, MRC0418-288. This source is likely to be reddened or
dusty, which means that it has a higher chance of being inclined at a
large angle to the line of sight. MRC1256-243 is an extra
core-dominated source which was not observed due to scheduling
limitations; this source is likely to have been boosted into the
sample by virtue of its strong core emission in any case.

Throughout the paper, a cosmology of $H_0 = 70$ km s$^{-1}$
Mpc$^{-1}$, $\Omega_{\mathrm{M}} = 0.3$ and $\Omega_{\Lambda} = 0.7$
is assumed.

\section{Infrared Spectra}
\label{sec:irdata}

\subsection{Data acquisition}

Near infrared spectra of \ha{} were obtained for the Molonglo quasar
sub-sample with the Infrared Spectrometer And Array Camera (ISAAC)
spectrograph \citep{moorwood98} at ESO's VLT UT1, in service mode,
between October 2001 and February 2002. The spectra were taken in
short-wavelength, low-resolution mode, using a 1-arcsec-wide
slit. The observational details, including exposure times, seeing and
airmass, are given in Table \ref{tab:obsdata}.

\begin{table*}
\centering
\begin{minipage}{160mm}
\begin{tabular}{lllllllllll}
\hline
\multicolumn{2}{|l|}{Quasar} & R.A.   & Dec.   & $z$  & $b_{\mathrm{J}}$ & Date     & Wave- 
& Exposure         & Seeing        & Airmass  \\ 
\multicolumn{2}{|l|}{}       & J(2000) & J(2000)& {}  & {}             & observed & band        
& time (s)         & (arcsec) & {}       \\
\multicolumn{2}{|l|}{(1) \,\,\,\, (2) }  & (3) & (4) & (5) & (6) & (7) & (8) & (9) & (10) & (11) \\
\hline
\hline
1    & MRC0222-224  &   02 25 16.6 & -22 15 22	& 1.603 & 19.1 & 2001-10-07  & H & 180 $\times$ 8  & 0.9	& 1.015 \\
2    & MRC0327-241  &   03 29 54.1 & -23 57 09	& 0.895 & 19.4 & 2001-10-12  & J & 180 $\times$ 8  & 0.6	& 1.209 \\	
3    & MRC0346-279  &   03 48 38.1 & -27 49 14	& 0.989 & 20.5 & 2001-10-12  & J & 180 $\times$ 8  & 0.7	& 1.077 \\
4    & MRC0413-210  &   04 16 04.3 & -20 56 28	& 0.807 & 18.4 & 2001-10-12  & J & 180 $\times$ 8  & 0.9	& 1.053 \\
5    & MRC0413-296  &   04 15 08.7 & -29 29 03	& 1.614 & 18.6 & 2001-10-12  & H & 180 $\times$ 8  & 0.7	& 1.004 \\
6    & MRC0430-278  &   04 32 17.7 & -27 46 24	& 1.633 & 21.3 & 2001-12-23  & H & 180 $\times$ 10 & 1.2	& 1.138 \\
7    & MRC0437-244  &   04 39 09.2 & -24 22 08	& 0.834 & 17.5 & 2001-10-06  & J & 180 $\times$ 8  & 0.6	& 1.012 \\
8    & MRC0450-221  &   04 52 44.7 & -22 01 19	& 0.900 & 17.8 & 2001-11-20  & J & 180 $\times$ 8  & 0.5	& 1.214 \\
9    & MRC0549-213  &   05 51 58.3 & -21 19 49	& 2.245 & 19.1 & 2001-12-22  & K & 100 $\times$ 28 & 0.5	& 1.075 \\
10   & MRC1019-227  &   10 21 27.6 & -23 01 54	& 1.542 & 21.1 & 2001-12-25  & H & 180 $\times$ 10 & 0.7	& 1.144 \\
11   & MRC1114-220  &   11 16 54.5 & -22 16 53	& 2.286 & 20.2 & 2002-01-01  & K & 100 $\times$ 28 & 0.7	& 1.137 \\
12   & MRC1208-277  &   12 10 43.6 & -27 58 55	& 0.828 & 18.8 & 2002-01-27  & J & 180 $\times$ 8  & 0.5	& 1.034 \\
13   & MRC1217-209  &   12 20 22.3 & -21 13 09	& 0.814 & 20.2 & 2002-02-05  & J & 180 $\times$ 8  & 1.1	& 1.106 \\
14   & MRC1222-293  &   12 25 01.2 & -29 38 17	& 0.816 & 18.5 & 2002-01-27  & J & 180 $\times$ 8  & 0.6	& 1.006 \\
15   & MRC1301-251  &   13 04 14.7 & -25 24 37	& 0.952 & 21.0 & 2002-02-12  & J & 180 $\times$ 8  & 0.4	& 1.011 \\
16   & MRC1349-265  &   13 52 10.3 & -26 49 28	& 0.924 & 18.4 & 2002-02-11  & J & 180 $\times$ 8  & 0.5	& 1.051 \\
17   & MRC1355-215  &   13 58 38.2 & -21 48 54	& 1.607 & 19.9 & 2002-02-13  & H & 180 $\times$ 8  & 0.5	& 1.002 \\
18   & MRC1355-236  &   13 58 32.7 & -23 52 20 & 0.832 & 17.8 & 2002-02-11  & J & 180 $\times$ 8  & 0.4	& 1.013 \\
19   & MRC1359-281  &   14 02 02.4 & -28 22 25	& 0.802 & 18.7 & 2002-02-14  & J & 180 $\times$ 16 & 0.6	& 1.048 \\
\hline
\end{tabular}
\caption{Observational details: \textit{Columns 1 and 2}: Quasar index and Molonglo Reference Catalog (MRC) name; 
\textit{Columns 3 and 4}: J(2000) right ascension and declination;
\textit{Column 5}: Redshift;
\textit{Column 6}: Optical $b_{\mathrm{J}}$ magnitude from the UK Schmidt IIaJ plates, where 
$b_{\mathrm{J}} = B - 0.23(B-V)$ \citep{bahcall80};
\textit{Column 7}: Date of observation;
\textit{Column 8}: Waveband of observation;
\textit{Column 9}: Exposure time in seconds multiplied by the total number of exposures
in the observation;
\textit{Column 10}: Seeing in arcseconds, measured from the images;
\textit{Column 11}: Airmass.}
\label{tab:obsdata}
\end{minipage}
\end{table*}

\subsection{Data reduction}
\label{sec:datared}

The raw images were cleaned of cosmic rays in two stages:
\textit{crmedian} in \textit{IRAF}\footnote{\textit{IRAF} is
  distributed by the National Optical Astronomy Observatories, which
  are operated by the Association of Universities for Research in
  Astronomy, Inc., under cooperative agreement with the National
  Science Foundation.} was used to automatically remove cosmic rays
which fell more than $10 \sigma$ below or $3 \sigma$ above the median
value, replacing these pixels with the median value. The regions of
the image including the spectrum and the sky subtraction zone were
then cleaned by hand using \textit{credit} in \textit{IRAF}, replacing
bad pixels with local sky values. Known detector effects
\citep{isaacdrg} were corrected as follows.  ``Electrical ghosts'' are
additional signals in the image, and were removed using the dedicated
\textit{eclipse} \citep{eclipse} recipe \textit{ghost}. The ``odd-even
column effect'' causes an intensity difference in the alternate rows
of the image; this was cleaned from images in which it was apparent
using a script which masked the two pixels corresponding to the
highest frequency variations in Fourier space for each quadrant of the
image, following the method of \cite{isaacdrg}.

The images were flatfielded, corrected for distortion effects and
wavelength calibrated using arc lamp frames for the relative
calibration and OH sky emission lines to zero-point correct the
calibration; these procedures were carried out using standard
techniques in \textit{IRAF}. The sky background is strong and
time-variable in the infrared, and so ISAAC spectra are observed in
nod-and-jitter mode, producing pairs of images with the spectra
located on different regions of the CCD. For each image, the sky
background was subtracted using the neighbouring image, before all the
spectra were stacked using a script in \textit{IDL}.

The spectra were extracted with the \textit{IRAF} routine
\textit{apall}, and telluric features were removed and a flux
calibration applied simultaneously using one reference star for each
quasar, with \textit{IDL} procedures. Finally, the spectra were
corrected for the dust reddening of the Milky Way, using the
\citet{schlegel98} maps of Galactic dust emission, and the spectra
were corrected to the heliocentric rest-frame \footnote{The correction
  was calculated using ephemerides from the Markwardt \textit{IDL} library.}.

\subsection{Molonglo infrared spectra}
\label{sec:spectra}

\begin{figure*}
\begin{minipage}{160mm}
\begin{center}
\includegraphics[width = 1.0\textwidth]{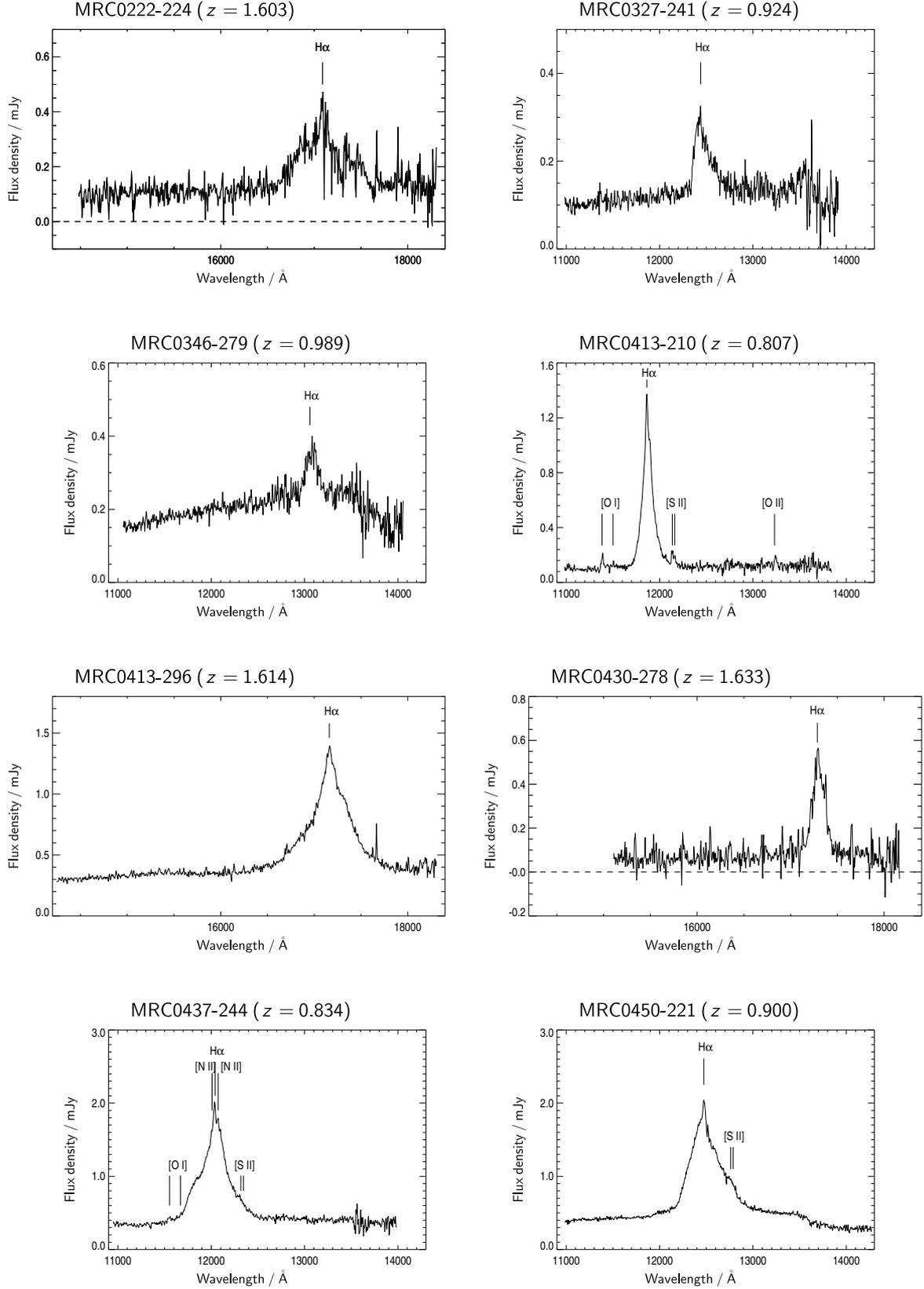}
\caption{Observed near-infrared spectra, including the \ha{} emission lines, for the Molonglo sub-sample quasars, observed with ISAAC on ESO's VLT UT1. The observed wavelength is in \AA, and the flux density scale is in mJy. The spectra have been slightly smoothed.}
\label{fig:isaacspec1}
\end{center}
\end{minipage}
\end{figure*}

Spectra of the quasars are presented in Figure \ref{fig:isaacspec1}.
For the purposes of this figure, these have been smoothed in
\textit{DIPSO}\footnote{\textit{DIPSO} is a Starlink program} with a
Gaussian filter of FWHM 4.7 pixels (pixel scales vary slightly between
spectra, but are typically 2 \AA{} pixel$^{-1}$) to reduce the noise.

\begin{figure*}
\begin{minipage}{160mm}
\begin{center}
\includegraphics[width = 1.0\textwidth]{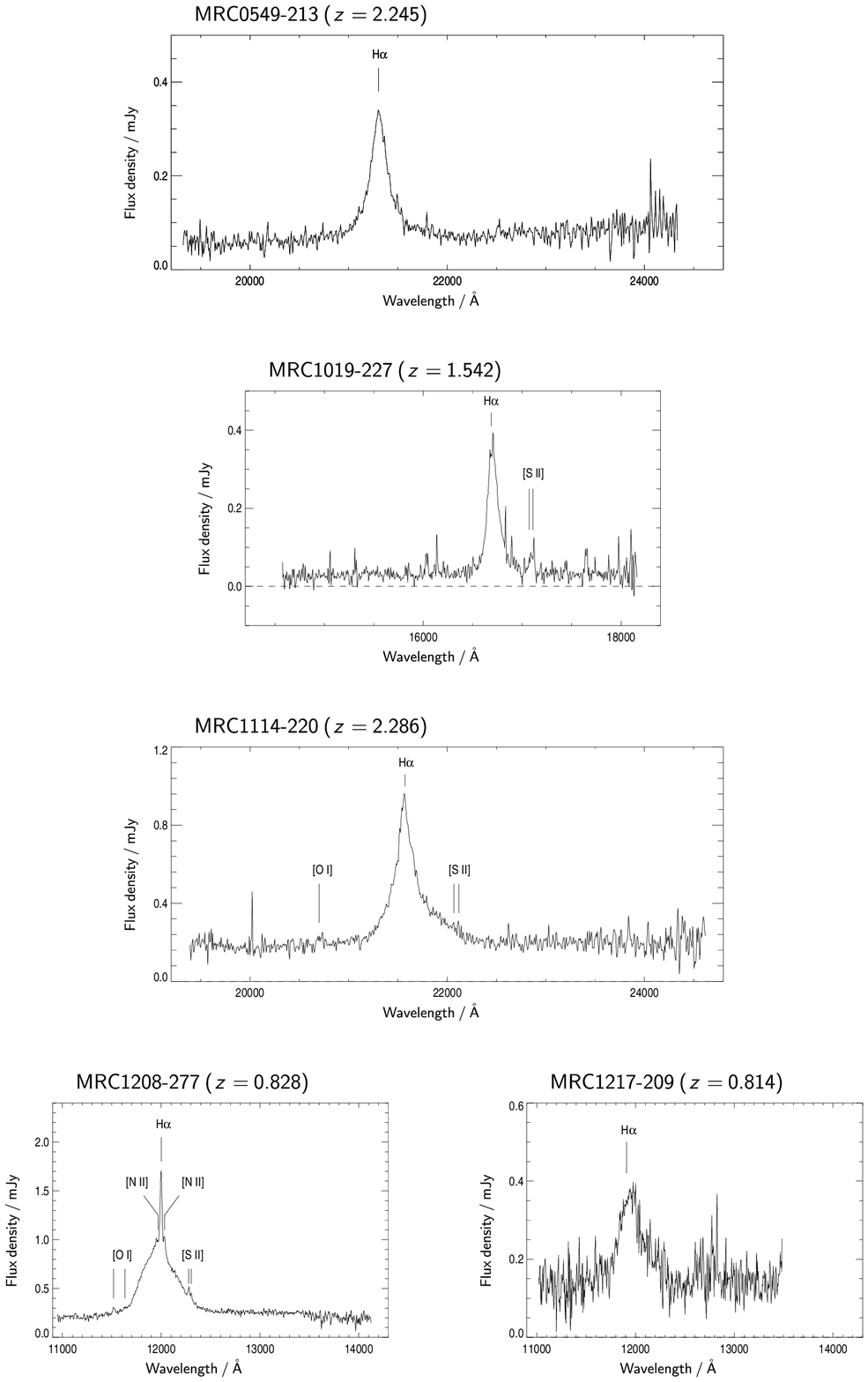}
\label{fig:isaacspec2}
\end{center}
\end{minipage}
\end{figure*}

\begin{figure*}
\begin{minipage}{160mm}
\begin{center}
\includegraphics[width = 1.0\textwidth]{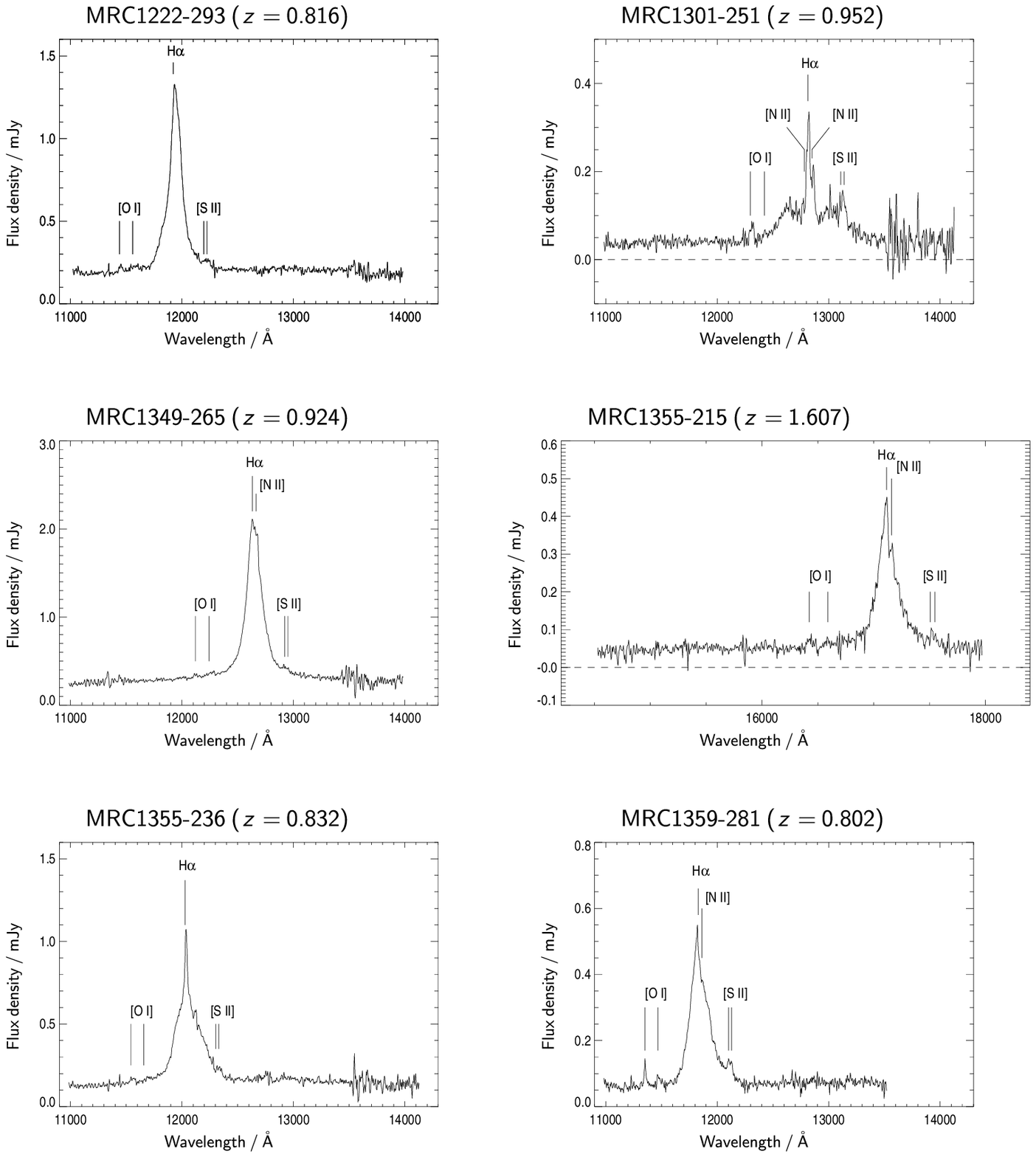}
\label{fig:isaacspec3}
\end{center}
\end{minipage}
\end{figure*}

\subsection{Notes on infrared spectra}

\noindent\textbf{MRC0222-224}: This spectrum contains noise spikes from
poor sky subtraction which affects most of the H-band spectra. The
apparent absorption line in the broad \ha{} line is an artifact, and
is excluded from the emission line fitting.

\noindent\textbf{MRC0327-241}: The continuum slopes with an index of
$\alpha \sim 1.7$, where $F_{\nu} \propto \nu^{-\alpha}$. This is a BL-Lac
type continuum, as these have spectral indices of $\alpha > 1$ (e.g. \citet{brown89}).

\noindent\textbf{MRC0346-279}: The continuum is strongly sloped, with $\alpha \sim 2.7$; this is a BL-Lac type continuum.

\noindent\textbf{MRC0413-210}: The narrow line at a wavelength of 
7322 \AA{} (observed 13230 \AA) is \oii. This line was not fitted in 
the analysis, as it is well-separated from the \ha{} region.

\noindent\textbf{MRC0413-296}: The apparent narrow features to the right of the broad \ha{} profile are artifacts of the sky subtraction.

\noindent\textbf{MRC0430-278}: This spectrum has a large number of noise spikes remaining from the sky subtraction process.

\noindent\textbf{MRC0450-221}: The drop in flux longwards of 1350
\AA{} is not a real feature, but an artifact introduced during the
flux calibration. This part of the spectrum is excluded from the
emission line fitting.

\noindent\textbf{MRC1019-227}: The sky subtraction is poor, leading to artifacts which resemble narrow lines. The apparent narrow 
line at the wavelength of \ha{} is a sky line defect, and is excluded
from the emission line fitting.

\noindent\textbf{MRC1217-209}: This spectrum is very noisy due to poor seeing. The
structure of \ha{} is not readily apparent.

\noindent\textbf{MRC1301-251}: Broad emission is weak in this spectrum compared to the strong narrow lines. 

\noindent\textbf{MRC1355-215}: The apparent narrow absorption features are artifacts from imperfect sky subtraction.

\subsection{Redshift measurements}

Improved redshifts for these quasars were measured from the ISAAC
spectra, or from new optical spectra (Janssens et al., in preparation)
in cases that strong narrow lines were available (\oiii{} for
preference, followed by Balmer lines). If no strong narrow lines were
present in the new spectra, then the most recent measurements from the
literature were selected. The only exception to this was MRC1349-265,
whose redshift was given as $z = 0.934$ in \citet{molonglo4}, but for
which $z = 0.924$ was a solid measurement, even in the absence of
narrow \oiii; a change of this size is likely to result from a
typographical error in the original paper. Table \ref{tab:properties}
lists the redshifts and their origins, in addition to measured FWHMs
of the broad lines, the integrated fluxes of the broad and narrow
\ha{} lines, and core-to-lobe flux ratios at 10 GHz in the rest frame
(\rten).

\begin{table*}
\centering
\begin{minipage}{170mm}
\begin{tabular}{llllllllll}
\hline

\multicolumn{2}{|l|}{Quasar} & $z$  & Source & FWHM   & $F_{\mathrm{Narrow} \, H_{\alpha}}$ & $F_{\mathrm{Broad} \, H_{\alpha}}$ & Measured & \rten{} from & Source of   \\ 
\multicolumn{2}{|l|}{}       & {}   & of $z$ & (\kms) & ($10^{-19}$ W m$^{-2}$) & ($10^{-19}$ W m$^{-2}$) & \rten    & literature   & \rten \\
\multicolumn{2}{|l|}{(1) \,\,\,\, (2) }  & (3) & (4) & (5) & (6) & (7) & (8) & (9) & (10) \\
\hline
\hline
1    & MRC0222-224  & 1.603 & 2   & 2230  & 46 & 740  & 0.0065 & $<$ 0.37 & 1  \\
2    & MRC0327-241  & 0.895 & 3   & 3420  & 5  & 255  & 10.0   & $>$ 1    & 1  \\	
3    & MRC0346-279  & 0.989 & 2,3 & 2320  & ND & 144  & 10.2   & $>$ 5    & 1  \\
4    & MRC0413-210  & 0.807 & 2,3 & 2250  & 58 & 1054 & 0.78   & 0.71     & 1  \\
5    & MRC0413-296  & 1.614 & 1   & 5530  & 71 & 2573 & 0.020  & 0.031    & 1  \\
6    & MRC0430-278  & 1.633 & 1   & 2480  & 4  & 581  & 0.46   & --       & {} \\
7    & MRC0437-244  & 0.834 & 2   & 4950  & 30 & 3477 & 0.10   & 0.098    & 1  \\
8    & MRC0450-221  & 0.900 & 2   & 6620  & 57 & 3739 & 0.060  & 0.086    & 1  \\
9    & MRC0549-213  & 2.245 & 2,3 & 2630  & ND & 561  & 0.31   & $<$ 0.55 & 1  \\
10   & MRC1019-227  & 1.542 & 1   & 1910  & 6  & 377  & 0.051  & --       & {} \\
11   & MRC1114-220  & 2.286 & 1   & 2650  & 55 & 1715 & 0.078  & 0.08     & 2  \\
12   & MRC1208-277  & 0.828 & 3   & 3490  & 144& 1886 & 0.049  & 0.088    & 1  \\
13   & MRC1217-209  & 0.814 & 2,3 & 5850  & ND & 537  & 0.078  & 0.057    & 1  \\
14   & MRC1222-293  & 0.816 & 3   & 2840  & 77 & 935  & 0.98   & 0.38     & 1  \\
15   & MRC1301-251  & 0.952 & 3   & 1530  & 52 & 343  & 0.014  & 0.020    & 1  \\
16   & MRC1349-265  & 0.924 & 1   & 3350  & 52 & 2366 & 0.25   & --       & {} \\
17   & MRC1355-215  & 1.607 & 1   & 2600  & 25 & 563  & 0.36   & 0.84     & 1  \\
18   & MRC1355-236  & 0.832 & 3   & 2180  & 76 & 1007 & 0.097  & 0.098    & 1  \\
19   & MRC1359-281  & 0.802 & 2,3 & 3820  & 25 & 575  & 0.12   & --       & {} \\
\hline
\end{tabular}
\caption{Measured properties of the quasar sub-sample: \textit{Columns 1 and 2}: Quasar index and MRC name; 
\textit{Column 3}: Redshift;
\textit{Column 4}: Origin of redshift measurement -- 1 = New measurement presented in this paper, 2 = \citet{molonglo4},
3 = \citet{silva2}; 
\textit{Column 5}: FWHM in \kms;
\textit{Column 6}: Narrow \ha{} line flux in units of $10^{-19}$ W m$^{-2}$. ND in this column 
indicates that narrow \ha{} was not detected.
\textit{Column 7}: Broad \ha{} line flux in units of $10^{-19}$ W m$^{-2}$.
\textit{Column 8}: Measured core-to-lobe flux ratio at rest-frame 10 GHz (\rten);
\textit{Column 9}: Core-to-lobe flux ratio at rest-frame 10 GHz (\rten) from the literature;
\textit{Column 10}: Origin of the core-to-lobe flux ratio from the literature -- 1 = \citet{molonglo3}, 2 = \citet{silva1}.}
\label{tab:properties}
\end{minipage}
\end{table*}

\section{Models of Emission}
\label{sec:models}

\subsection{Set of models}
\label{sec:modelset}

\begin{figure}
\begin{center}
\includegraphics[width = 0.4\textwidth]{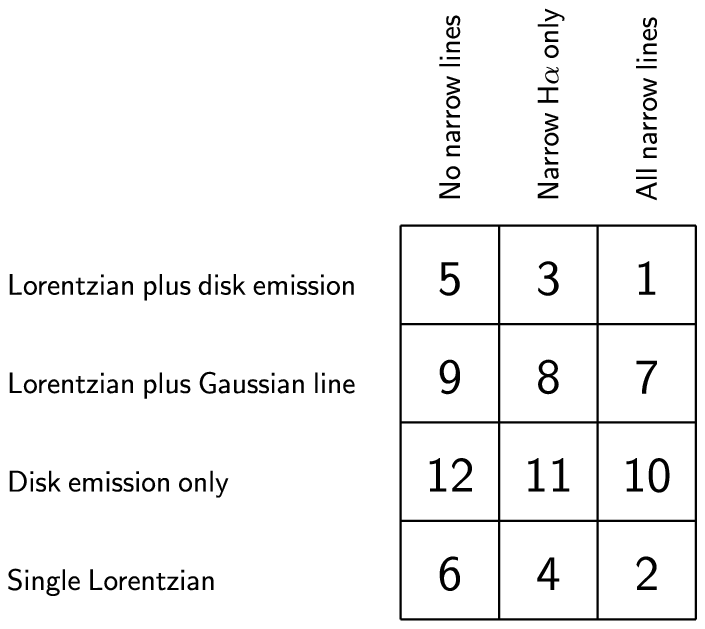}
\caption{An illustration of the components included in each of the models, 1 to 12.}
\label{fig:paramspace}
\end{center}
\end{figure}

The emission in the rest-frame range $\sim 6000$ -- $7000$ \AA{} was
modelled as the sum of narrow line emission and broad emission.  There
are four sets of models with different broad emission
contributions. One set of models has a single Lorentzian line
representing emission from a classical BLR of fast-moving clouds; one
set comprises a Lorentzian and a Gaussian line, to simulate two BLRs
at different temperatures, or a BLR plus an outflow. There are two
sets of models including accretion disks: one set has the accretion
disk plus a Lorentzian profile to represent a standard BLR, while the
other set includes broad emission from the accretion disk only. There
are three models in each set: one with narrow \ha{} plus \oi, \sii{}
and \nii{} lines; one with narrow \ha{} but none of the forbidden
lines; and one with no narrow line emission. There are therefore
twelve models constructed in this modular way, as shown in Figure
\ref{fig:paramspace}. The parameters included in each of the models
are detailed in Table \ref{tab:models}, and the prior probability
ranges for these parameters are shown in Table \ref{tab:priors}.

\begin{table*}
\begin{minipage}{160mm}
\begin{center}
\begin{tabular}{lp{6cm}cccccccccccc}
\hline
\multicolumn{2}{|c|}{Parameter} & \multicolumn{12}{|c|}{Model}\\ 
\multicolumn{2}{|c|}{} & 1 & 2 & 3 & 4 & 5 & 6 & 7 & 8 & 9 & 10 & 11 & 12 \\
\hline
\hline
1  & Flat continuum
& $\bullet$ & $\bullet$ & $\bullet$ & $\bullet$ & $\bullet$ & $\bullet$ & $\bullet$ & $\bullet$ & $\bullet$ & $\bullet$ & $\bullet$ & $\bullet$ \\	
{} & {} & {} & {} & {} & {} & {} & {} & {} & {} & {} & {} & {} & {} \\	      
2  & Broad \ha{} shift with respect to narrow \ha 
& $\bullet$ & $\bullet$ & $\bullet$ & $\bullet$ & {} & {} & $\bullet$ & $\bullet$ & {} & {} & {} & {} \\	      
3  & Broad \ha{} central wavelength  
& {} & {} & {} & {} & $\bullet$ & $\bullet$ & {} & {} & $\bullet$ & {} & {} & {} \\       
4  & Broad \ha{} Lorentzian width 
& $\bullet$ & $\bullet$ & $\bullet$ & $\bullet$ & $\bullet$ & $\bullet$ & $\bullet$ & $\bullet$ & $\bullet$ & {} & {} & {} \\	      
5  & Broad \ha{} intensity	
& $\bullet$ & $\bullet$ & $\bullet$ & $\bullet$ & $\bullet$ & $\bullet$ & $\bullet$ & $\bullet$ & $\bullet$ & {} & {} & {} \\	      
{} & {} & {} & {} & {} & {} & {} & {} & {} & {} & {} & {} & {} & {} \\
6  & Narrow \ha{} central wavelength	      
& $\bullet$ & $\bullet$ & $\bullet$ & $\bullet$ & {} & {} & $\bullet$ & $\bullet$ & {} & $\bullet$ & $\bullet$ & {} \\	
7  & Narrow \ha{} Gaussian width	      
& $\bullet$ & $\bullet$ & $\bullet$ & $\bullet$ & {} & {} & $\bullet$ & $\bullet$ & {} & $\bullet$ & $\bullet$ & {} \\	
8  & Narrow \ha{} intensity
& $\bullet$ & $\bullet$ & $\bullet$ & $\bullet$ & {} & {} & $\bullet$ & $\bullet$ & {} & $\bullet$ & $\bullet$ & {} \\	      
{} & {} & {} & {} & {} & {} & {} & {} & {} & {} & {} & {} & {} & {} \\
9  & ${\nii_{6550}}/{\mathrm{Narrow\, H}\,\alpha}$ intensity ratio
& $\bullet$ & $\bullet$ & {} & {} & {} & {} & $\bullet$ & {} & {} & $\bullet$ & {} & {} \\	         
10 & ${\sii_{6718}}/{\sii_{6732}}$ intensity ratio    	
& $\bullet$ & $\bullet$ & {} & {} & {} & {} & $\bullet$ & {} & {} & $\bullet$ & {} & {} \\	   
11 & ${\sii_{6732}}/{\mathrm{Narrow\, H}\,\alpha}$ intensity ratio
& $\bullet$ & $\bullet$ & {} & {} & {} & {} & $\bullet$ & {} & {} & $\bullet$ & {} & {} \\	
12 & ${\oi_{6302}}/{\mathrm{Narrow\, H}\,\alpha}$ intensity ratio	      
& $\bullet$ & $\bullet$ & {} & {} & {} & {} & $\bullet$ & {} & {} & $\bullet$ & {} & {} \\	  
{} & {} & {} & {} & {} & {} & {} & {} & {} & {} & {} & {} & {} & {} \\
13 & Second broad \ha{} component shift with respect to narrow \ha
& {} & {} & {} & {} & {} & {} & $\bullet$ & $\bullet$ & $\bullet$ & {} & {} & {} \\      
14 & Second broad \ha{} component Gaussian width	      
& {} & {} & {} & {} & {} & {} & $\bullet$ & $\bullet$ & $\bullet$ & {} & {} & {} \\	
15 & Second broad \ha{} component intensity     
& {} & {} & {} & {} & {} & {} & $\bullet$ & $\bullet$ & $\bullet$ & {} & {} & {} \\	
{} & {} & {} & {} & {} & {} & {} & {} & {} & {} & {} & {} & {} & {} \\
16 & Disk intensity normalisation
& $\bullet$ & {} & $\bullet$ & {} & $\bullet$ & {} & {} & {} & {} & $\bullet$ & $\bullet$ & $\bullet$ \\      
17 & Disk shift	with respect to 6564.61 \AA	      
& $\bullet$ & {} & $\bullet$ & {} & $\bullet$ & {} & {} & {} & {} & $\bullet$ & $\bullet$ & $\bullet$ \\	
18 & Sine of the disk angle 
& $\bullet$ & {} & $\bullet$ & {} & $\bullet$ & {} & {} & {} & {} & $\bullet$ & $\bullet$ & $\bullet$ \\	
19 & Local velocity dispersion of the disk material 
& $\bullet$ & {} & $\bullet$ & {} & $\bullet$ & {} & {} & {} & {} & $\bullet$ & $\bullet$ & $\bullet$ \\	
20 & Inner disk radius 
& $\bullet$ & {} & $\bullet$ & {} & $\bullet$ & {} & {} & {} & {} & $\bullet$ & $\bullet$ & $\bullet$ \\	
21 & Multiplication factor for disk outer radius 
& $\bullet$ & {} & $\bullet$ & {} & $\bullet$ & {} & {} & {} & {} & $\bullet$ & $\bullet$ & $\bullet$ \\
\hline	
\multicolumn{2}{|c|}{Number of free parameters} & 17 & 11 & 13 & 7 & 10 & 4 & 14 & 10 & 7 & 14 & 10 & 7 \\	
\hline
\end{tabular}
\caption{Summary of the components included in each model. \textit{Columns 1
    and 2}: Index and description of the free parameters;
  \textit{Columns 3 -- 12}: points indicating that a parameter is
  included in a particular model. Figure \ref{fig:paramspace} gives a
  simplified summary of this information.}
\label{tab:models}
\end{center}
\end{minipage}
\end{table*}

\begin{table*}
\begin{minipage}{160mm}
\begin{center}
\begin{tabular}{lp{7cm}p{4.25cm}l}
\hline
{} & Parameter & Range & Log? \\
\hline
\hline
1  & Flat continuum 		      
& $-0.5 \,$ to $\, 0.5 \,\,\times$ \swmc     & No  \\
2  & Broad \ha{} shift with respect to narrow \ha	      
& $0.985 \,$ to $\, 1.015 \, \lambda_{\mathrm{ref}}$                   & No  \\
3  & Broad \ha{} central wavelength         
& $6540 \,$ to $\, 6590$ \AA                & No  \\
4  & Broad \ha{} Lorentzian width 	      
& $20 \,$ to $\, 1000$ \AA             & Yes \\
5  & Broad \ha{} intensity	      
& $0.05 \,$ to $\, 20 \,\,\times$ \swmc            & Yes \\
6  & Narrow \ha{} central wavelength	      
& $6540 \,$ to $\, 6590$ \AA                & No  \\
7  & Narrow \ha{} Gaussian width	      
&	$1 \,$ to $\, 30$ \AA               & Yes \\
8  & Narrow \ha{} intensity	      
& $0.01 \,$ to $\, 10 \,\,\times$ \swmc            & Yes \\
9  & ${\nii_{6550}}/{\mathrm{Narrow\, H}\,\alpha}$ intensity ratio	      
& $0.003 \,$ to $\, 10$                     & Yes \\
10 & ${\sii_{6718}}/{\sii_{6732}}$ intensity ratio    		      
& $0.2 \,$ to $\, 3.0$                      & No  \\
11 & ${\sii_{6732}}/{\mathrm{Narrow\, H}\,\alpha}$ intensity ratio	      
& $0.003 \,$ to $\, 10$                     & Yes \\
12 & ${\oi_{6302}}/{\mathrm{Narrow\, H}\,\alpha}$ intensity ratio	      
& $0.003 \,$ to $\, 10$                     & Yes \\
13 & Second broad \ha{} component shift with respect to narrow \ha	      
& $0.985 \,$ to $\, 1.015 \, \lambda_{\mathrm{ref}}$                  & No  \\
14 & Second broad \ha{} component Gaussian width	      
&	$20  \,$ to $\, 1000$ \AA              & Yes \\
15 & Second broad \ha{} component intensity     
& $0.05 \,$ to $\, 20 \,\,\times$ \swmc            & Yes \\

16 & Disk intensity normalisation	      
& $10 \,$ to $\, 10^5$                    & Yes \\
17 & Disk shift	with respect to 6564.61 \AA	      
& $-100 \,$ to $\, 100$ \AA                 & No  \\
18 & Sine of the disk angle 
& $0 \,$ to $\, 0.9994$                     & No  \\
19 & Local velocity dispersion of the disk material 
& $10^{-3} \,$ to $\, 10^{-2} c$         & Yes \\
20 & Inner disk radius ($R_{\mathrm{inner}}$)
& $100 \,$ to $\, 1000$ R$_G$               & Yes \\
21 & Multiplication factor for disk outer radius 
& $2 \,$ to $\, 100$ $R_{\mathrm{inner}}$ & Yes \\
\hline
\end{tabular}
\caption{Prior ranges of input parameters for the emission
  models. \textit{Columns 1 and 2}: Index and description of the free
  parameters; \textit{Column 3}: Prior ranges of the parameter values;
  \textit{Column 4}: Indication as to whether the coverage of the
  parameter space is logarithmic. \textit{Notes on units}: \mbox{$1
    \mathrm{ W m}^{-3} \equiv 10^{-10} \mathrm{ W m}^{-2}
    $\AA$^{-1}$}.  $\lambda_{\mathrm{ref}}$ is the wavelength of
  narrow \ha{} if present; the wavelength of the Lorentzian broad line
  component for models 5, 6 and 9; and the laboratory wavelength of
  \ha{} in the case of model 12.}
\label{tab:priors}
\end{center}
\end{minipage}
\end{table*}

\subsection{Continuum emission}
\label{sec:continuumemission}

Broadband emission from the central engine contributes a smooth
continuum to the spectrum; this part of the emission was not included
in the models, since it is dominated by the processes occurring
immediately around the black hole, and not the distribution of gas and
dust outside the central region. Instead, it was subtracted from the
spectrum, using either a linear or a quadratic fit made in
\textit{DIPSO}. The necessity of the quadratic term was judged by eye.
In order to allow for a small residual component of continuum
emission, a constant term was included in each of the models.

\subsection{Narrow-line emission}
\label{sec:nlemission}

All narrow lines were modelled with Gaussian distributions. The
relative intensities of some of the forbidden lines were constrained,
either because they are fixed by transition probabilities, or because
they depend on temperature and electron density, which can be assumed
to be approximately constant within the NLR. The \nii{}
and \oi{} emission lines are temperature sensitive, and may depend on
other factors such as reddening; however, since these emission lines
make only a modest contribution to the spectra, the line ratios of
$\oi_{6300}/\oi_{6364}$ and $\nii_{6583}/\nii_{6548}$ were fixed at a
value of 3.0, with reference to \citet{koski78}. The
$\sii_{6716}/\sii_{6731}$ ratio depends both on the square root of the
temperature and on the electron density; this ratio was therefore left
as a free parameter, allowed to vary from 0.2 -- 3.0
\citep{peterson97}.  Reasonable prior ranges of 0.003 - 10 for the
relative intensities of \sii$_{6732}$, \nii$_{6550}$ and \oi$_{6302}$
with respect to narrow \ha{} were estimated from \citet{veilleux87}.

The widths of the observed narrow lines depend upon the effective
spectral resolution. The intrinsic line width due to Doppler
broadening of narrow \ha{} was left as a free parameter in the model,
and assumed to be the same for all narrow lines, since they are all
low-ionisation lines formed in approximately the same region. This
intrinsic width was convolved with the line width due to the spectral
resolution, which is a wavelength-dependent quantity. The spectral
resolution was measured from the night sky OH lines for each waveband,
and was found to be in the range 400 -- 600. In cases where the quasar
did not fill the spectrograph slit, this resolution was scaled down by
the ratio of the seeing (estimated for each spectrum by its spatial
extent) to the slit width.

\subsection{Broad-line emission}
\label{sec:blemission}

Emission from the BLRs was modelled as a Lorentzian line,
which is collisionally broadened with a width proportional to $P/T$,
where $P$ is the pressure and $T$ the temperature. This line profile
was chosen since AGN broad lines have been found observationally to have
broader wings than Gaussians (e.g. \citet{peterson97}), and Lorentzian
profiles fulfil this requirement, whilst being simple and smooth.

One set of the models has an additional broad line component, which
can represent a range of different physical processes, including two
separate BLRs with different temperatures, or outflows
from the central region. A red wing on the emission line may be caused
by an outflow of optically thick clouds, in which case only those on
the far side of the quasar from the central source would be visible
(\citet{capriotti79}, \citet{smith81}). This broad component is
modelled by a Gaussian profile, which arises from Doppler or thermal
broadening, and so has a width proportional to $\sqrt{T}$. The
difference between Lorentzian and Gaussian profiles is minimal near
the line centroid, though the Lorentzian has much broader wings;
combining the two different lines allowed the maximum degree of
flexibility in the two-component BLR models.

\subsection{Accretion disk emission}
\label{sec:diskemission}

The template used for the accretion disk emission was taken from
\citet{chen89}, and describes emission from a geometrically-thin,
optically-thick disk, which is illuminated by a thick, hot inner
disk. The model assumes that the disk is circular and Keplerian. In a
disk such as this, viewed at a non-zero angle to the line of sight,
the receding material is redshifted, and the approaching material
blueshifted; additionally, the blue peak has a higher intensity due to
Doppler boosting. If the accretion disk is viewed face on (at an
inclination angle of zero), there is no velocity difference along the
line of sight, and the disk emission is single-peaked. As the angle
increases, the velocity of the disk material along the line of sight
increases, and the red and blue peaks move further apart.

Following \citet{chen89}, the expression for the disk emission per
unit frequency interval, $F_{\nu}$, is

\begin{equation}
F_{\nu} = K \int_{R_{\mathrm{inner}}}^{R_{\mathrm{outer}}} \int_{-\frac{\pi}{2}}^{\frac{\pi}{2}} \exp{-\frac{(\frac{\nu}{\nu_0} - D)^2}{2 D^2 {\Delta v}^2}} D^3 R^{1 - q} g(D) \ud R \ud \phi
\end{equation}
where 
\begin{equation}
K = \frac{\mathrm{G}^2}{\nu_0 \mathrm{c}^4} \frac{2 \epsilon_0 M^2 \cos{\theta_{\mathrm{disk}}}}{4 \pi d^2} \frac{\Delta v}{(2 \pi)^{\frac{1}{2}}} 
\end{equation}
and
\begin{equation} 
g(D) = 1 + \frac{1}{R} \Big\{ \frac{2 D^2}{D^2 \cos^2\theta_{\mathrm{disk}} + R(D - (1 - \frac{3}{R})^{\frac{1}{2}})^2} - 1 \Big\} 
\label{eq:chen}
\end{equation}
and G is the gravitational constant, c is the speed of light; $M$ is
the black hole mass; $\epsilon_0$ is the normalisation and $q$ is the
radial exponent of the disk emissivity, defined by $\epsilon = (
\epsilon_0 / 4\pi ) R^{-q}$; $d$ is the luminosity distance of the
quasar; $\theta_{\mathrm{disk}}$ is the disk angle, defined as the
angle between the accretion disk rotation axis and the line of sight;
$\nu_0$ is the rest-frame frequency of the line emission; $\Delta v$
is the dimensionless local velocity dispersion of the disk material in
units of c; $R$ is the dimensionless disk radius in units of the
gravitational radius, $\rg$, integrated between characteristic inner
and outer radii of $R_{\mathrm{inner}}$ and $R_{\mathrm{outer}}$;
$\phi$ is the azimuthal angle of the disk to be integrated over (note
that in the weak field approximation, a linear perturbation to the
Special Relativity metric, the photons emitted at $\phi$ make the same
contribution as those emitted at $\pi - \phi$, and therefore the
emission is integrated between $-\frac{\pi}{2}$ and $\frac{\pi}{2}$,
with the contribution to the emission from the back half of the disk
accounted for by a factor of 2 in the normalisation); $\nu_0$ is the
rest frequency and $\nu$ is the observed frequency; and $D =
\frac{\nu}{\nu_e}$, the Doppler factor, where $\nu_e$ is the emission
frequency.

\noindent In the weak field approximation, the Doppler factor is
\begin{equation}
D = \frac{(1 - \frac{3}{R})^{\frac{1}{2}}}{(1 + R^{-\frac{1}{2}} \sin{\theta_{\mathrm{disk}}} \sin{\phi})} .
\end{equation}
Incorporating this into Equation (\ref{eq:chen}), converting from frequency to wavelength, and simplifying:
\begin{eqnarray}
F_{\nu} & = & K \int_{R_{\mathrm{inner}}}^{R_{\mathrm{outer}}} \int_{-\frac{\pi}{2}}^{\frac{\pi}{2}} \exp{-\frac{(\frac{\lambda_0}{\lambda D} - 1)^2 }{2{\Delta v}^2}} D^3 R^{-q} \nonumber \\
& & \bigg[R - 1 + \frac{2}{(1 - \sin^2{\theta_{\mathrm{disk}}} \cos^2{\phi})^2}\bigg] \ud R \ud \phi \,\,.
\label{eq:diskeq}
\end{eqnarray}

The parameters $R_{\mathrm{inner}}$, $R_{\mathrm{outer}}$,
$\theta_{\mathrm{disk}}$ and $\Delta v$ were found from fitting to the
observed line profile, while the normalisation fixes $\epsilon_0
M^2$. The radial exponent of the disk emissivity $q$ can also be
fitted from the line shape; however, to reduce the number of free
parameters and simplify the model, this parameter was fixed at a
fiducial value of $q = 3$ for \ha{} \citep{eracleous03}. This was a
reasonable approximation to make, since the emission line flux
re-radiated by the disk is proportional to the illuminating flux,
which is predicted to vary as $r^{-3}$ for a wide range of radii: the
illuminating flux falls as $r^{-2}$ from the central source, and the
flux falling per radius increment on the disk decreases as $r^{-1}$
due to geometric effects.

The disk emission was calculated computationally using a
multi-dimensional Monte Carlo integration routine,
\textit{gsl\_monte\_vegas}, from the GNU Scientific Library (GSL), for
the grid of input parameters shown in Table \ref{tab:intparams}.

\begin{table}
\begin{center}
\begin{tabular}{p{3.5cm}lp{0.8cm}l}
\hline
Parameter & Range & No. of points & Log? \\
\hline
\hline
Wavelength & 6064 to 7064 \AA & 101 & No \\
Sine of the disk angle & 0 to 0.9994 & 21 & No \\
Velocity dispersion of disk material & $10^{-3}$ to $10^{-2}$ c & 21 & Yes \\
Inner disk radius ($R_{\mathrm{inner}}$) & 100 to 1000 R$_G$ & 21 & Yes \\
\mbox{Multiplication factor for} disk outer radius & 2 to 100 $R_{\mathrm{inner}}$ & 26 & Yes \\
\hline
\end{tabular}
\caption{Ranges of input parameters for the accretion disk model. \textit{Column 1}: Parameter description; \textit{Column 2}: Range of parameter space covered by the models. The outer radius of the accretion disk is defined in terms of a multiplication factor for the inner radius; \textit{Column 3}: Number of points in the array covering the range; \textit{Column 4}: Indication as to whether the coverage of the parameter space is logarithmic.}
\label{tab:intparams}
\end{center}
\end{table}

The wavelength coverage of the disk models is 6064 -- 7064 \AA, chosen
as the \ha{} line emission was observed to be negligible outside this
region for all quasars in the sub-sample. The emission models were
calculated at a resolution of 10 \AA, which is adequate, as the
resolution of the spectra themselves is 12 -- 16 \AA{}pixel$^{-1}$,
and the accretion disk emission is smooth. The sine of the angle of
the disk axis to the line of sight was allowed to vary across all of
the possible parameter range, stopping just short of a value of unity
(an exactly edge-on disk), since the disk has zero thickness in the
model. The local velocity dispersion of the material in the disk
covers the range $10^{-3}$ to $10^{-2}$ c, which are typical values
for velocity dispersion in the BLR. The inner and outer
disk radii were defined in terms of gravitational radii: the inner
disk radius has a logarithmic range, and the outer disk radius was
defined in terms of a multiplication factor for the inner radius.

The disk emission models were checked both by eye and specially
written automated routines, and those models found to have artifacts
due to poor integration were recalculated using a larger number of
steps in the integration routine.  During the Bayesian fitting
process, the disk parameters (sine of the disk angle, $\sin
\theta_{\mathrm{disk}}$, local velocity dispersion of the disk
material, $\Delta v$, and the inner and outer radii,
$R_{\mathrm{inner}}$ and $R_{\mathrm{outer}}$) were allowed to take
continuous values over the prior ranges. The model for each set of
parameters was calculated by interpolating between the sixteen disk
emission templates which bracketed the required parameter values. This
was a reasonable approximation, since the disk emission varies slowly
over each of the parameter ranges.

It should be emphasised that the analysis of this paper is strongly
dependent on the simplicity of the accretion disk model used, and the
choice to fix the radial exponent of the disk emissivity to 3. A
different disk morphology, such as an elliptical or warped disk, would
alter the emission line profiles, and may affect the results.

\section{Parameter Fitting and Model Selection}
\label{sec:fitting}

\subsection{Bayesian fitting}
\label{sec:bayesfit}

The emission models were fit to the spectra following the Bayesian
method, which uses a calculation of the likelihood of the recorded
data arising, given a certain model, in order to find the probability
distribution for each parameter. The Bayesian method is not discussed
in detail here (see \citet{sivia06}), but it should be noted
specifically that in the Bayesian context, the term ``model'' includes
both the equation which describes the fit in terms of the free
parameters, and the prior distributions of those parameters.

The Bayesian fitting was carried out using a ``least squares'' method.
This folds in two important assumptions: first, that the prior is a
fixed value over the entire range; second, that the noise on the data
is well-approximated by a Gaussian distribution. There was limited
prior knowledge available as to the values of the model parameters, so
it was sensible to assign uniform priors over suitable ranges. The
assumption that the noise on the data was Gaussian was a reasonable
one; the noise on each data point was assumed to be Poisson-like sky
emission noise, and since the data values were much larger than the
error bars, this could be approximated as Gaussian noise.

Sky emission dominates the errors in infrared spectroscopy, so the
error, $\sigma_i$, on each data point was approximated as the
Poissonian noise resulting from sky emission, with an unknown scaling
factor:
\begin{equation}
\sigma_i = Q \sqrt{\ps_i} ,
\end{equation}
where $Q$ is the scaling factor, and S$_i$ are the values of the sky
spectrum at each data point. The absolute values of the errors were
not important to the analysis, and so the scaling factor $Q$ was
marginalised. Following the method of \citet{sivia06}, the
normalisation of the error bars was integrated out of the expression
for the likelihood function using a Jeffreys' prior on $Q$, which is
uniform in logarithmic space to encode ignorance as to the magnitude
of the errors.

\subsection{BayeSys3}
\label{sec:bayesys}

The Bayesian optimisation was carried out using \textit{BayeSys3} by
\citet{bayesys}\footnote{The \textit{BayeSys3} program and user guide
  are available at: http://www.inference.phy.cam.ac.uk/bayesys/ },
which explores the parameter space using a range of Monte Carlo
engines. \textit{BayeSys3} was called through a wrapper written by
D. Sivia. For each model, a C program was used to calculate the
likelihood function from each set of parameters provided by
\textit{BayeSys3}.

\textit{BayeSys3} was initialised with an ensemble (i.e. the number of
parallel explorations of the parameter space) of 20; this was
increased from an ensemble of 10 following the discovery of unstable
results with smaller values (see Appendix \ref{sec:stabtests} for an
overview of the tests performed to determine the stability of the
Bayesian fitting process). All available Monte Carlo exploration
engines were switched on, to minimise the risk of probability density
accumulating in local minima. The annealing rate, which controls the
speed at which the simulation switches from exploring the entire
parameter space to exploring the posterior parameter space, was set at
what is suggested to be a reasonably slow value for \textit{BayeSys3}
(0.1 in arbitrary units, see \citet{bayesys}).

\subsection{Model selection}
\label{sec:selection}

The Bayesian evidence is the probability of obtaining a certain data
set given a model, naturally weighted against models with a larger
prior parameter space. Models with unwarranted complexity are
penalised in comparison to simpler models which fit the data equally
well, according to Occam's Razor. The Bayesian evidence values from
the twelve models were compared for each quasar, to find the most
likely model. Since the spectra have different resolutions, the
evidence values of fits to different quasars are not compared.

The Bayes factor $B_{\mathrm{AB}}$ is written
\begin{equation}
B_{\mathrm{AB}} = \frac{\pprob(\pd|\pmm_{\mathrm{A}})}{\pprob(\pd|\pmm_{\mathrm{B}})} ,
\end{equation}
where D is the data and M the model \citep{trotta08}. The Bayes factor
gives a statistical measure of the degree to which Model A has gained
or lost support compared to Model B, given the data. The ``Jeffreys'
scale'' shown in Table \ref{tab:odds} \citep{jeffreys39} provides an
empirical scale for translating the relative Bayesian evidence of two
models into the more intuitive scale of odds, binning this into four
bands of evidence: strong, moderate, weak and inconclusive.

\begin{table}
\begin{center}
\begin{tabular}{cccc}
\hline
$|$ln B$_{\mathrm{AB}}|$ & Odds radio & Probability & Strength of evidence  \\
\hline
\hline
$<$1.0 & $\lesssim$ 3:1 & $<$ 0.750 & Inconclusive \\
1.0 & $\sim$3:1 & 0.750 & Weak evidence \\
2.5 & $\sim$12:1 & 0.923 & Moderate evidence \\
5.0 & $\sim$150:1 & 0.993 & Strong evidence \\
\hline
\end{tabular}
\caption{``Jeffreys' scale'' for comparing the strength of the evidence for two models \citep{jeffreys39}. \textit{Column 1}: Absolute value of the natural logarithm of the Bayes factor, which is a statistical measure of how Model A has increased or decreased in likelihood compared to Model B, given some data; \textit{Column 2}: Odds ratio, an expression of the probability that Model A is true over Model B; \textit{Column 3}: Probability of Model A compared to Model B; \textit{Column 4}: Empirical statement of the strength of the evidence.}
\label{tab:odds}
\end{center}
\end{table}

Based on the Jeffreys' scheme of Table \ref{tab:odds}, a quasar is
considered to have strong evidence for the presence of a disk (denoted
by ``SD'') if the model with the highest evidence is a disk model, and
if no models without disks fall within one ``Jeffreys' criterion''
($\Delta \ln(\mathrm{Evidence}) = 5$) of the preferred model. Moderate
evidence for a disk (MD) and weak evidence for a disk (WD) are defined
analogously, with the corresponding odds ratios. The quasar possesses
a possible disk (PD) if the evidence is inconclusive, or if there is
only weak or moderate evidence against the presence of a disk. The
category of non-disk (ND) is assigned in cases where there is strong
evidence against the presence of a disk, i.e. an accretion disk is
apparently excluded by Jeffreys' criterion.

\begin{figure}
\begin{center}
\includegraphics[width = 0.47\textwidth]{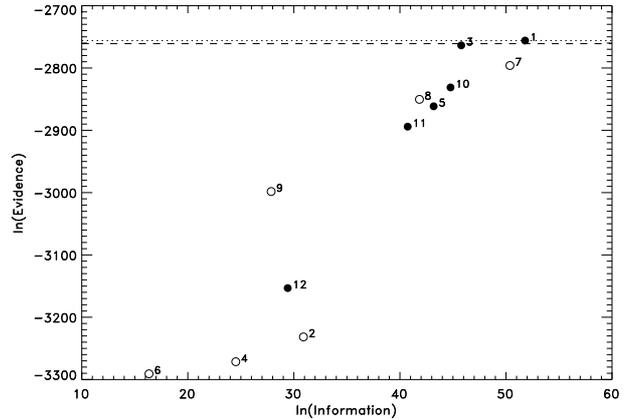}
\caption{Natural logarithmic evidence plotted against natural
  logarithmic information for all model fits to the spectrum of
  MRC0450-221. The dotted line shows the evidence of the preferred
  model, while the dashed line is plotted at an evidence difference of
  one Jeffreys' criterion (i.e. $\Delta \ln(\mathrm{Evidence}) = 5$)
  from the best fit. In this case, Model 1 is selected, with no other
  models falling within the Jeffreys' criterion.}
\label{fig:evinfplots}
\end{center}
\end{figure}

Figure \ref{fig:evinfplots} illustrates the selection procedure with
a plot of the natural logarithmic evidence versus the natural
logarithmic information for one quasar, although it should be noted
that only the Bayesian evidence was used in the selection process. The
quality of the data is a large factor in the model selection. Broadly
speaking, high signal-to-noise spectra have very high ranges of
evidence, as the difference between the best and worst fits is more
apparent than for the lower signal-to-noise spectra, so the model
selection procedure is more conclusive.

The Bayesian information is a measure of the ratio of the volume of
the prior parameter space to the volume of the posterior parameter
space. It therefore provides an indication of how much the data has
increased knowledge of the parameter values, for a given model. There
is a correlation between logarithmic evidence and logarithmic
information for the high signal-to-noise cases: higher evidence is
linked to better fits, which constrain the posterior parameters more
tightly. For low signal-to-noise spectra, however, such as those of
MRC0346-279 or MRC1217-209, there is no correlation apparent between
the evidence and information, and one Jeffreys' criterion can
encompass most of the models, so it is impossible to discriminate
reliably between them; in these cases, accretion disk emission may be
present, but there is no evidence for it.

\subsection{Notes on individual quasars}
\label{sec:selectionnotes}

\noindent\textbf{MRC0222-224}: The model including the Lorentzian
line, accretion disk and all narrow lines (Model 1) was selected with
higher evidence by $\Delta \ln(\mathrm{Evidence}) = 6.4$ than the next
best model (Model 10), which includes the accretion disk and narrow
lines only.

\noindent\textbf{MRC0327-241}: The best-fit model includes the 
Lorentzian line, the accretion disk emission and narrow \ha{} (Model
3). Four other models are within the $\Delta \ln(\mathrm{Evidence}) =
5$ bound of the selected model; however, all these models include disk
emission, so this quasar has strong evidence for disk emission.

\noindent\textbf{MRC0346-279}: This spectrum suffers from low
signal-to-noise, and so there is little evidence to discriminate
between the models: all but one of the models fall within the
Jeffreys' criterion of the best-fit model. Model 9, the fit with the
broad Lorentzian and the broad Gaussian, was selected by the Bayesian
evidence, which is the next simplest model after the single
Lorentzian. The selected model has greater evidence by $\Delta
\ln(\mathrm{Evidence}) = 0.81$ than the next best fit. The low
signal-to-noise ratio is likely to be the reason for the lack of
fitted narrow emission lines in this spectrum.

\noindent\textbf{MRC0413-210}: Model 1, which includes the Lorentzian
broad line, the disk emission and all narrow lines, is preferred, with
$\Delta \ln(\mathrm{Evidence}) = 6.3$ over the next best-fit model.

\noindent\textbf{MRC0413-296}: This quasar is a special case, since
there are obviously narrow lines present in the spectrum, but these
did not constrain properly in the fit as the narrow \ha{} parameters
attempted to fit a broad component of emission. There appears to be
more than two components of broad emission in this spectrum. From the
evidence, the best model is Model 7, which has the broad Lorentzian,
the broad Gaussian and all the narrow lines, although the narrow \ha{}
fitted a third broad component. $\Delta \ln(\mathrm{Evidence}) = 10.8$
between the best fit, and the next best-fit model which includes a
disk, so there is strong evidence against the presence of a disk.

\noindent\textbf{MRC0430-278}: This spectrum has a
reasonably low signal-to-noise ratio. The preferred model is Model 7,
with the Lorentzian broad line plus the Gaussian broad line and all
the narrow lines, though many models fall within the Jeffreys'
criterion. The margin in logarithmic evidence of Model 7 over Model 1,
which includes the accretion disk emission, is only $\Delta
\ln(\mathrm{Evidence}) = 2.2$, so this is a ``possible disk'' quasar.

\noindent\textbf{MRC0437-244}: Model 1, with the Lorentzian broad
line, accretion disk emission and narrow lines, is preferred over the
next best-fit model by $\Delta \ln(\mathrm{Evidence}) = 1.4$, though
the second best-fit model also includes accretion disk emission. There
is strong evidence ($\Delta \ln(\mathrm{Evidence}) = 92.0$) for the
presence of a disk.

\noindent\textbf{MRC0450-221}: The evidence is strong 
($\Delta \ln(\mathrm{Evidence}) = 40.1$) for the presence of an
accretion disk, with the selected best-fit model being Model 1,
including the accretion disk plus Lorentzian line and the full
complement of narrow lines.

\noindent\textbf{MRC0549-213}: Most of the models fall within the 
Jeffreys' criterion of the best-fit model, including all of the models
with an accretion disk plus a standard BLR, so this
quasar is a ``possible disk'' source. The preferred model for this low
signal-to-noise ratio spectrum is Model 6, with one single Lorentzian
line only. This is preferred over a disk emission model by $\Delta
\ln(\mathrm{Evidence}) = 2.3$.

\noindent\textbf{MRC1019-227}: This spectrum contains artifacts from
poor sky subtraction which were excluded from the Bayesian fitting
process, including the region surrounding narrow \ha. As a result, the
sub-set of models with the narrow \ha{} line only were not
constrained, although the models with all narrow lines present could
converge, since the narrow \ha{} parameters were fixed with reference
to the other narrow lines. The best-fit model is Model 1, with the
broad Lorentzian, the accretion disk emission and all the narrow
lines, though this is only favoured over Model 7 (which includes the
Lorentzian, the Gaussian and all the narrow lines) by $\Delta
\ln(\mathrm{Evidence}) = 2.5$, so this source is a ``weak disk''
candidate.

\noindent\textbf{MRC1114-220}: The selected model is Model 1, with 
the Lorentzian line plus accretion disk emission and the full
complement of narrow lines. There is strong evidence for a disk, since
the best-fit model without an accretion disk has lower evidence by
$\Delta \ln(\mathrm{Evidence}) = 7.9$.

\noindent\textbf{MRC1208-277}: There is strong evidence supporting
Model 1, with all narrow lines and the Lorentzian line plus the
accretion disk component. This has higher evidence than the best model
without a disk by $\Delta \ln(\mathrm{Evidence}) = 34.8$.

\noindent\textbf{MRC1217-209}: Model 5, which includes the Lorentzian 
line and the accretion disk, is preferred, but the evidence for this
model is weak. Most of the models fall within the Jeffreys' criterion
of the best-fit model, with an evidence difference between Model 5 and
the best-fit model without a disk of $\Delta \ln(\mathrm{Evidence}) =
1.4$. The lack of fitted narrow lines is likely to be due to the poor
signal-to-noise ratio of the spectrum.

\noindent\textbf{MRC1222-293}: Model 1, which includes the Lorentzian 
line plus the accretion disk emission and the full set of narrow lines,
was selected as the best model with strong evidence ($\Delta
\ln(\mathrm{Evidence}) = 10.8$).

\noindent\textbf{MRC1301-251}: Model 1 (the Lorentzian and
the disk emission plus narrow lines), Model 7 (the Lorentzian and
Gaussian plus narrow lines), and Model 10 (accretion disk emission
plus narrow lines), all have probabilities within the Jeffreys'
criterion of each other and are clearly preferred above the other
models. Model 7 is preferred by a margin of $\Delta
\ln(\mathrm{Evidence}) = 0.5$ over Model 1, so the results are
inconclusive. The wavelength shifts of the two broad components
relative to the narrow \ha{} line for Model 7 are $-4200$ \kms{} for
the Lorentzian line and $+4500$ \kms{} for the Gaussian line; these
are within a plausible range for opposing outflows. It is clear,
however, that more complex broad emission than a single broad line is
required to fit this spectrum.

\noindent\textbf{MRC1349-265}:The preferred fits are those with two 
broad components plus the full set of narrow lines. Of these, Model 7
with the Lorentzian line plus Gaussian line and narrow lines is
preferred, but only with weak to moderate evidence ($\Delta
\ln(\mathrm{Evidence}) = 2.5$) over Model 1, which includes emission
from an accretion disk in addition to the Lorentzian line and narrow
lines.

\noindent\textbf{MRC1355-215}: The preferred models for this spectrum 
are overwhelmingly those with two components of broad emission and
with all the narrow lines. Of these, Model 7, which includes the
Lorentzian line and the Gaussian line in addition to the narrow lines,
is weakly preferred to Model 1, with the Lorentzian line and the
accretion disk emission as well as the narrow lines, by $\Delta
\ln(\mathrm{Evidence}) = 1.61$.

\noindent\textbf{MRC1355-236}: There is strong evidence 
($\Delta \ln(\mathrm{Evidence}) = 6.9$) that Model 1 with the
Lorentzian line, the disk emission, and all the narrow lines is
preferred over the next best model.

\noindent\textbf{MRC1359-281}: There is strong evidence ($\Delta
\ln(\mathrm{Evidence}) = 6.6$) for Model 1 with the Lorentzian line,
accretion disk emission and all narrow lines.

\subsection{Correlations between the parameters}
\label{sec:correlations}

\begin{figure}
\begin{center}
\includegraphics[width = 0.47\textwidth]{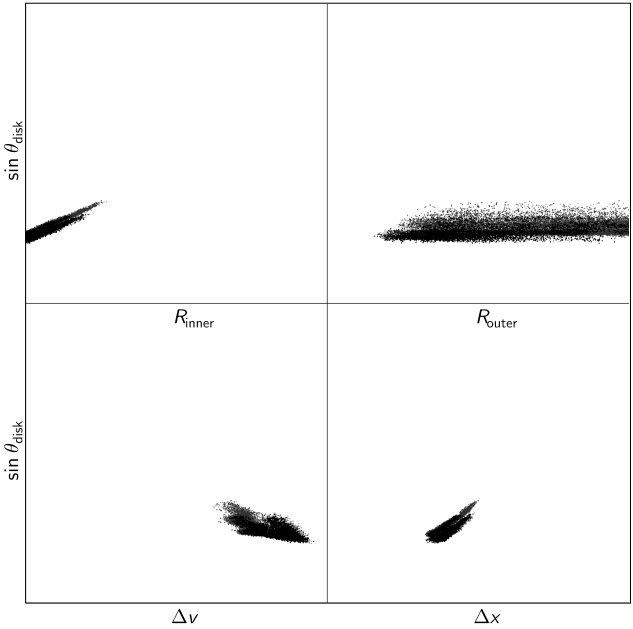}
\caption{A plot of the correlations between the sine of the disk angle
  ($\sin \theta_{\mathrm{disk}}$) and the other accretion disk
  parameters for MRC0450-221: the inner disk radius
  ($R_{\mathrm{inner}}$), the outer disk radius
  ($R_{\mathrm{outer}}$), the local velocity dispersion of the disk
  material ($\Delta v$), and the shift of the disk emission with
  respect to 6564.61 \AA{} ($\Delta x$). There are distributions for
  five different random seeds plotted here: for all seeds, the
  probability distributions converge on similar regions (but see
  discussion in Appendix \ref{sec:stabtests}).}
\label{fig:cosmoseed}
\end{center}
\end{figure}

Figure \ref{fig:cosmoseed} shows the correlations between the
posterior probability for the accretion disk angle and the other disk
fitting parameters of the selected model of the quasar MRC0450-221
(note that there are probability distributions from fits with five
different random seeds marked on this plot). Similar plots were
examined for all fit parameters. There are no strong correlations of
the local velocity dispersion of the disk material or the outer disk
radius with any of the other disk parameters. There is a correlation
such that when the sine of the disk axis angle increases, the inner
radius increases and the wavelength shift of the disk component with
respect to 6564.61 \AA{} increases.

In the model of \citet{chen89} for an optically-emitting accretion
disk, the emission line widths increase with decreasing inner disk
radius. The inner disk radius, which is defined as a dimensionless
quantity in units of the gravitational radius $\rg$, anticorrelates
with the black hole mass. In the model for double-peaked lines, the
line width therefore correlates positively with the black hole
mass. Observationally, the widths of broad, low-ionisation emission
lines are known to correlate with black hole mass
(\citet{vestergaard02}, \citet{mclure02}), which is postulated to be
due to virialisation of the emitting material.

An increase in the disk angle causes the two peaks of the disk
emission to move further apart, as the rotating material in the disk
has a higher velocity component along the line of sight, as explained
in Section \ref{sec:diskemission}. This effect has been observed
observationally by \citet{jarvis06}, who found that using radio
spectral indices as a proxy for source orientation, the sources at
greater angles to the line of sight have larger broad-line widths.

The correlation between the disk angle and the inner radius of the
disk can therefore be interpreted in terms of the observed width of
the broad emission; these two parameters act upon the emission line
width in opposing senses, and so the fitted model is a trade-off
between the two. 

The correlation of the disk angle with the shift of the disk component
is caused by the stronger (blueward) line peak aligning with the
strongest component in the spectrum, whereas the weaker (redward) peak
of the line does not impose such a strong constraint on the fitting;
the wavelength shift required to fit the line profile depends on the
separation between the two peaks, which is most strongly dependent on
the angle of the accretion disk to the line of sight.

The disk angle has by far the greatest effect on the profile of the
emission, since this parameter is most strongly correlated with the
line width and the separation between the red and blue peaks of the
line. The disk angle therefore constrains most strongly of all the
disk fitting parameters, and so under the assumption of this given
model, the double-peaked emission provides a reasonably robust method
for measuring orientation. It should be emphasised that the model and
input parameters chosen here (see Section \ref{sec:diskemission})
affect the range of disk emission profiles available during the
fitting process, and hence may affect both the disk angles fitted and
whether there is Bayesian evidence for a disk.

\section{Results and Discussion}
\label{sec:results}

\subsection{Fits to the emission spectra}
\label{sec:fits}

\begin{figure*}
\begin{minipage}{160mm}
\begin{center}
\includegraphics[width = 1.0\textwidth]{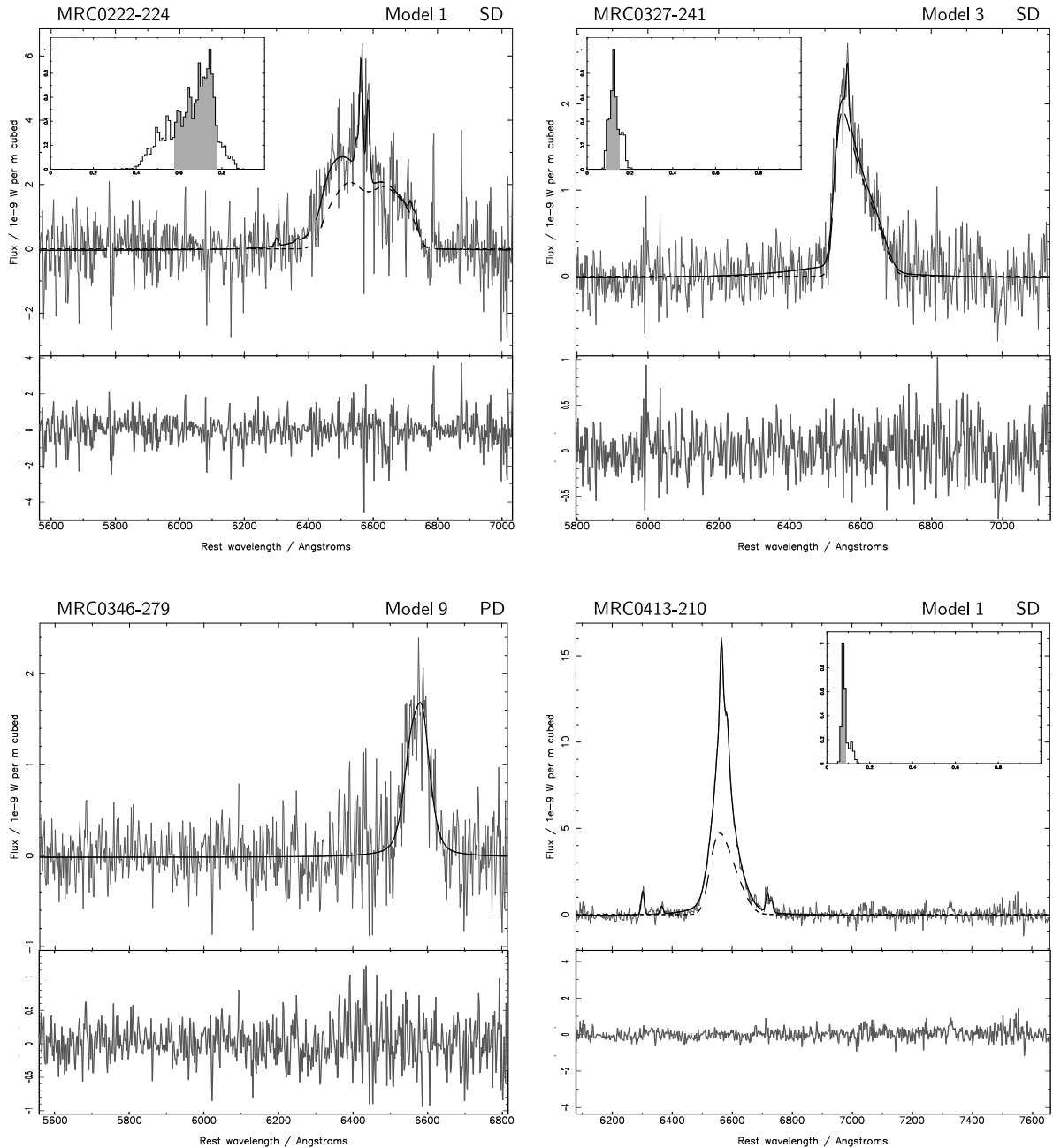}
\caption{The best Bayesian fits to each quasar spectrum, and the
  residuals from the fit. The flux density scale of the residuals is
  the same as for the spectrum in most cases; for a few sources it is
  compressed, but the units remain the same. Wavelengths are
  rest-frame.  The best fit is plotted as a solid black line. In cases
  where a disk is included in the selected model, the disk component
  is shown with a dashed line, and the posterior distribution of the
  sine of the disk angle is shown in an inset. In the inset plots, the
  y-axis shows the probability normalised to unity, and the x-axis
  covers the range $0 < \sin{\theta} < 1$. The shaded area of the
  inset plots shows the $1 \sigma$ error bounds on the posterior
  distribution of the sine of the disk angle. Each plot is labelled
  with the index of the model with which it was fit, and with a code
  according to whether there is evidence for an accretion disk: SD =
  strong evidence for a disk; MD = moderate evidence for a disk; WD =
  weak evidence for a disk; PD = a possible disk (i.e. the evidence is
  inconclusive, or there is only weak or moderate evidence against the
  presence of a disk); ND = strong evidence against the presence of a
  disk.}
\label{fig:diskfits}
\end{center}
\end{minipage}
\end{figure*}

The best-evidence Bayesian fits to the quasar spectra and the
residuals from these fits are shown in Figure \ref{fig:diskfits},
where the fit is marked as a solid black line on the spectrum. In
cases where disk models were selected, the disk component is shown by
a dashed line, and the posterior distribution of the sine of the disk
angle is shown in an inset plot. Each plot is labelled with the model
with which it was fitted, and whether there is strong evidence for an
accretion disk according to the Jeffreys' criterion (SD), moderate
evidence (MD), weak evidence (WD), whether the results are
inconclusive but do not rule out a disk (PD), or whether there is
strong evidence against the presence of a disk (ND) (see Table
\ref{tab:odds}). Figure \ref{fig:diskfitsnoev} shows the best-fit
models from the subset which include a disk, for those quasars whose
selected best-fit models do not include an accretion disk.

\begin{figure*}
\begin{minipage}{160mm}
\begin{center}
\includegraphics[width = 1.0\textwidth]{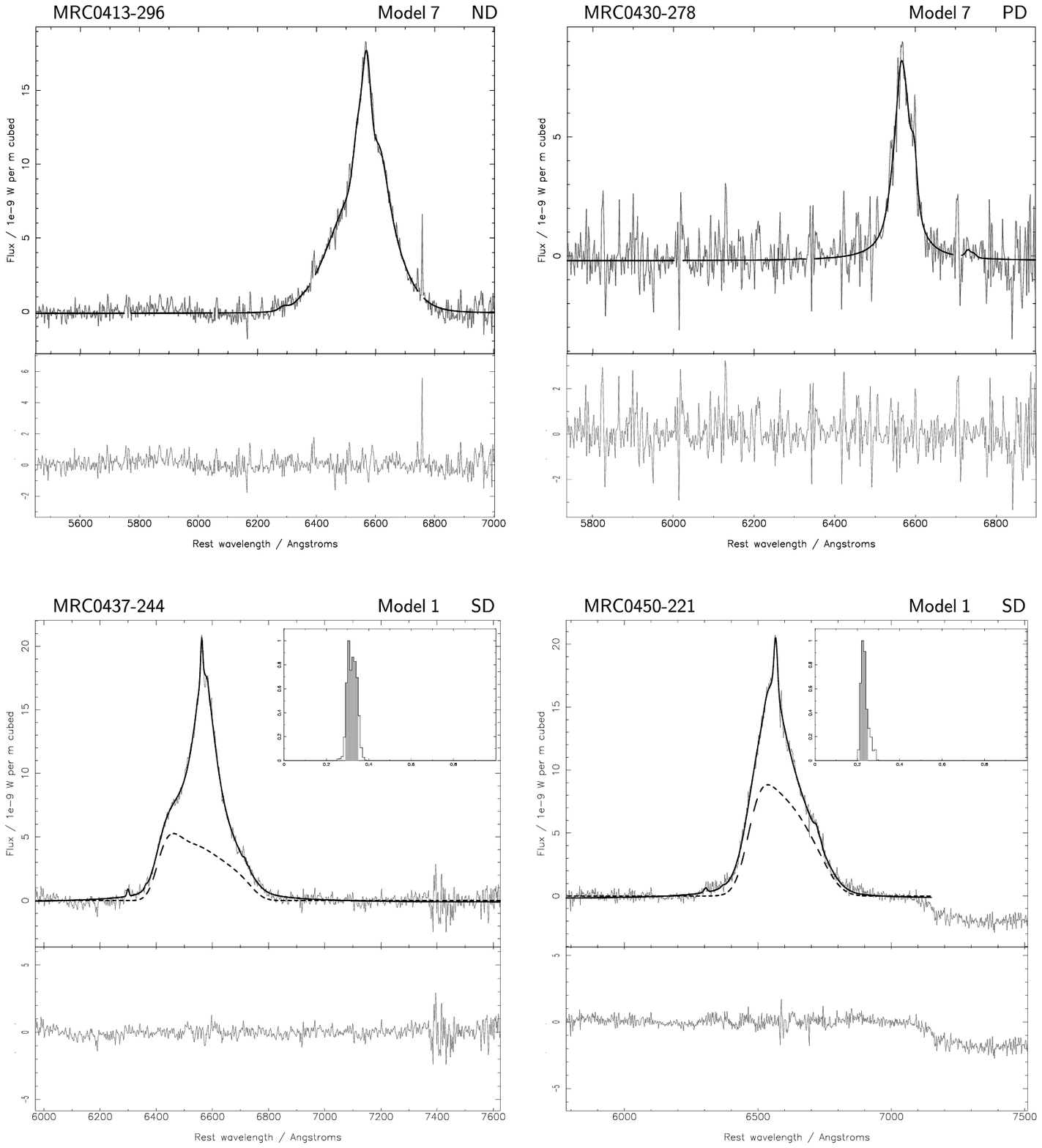}
\end{center}
\end{minipage}
\end{figure*}

\begin{figure*}
\begin{minipage}{160mm}
\begin{center}
\includegraphics[width = 1.0\textwidth]{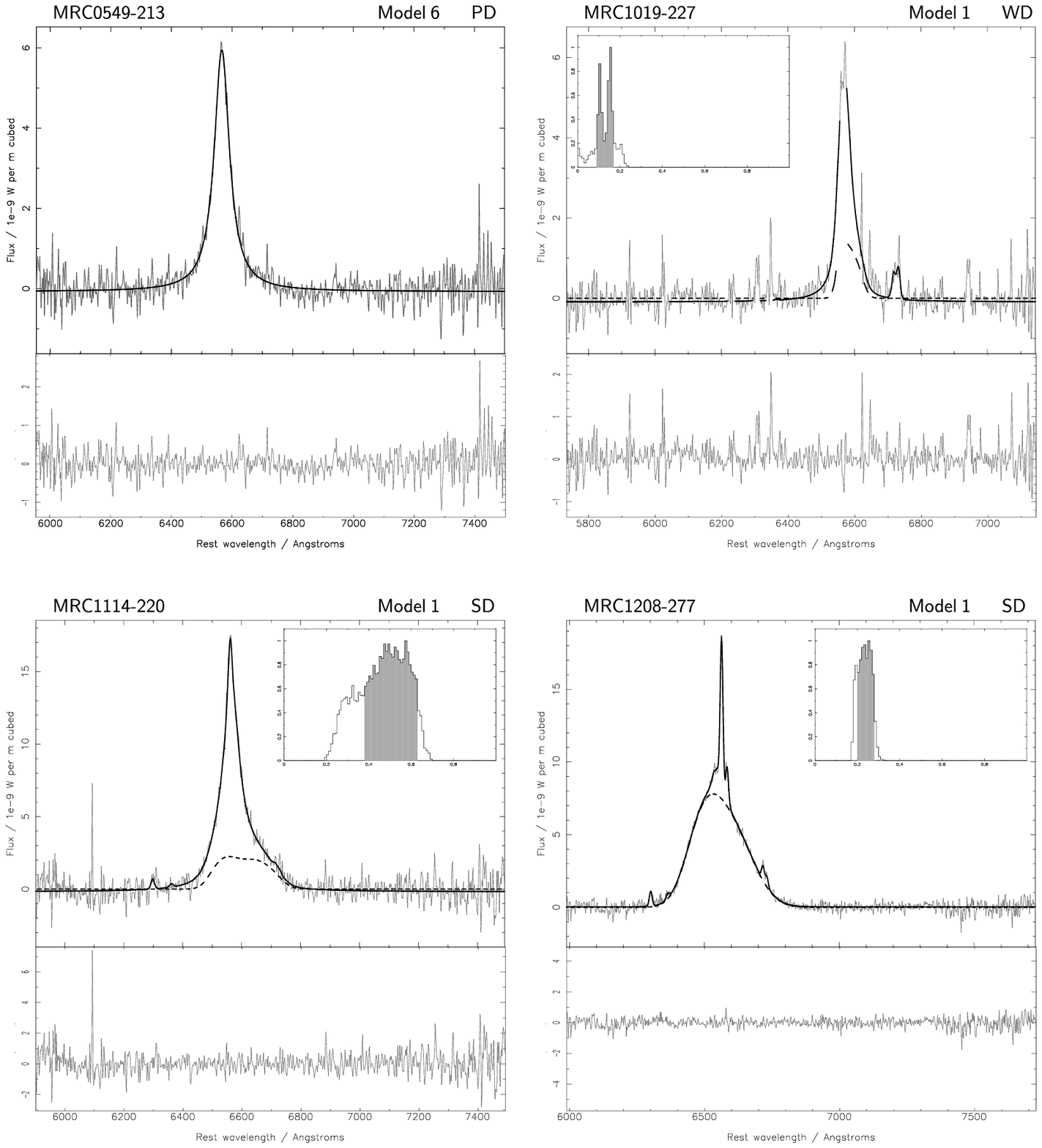}
\end{center}
\end{minipage}
\end{figure*}

\begin{figure*}
\begin{minipage}{160mm}
\begin{center}
\includegraphics[width = 1.0\textwidth]{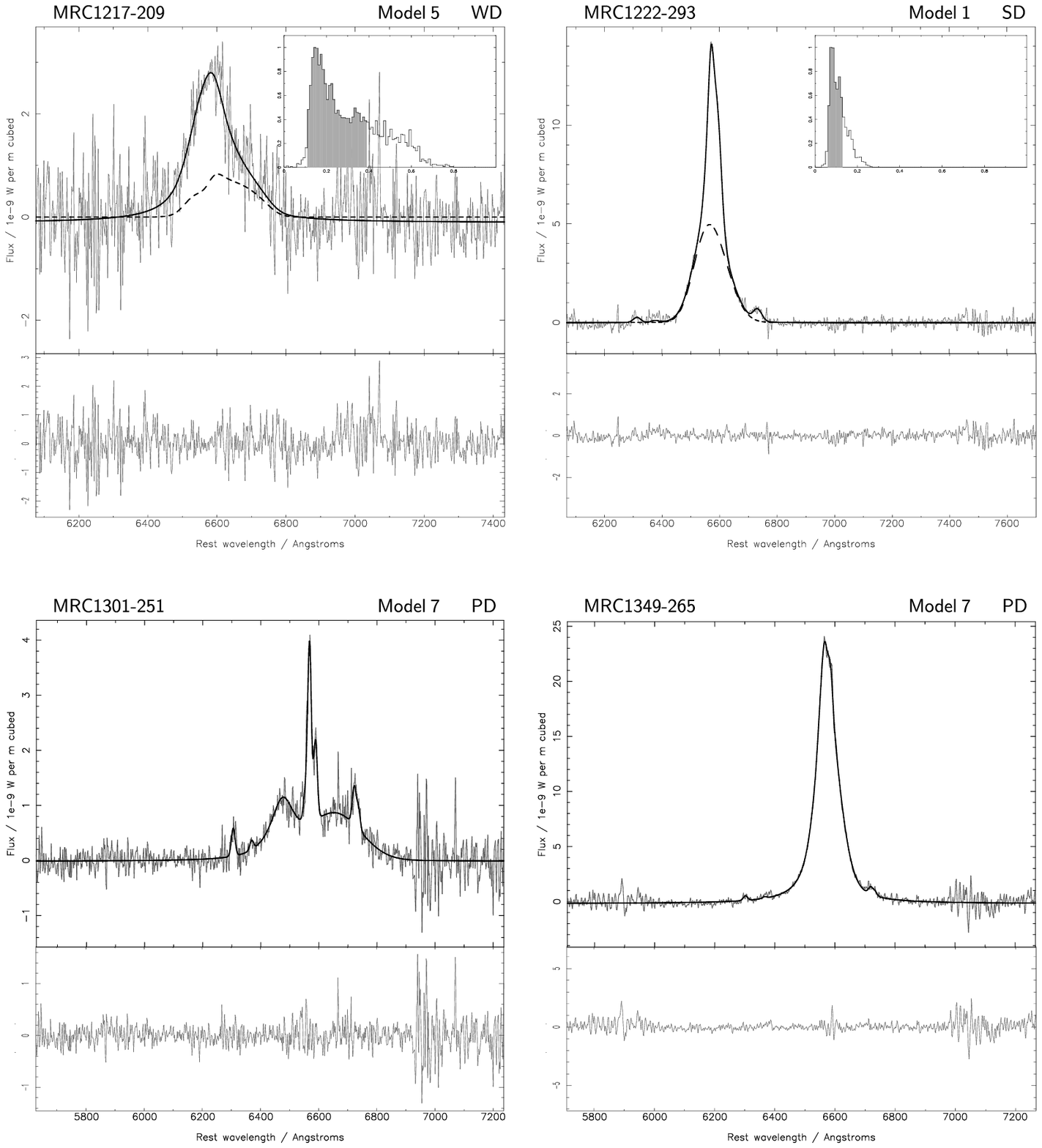}
\end{center}
\end{minipage}
\end{figure*}

\begin{figure*}
\begin{minipage}{160mm}
\begin{center}
\includegraphics[width = 1.0\textwidth]{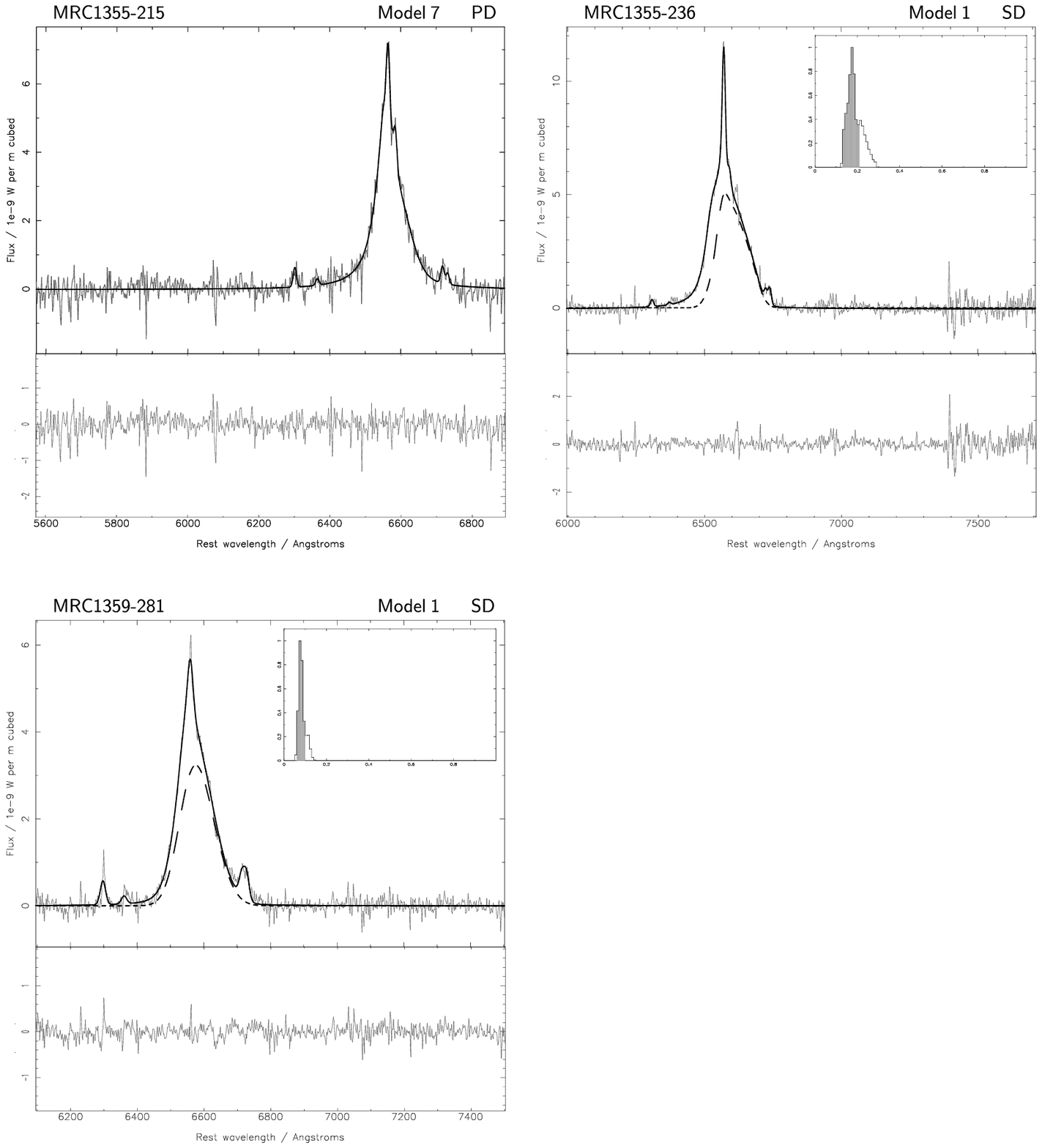}
\end{center}
\end{minipage}
\end{figure*}

\begin{figure*}
\begin{minipage}{160mm}
\begin{center}
\includegraphics[width = 1.0\textwidth]{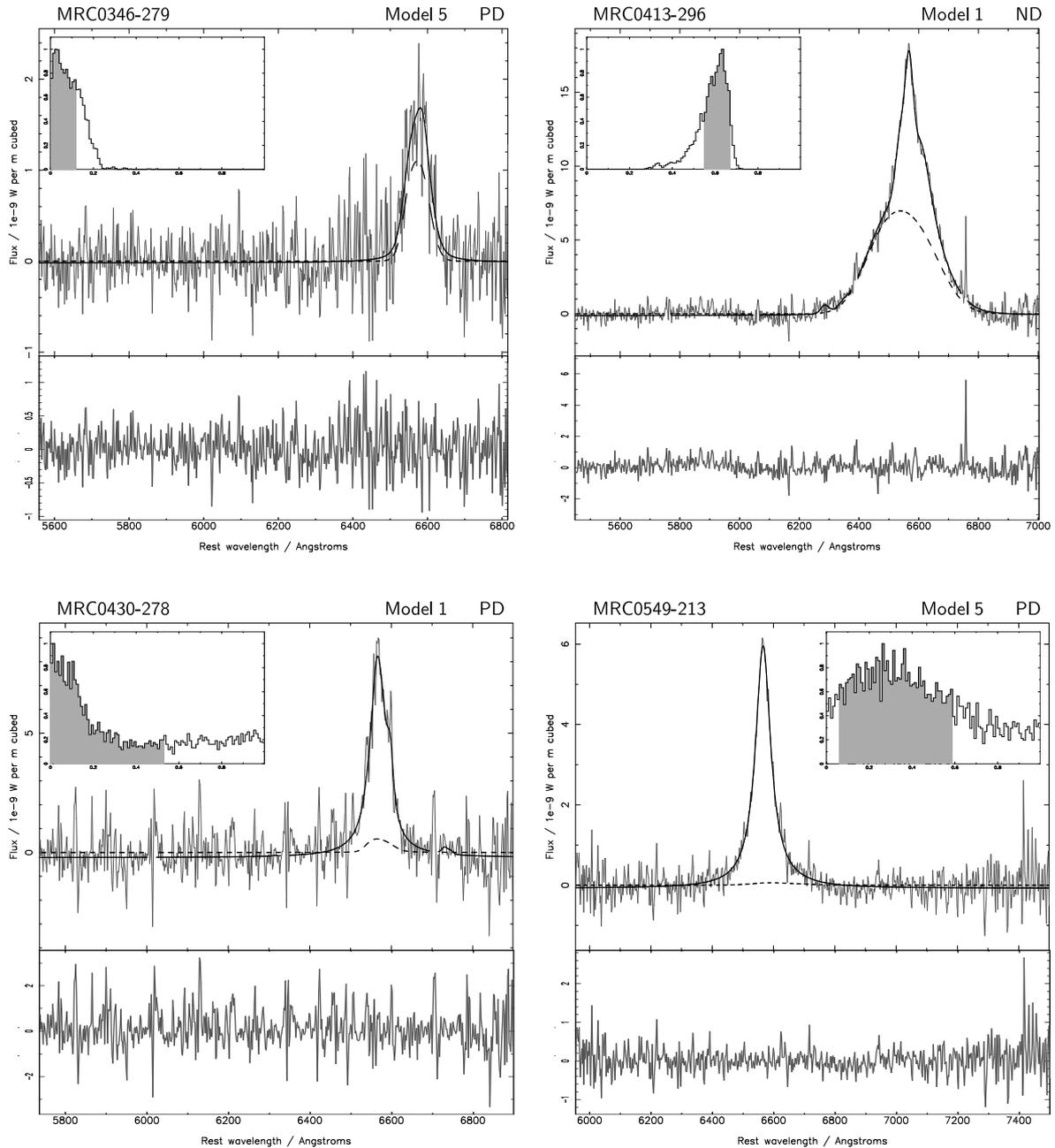}
\caption{The best Bayesian fits which include an accretion disk
  component, for each quasar for which the highest evidence fit did
  not include a disk. Residuals are plotted below each fit. The flux
  density scale of the residuals is the same as for the spectrum in
  most cases; for a few sources it is compressed, but the units remain
  the same. Wavelengths are rest-frame. The best fit is plotted as a
  solid black line. The disk contribution is shown with a dashed line,
  and the posterior distribution of the sine of the disk angle is
  shown in an inset. In the inset plots, the y-axis shows the
  probability normalised to unity, and the x-axis covers the range $0
  < \sin{\theta} < 1$. The shaded area of the inset plots shows the
  $1 \sigma$ error bounds on the posterior distribution of the sine of
  the disk angle. Each plot is labelled with the index of the model
  with which it was fit, and with a code according to whether there is
  evidence for a disk: SD = strong evidence for a disk; MD = moderate
  evidence for a disk; WD = weak evidence for a disk; PD = a possible
  disk (i.e. the evidence is inconclusive, or there is only weak or
  moderate evidence against the presence of a disk); ND = strong
  evidence against the presence of a disk.}
\label{fig:diskfitsnoev}
\end{center}
\end{minipage}
\end{figure*}

\begin{figure*}
\begin{minipage}{160mm}
\begin{center}
\includegraphics[width = 1.0\textwidth]{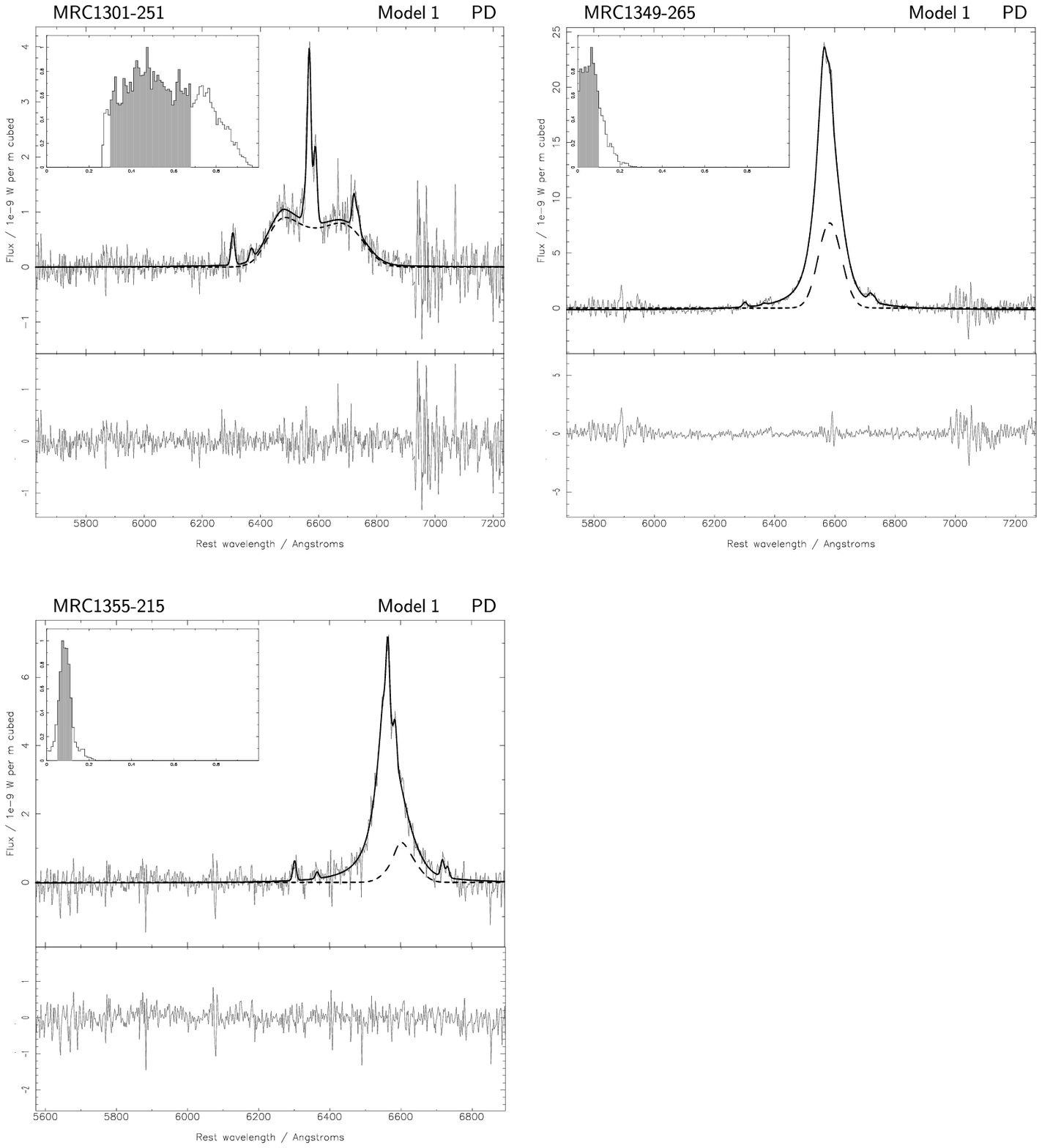}
\end{center}
\end{minipage}
\end{figure*}

Table \ref{tab:diskangles} summarises the results of the model
selection process. Of the nineteen quasars, ten have strong evidence
for disks according to Jeffreys' criterion; two have weak evidence for
a disk; six have possible disks, which means that either the results
are inconclusive, or that there is weak or moderate evidence against
a disk; and in only one case is there strong evidence against the
presence of a disk.

The exceptional case, MRC0413-296, shows strong Bayesian evidence that
there is no accretion disk emission in the spectrum. However, the
best-fit model (Model 7, which includes the Lorentzian and Gaussian
broad lines, in addition to both narrow \ha{} and the forbidden narrow
lines) did not fit in the expected manner: the width of the narrow
\ha{} emission was unconstrained in the fit, and this line was fitting
to a third broad component. This quasar appears to be anomalous within
the sub-sample in that two broad components are not sufficient to
describe the emission, and therefore it is entirely plausible that
this source requires a model not tested here, such as a
three-component model including emission from an accretion disk, a
BLR which gives rise to single-peaked lines, and an outflow.

It is very notable that for all but two sources (PD cases MRC0346-279
and MRC0549-213), the selected models include a complex BLR
of more than one component. The most basic model of a single emitting
region is not adequate to describe the complex profiles of the
majority of these lines. In most of the cases, the preferred models
were the ones with all the narrow lines and either the Lorentzian
broad line plus the accretion disk (Model 1), or the Lorentzian line
plus the broad Gaussian line (Model 7). In only four fits were models
with less than the full complement of narrow lines preferred, and of
these, three were spectra with low signal-to-noise ratios
(MRC0327-241, MRC0346-279 and MRC1217-209).  MRC0549-213 does appear
to be well-fitted with a single broad Lorentzian line, with the
possible presence of weak forbidden narrow lines, but very weak or
absent narrow \ha.

None of the models with the accretion disk emission only were
selected; in most cases, these models have extremely low
evidence. There is certainly a component of the BLR which gives rise
to single-peaked lines present.

The fitted posterior probability distributions for the sine of the
disk angle, shown as insets in Figures \ref{fig:diskfits} and
\ref{fig:diskfitsnoev}, are on the whole reasonably well-constrained
with slightly asymmetric distributions, though in some cases, these
are cut off by the zero-angle prior boundary. MRC1019-227 has a double
peak in the posterior probability distribution, though the two peaks
are closely spaced. MRC1217-209 has a poorly-constrained distribution,
due to the low signal-to-noise ratio of the spectrum. The MRC0430-278
spectrum has a low signal-to-noise ratio, and the fitted posterior
probability distribution for the sine of the disk angle is therefore
less tightly constrained in one region, although it is strongly peaked
at low inclinations. MRC0549-213 has a posterior probability
distribution for the sine of the disk angle which extends over the
entire range. This source appears to have weak or absent disk
emission, as it is well-fit by a Lorentzian line, but is classed as a
PD source, as there is no strong evidence against a disk.

\begin{table*}
\begin{minipage}{160mm}
\begin{center}
\begin{tabular}{lllllllll}
\hline
\multicolumn{2}{|l|}{}       & Best-fit  & Evidence & \multicolumn{2}{|l|}{For best-fit model} & \multicolumn{2}{|l|}{For best-fit model} & {} \\ 
\multicolumn{2}{|l|}{Quasar} & model   & for disk      & \multicolumn{2}{|l|}{with disk}    & \multicolumn{2}{|l|}{without disk} & Disk angle \\ 
\multicolumn{2}{|l|}{}       & {} & {}     & $\Delta \ln E^{2}_{1\mathrm{d}}$ & $\Delta \ln E^{\mathrm{n}}_{1\mathrm{d}}$ & $\Delta \ln E^{2}_{1\mathrm{n}}$ & $\Delta \ln E^{\mathrm{d}}_{1\mathrm{n}}$ & {}  \\
\hline
\hline
1  & MRC0222-224 & 1 & 	SD & 6.43  & 12.50 &      & 	 & 48 $\degree\pm {}^{2}_{12}$ \\	
2  & MRC0327-241 & 3 & 	SD & 0.36  & 5.48  &      & 	 &  7 $\degree\pm {}^{1}_{2}$ \\	
3  & MRC0346-279 & 9 (5) & PD &    &       & 0.80 & 1.91 &  1 $\degree\pm {}^{5}_{1}$ \\ 	
4  & MRC0413-210 & 1 & 	SD & 6.30  & --    &      & 	 &  4 $\degree\pm {}^{1}_{1}$ \\	
5  & MRC0413-296 & 7 (1) & ND &    &       & 10.79& --   & 39 $\degree\pm {}^{3}_{6}$ \\	
6  & MRC0430-278 & 7 (1) & PD &    &       & 1.40 & 2.17 &  1 $\degree\pm {}^{31}_{1}$ \\      
7  & MRC0437-244 & 1 & 	SD & 1.44  & 91.96 &      & 	 & 18 $\degree\pm {}^{2}_{1}$ \\ 	
8  & MRC0450-221 & 1 & 	SD & 7.74  & 40.07 &      & 	 & 13 $\degree\pm {}^{1}_{1}$ \\ 	
9  & MRC0549-213 & 6 (5) & PD &    &       & 0.90 & 2.34 & 15 $\degree\pm {}^{20}_{12}$ \\      
10 & MRC1019-227 & 1 &  WD & 2.48  & --    &      &      &  9 $\degree\pm {}^{1}_{4}$ \\	
11 & MRC1114-220 & 1 & 	SD & 4.40  & 7.93  &      & 	 & 35 $\degree\pm {}^{4}_{12}$ \\	
12 & MRC1208-277 & 1 & 	SD & 28.41 & 34.83 &      & 	 & 15 $\degree\pm {}^{1}_{3}$ \\	
13 & MRC1217-209 & 5 & 	WD & 0.64  & 1.37  &      & 	 &  8 $\degree\pm {}^{14}_{2}$ \\      
14 & MRC1222-293 & 1 & 	SD & 10.84 & --    &      & 	 &  4 $\degree\pm {}^{3}_{1}$ \\	
15 & MRC1301-251 & 7 (1) & PD &	   &       & 0.54 & --   & 28 $\degree\pm {}^{14}_{11}$ \\	
16 & MRC1349-265 & 7 (1) & PD &    &       & 2.50 & --	 &  4 $\degree\pm {}^{2}_{3}$ \\	
17 & MRC1355-215 & 7 (1) & PD &    &       & 1.61 & --   &  4 $\degree\pm {}^{2}_{1}$ \\	
18 & MRC1355-236 & 1 & 	SD & 6.93  & --    &      & 	 & 10 $\degree\pm {}^{2}_{2}$ \\	
19 & MRC1359-281 & 1 &  SD & 6.64  & --    &      &      &  4 $\degree\pm {}^{1}_{1}$ \\ 	
\hline
\end{tabular}
\caption{Summary of the model fitting results. \textit{Columns 1 and
    2}: Quasar index and MRC name; \textit{Column 3}: Index of the
  best-fit model. If the selected model does not include an accretion
  disk, the index of the highest-evidence model which includes an
  accretion disk is shown in brackets.  \textit{Column 4}: Indication
  of whether there is evidence for a disk according to the odds ratio,
  with codes as follows: SD = strong evidence for a disk; MD =
  moderate evidence for a disk; WD = weak evidence for a disk; PD = a
  possible disk, i.e. the best fit is a non-disk model, but there is
  only inconclusive, weak or moderate evidence for this; ND = strong
  evidence against the presence of a disk; \textit{Columns 5 and 6}:
  Natural logarithmic evidence difference between the best-fit model,
  in cases where this includes an accretion disk, and the second best
  model (\textit{Column 5}, $\Delta \ln E^{2}_{1\mathrm{d}}$), and in
  cases where the second best fit also includes a disk, between the
  best-fit model and the best-fit model without a disk (\textit{Column
    6}, $\Delta \ln E^{\mathrm{n}}_{1\mathrm{d}}$); \textit{Columns 7
    and 8}: Natural logarithmic evidence difference between the
  best-fit model, in cases where this does not include an accretion
  disk, and the second best model (\textit{Column 7}, $\Delta \ln
  E^{2}_{1\mathrm{n}}$), and in cases where the second best fit also
  does not include a disk, between the best-fit model and the best-fit
  model with a disk (\textit{Column 8}, $\Delta \ln
  E^{\mathrm{d}}_{1\mathrm{n}}$); \textit{Column 9}: Best-fit angle of
  the disk and the error in degrees.}
\label{tab:diskangles}
\end{center}
\end{minipage}
\end{table*}

The disks have fitted rotation axis angles between 1$\degree${} and
48$\degree$ to the line of sight; this range is consistent with the
definition of quasars as being objects viewed within the opening angle
of a dusty torus, where the opening angle is dependent on source
luminosity but is generally supposed to be roughly 45$\degree$ for
radio-luminous AGN \citep{lawrence91}.

Those sources with disk rotation axes at small angles to the line of
sight in general have less strong evidence for a disk. The reason for
this is that when the disk axis is at a small angle to the line of
sight, the disk emission is not distinctive and double-peaked, but
instead, rather similar to a Gaussian profile (see Section
\ref{sec:correlations}). In these cases, there is little to
distinguish the Lorentzian plus accretion disk model for the broad
emission from the Lorentzian plus Gaussian model, except that the fit
with the Gaussian has fewer parameters, and is therefore likely to
have a smaller prior parameter space and hence be favoured by Occam's
Razor. It is only those fits to sources with greater disk angles, or
those with very high signal-to-noise ratio spectra, that make it
possible to detect disk emission with high probability.

\begin{figure}
\begin{center}
\includegraphics[width = 0.5\textwidth]{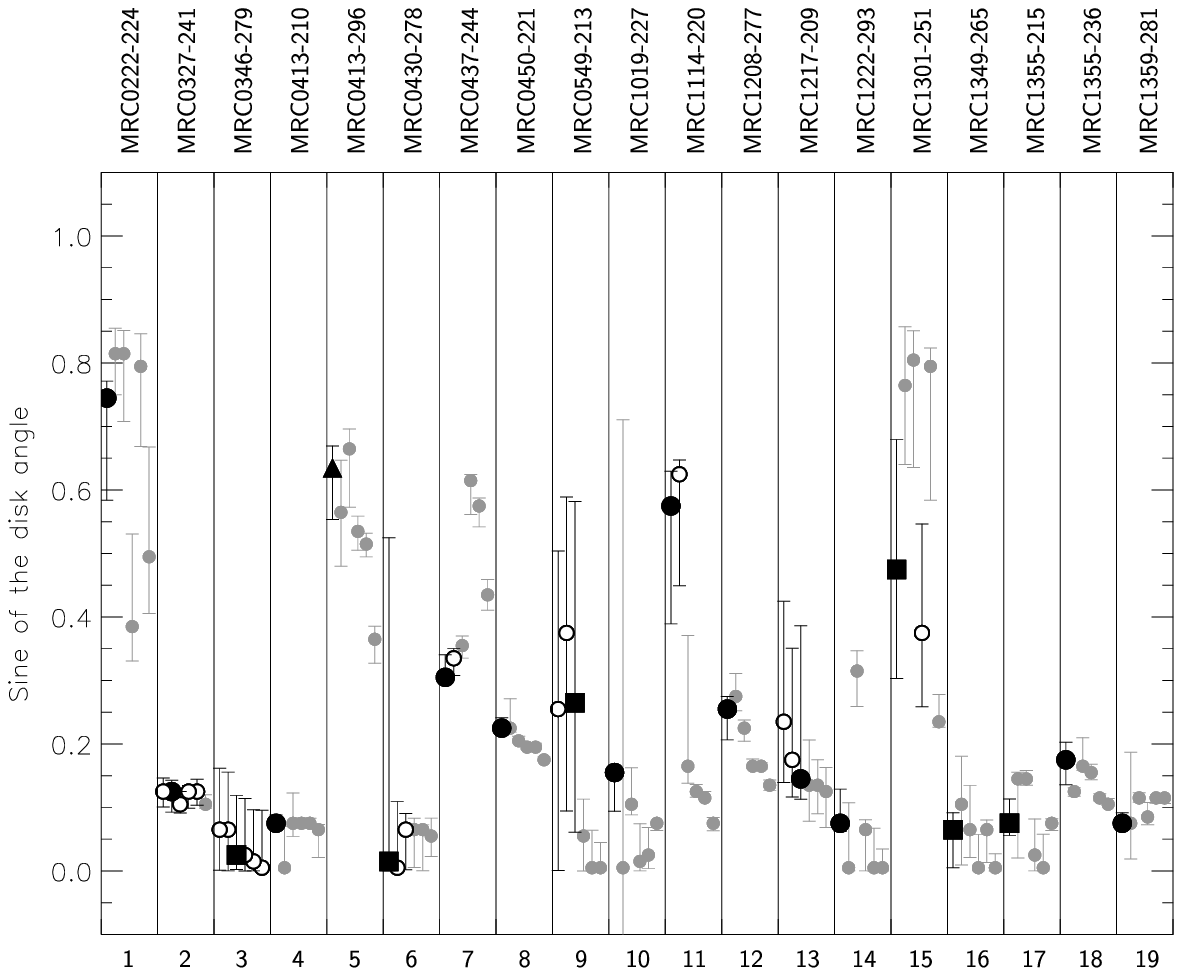}
\caption{Disk angles for model fits with an accretion disk. Circular
  symbols indicate evidence for a disk, square symbols indicate a
  possible disk (inconclusive evidence, or weak to moderate evidence
  against a disk), while triangular symbols indicate strong evidence
  against the presence of a disk. White points mark fits which fall
  within one Jeffreys' criterion of the best fit, and these can be
  seen in all cases to be consistent with the best-fit disk angle. The
  remainder of the fitted disk angles are plotted in pale grey. }
\label{fig:plotallangles}
\end{center}
\end{figure}

Figure \ref{fig:plotallangles} shows the disk angles from all model
fits for each quasar. In most cases, the angles are extremely
stable. In cases where there are differences in the inferred angles,
all angles from models within one Jeffreys' criterion of the best fit
are consistent with each other.

\subsection{Relationships with projected radio source size}
\label{sec:sizes}

\begin{figure}
\begin{center}
\includegraphics[width = 0.5\textwidth]{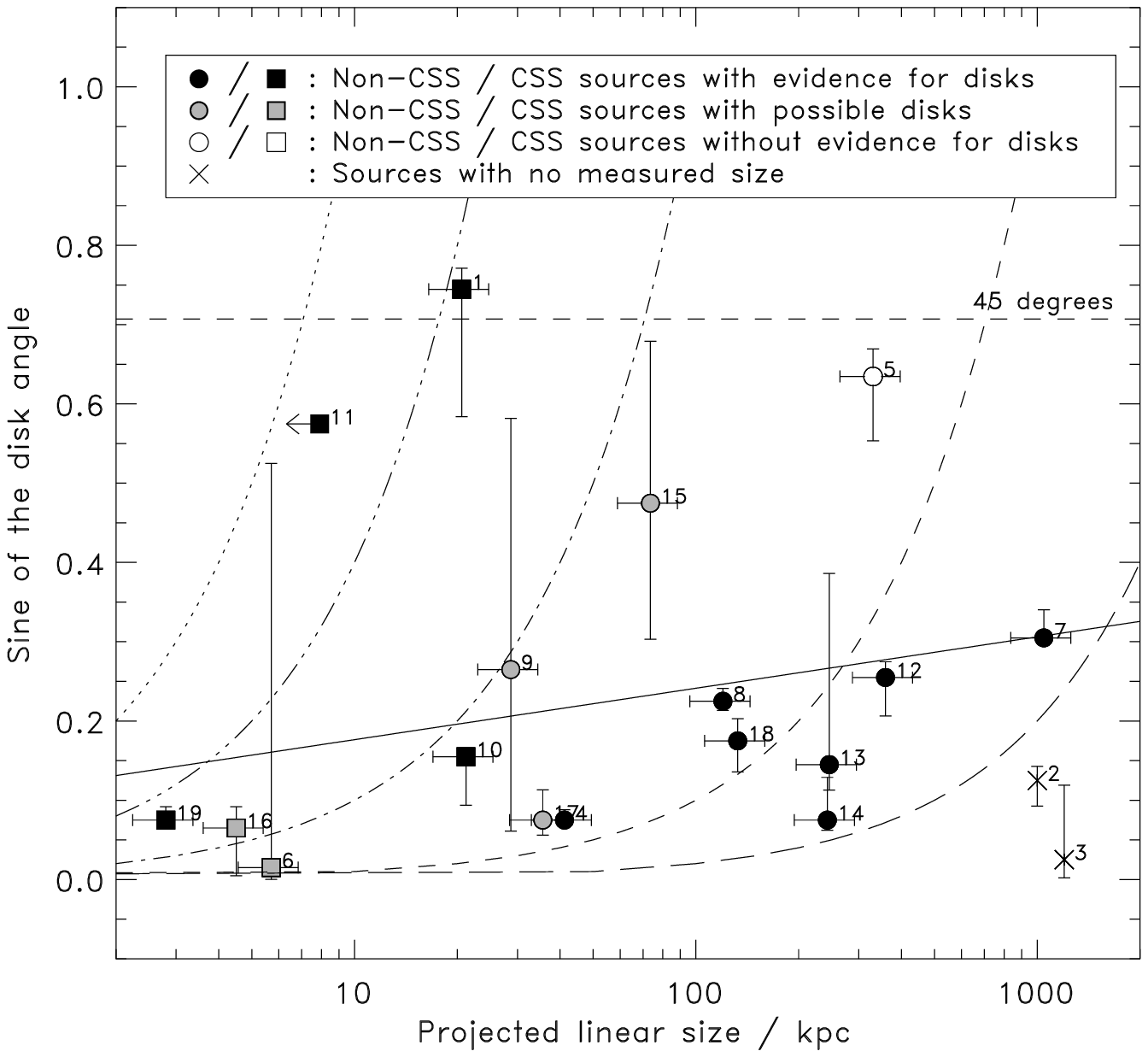}
\caption{Sine best-fit disk angles versus the projected linear size of
  the quasar in kpc. The disk angle error bars include both the errors
  returned from the individual fits and a systematic error calculated
  from the variation in results from using different random seeds to
  initialise the Monte Carlo engines, added in quadrature, but do not
  take into account any error due to model selection. The non-CSS FRII
  sources are plotted with circles, and CSS sources are plotted with
  squares. The two sources plotted with crosses are core-dominated
  sources without measured projected sizes. The black points have weak
  to strong evidence for a disk, grey points have possible disks, and
  the white point has Bayesian evidence against the presence of a
  disk. Index numbers are matched to source names in Table
  \ref{tab:obsdata}. The black line shows the best linear fit to all
  sources excluding the core-dominated sources. There is a highly
  tentative correlation (Kendall $\tau$ coefficient of 0.40,
  significant at the $1 \sigma$ level) between the sine of the disk
  angle and the projected linear size for all sources excluding the
  core-dominated sources, and for the 11 non-CSS FRII sources (Kendall
  $\tau$ coefficient of 0.44). The dotted, dot-dot-dot-dashed,
  dot-dashed, dashed and long-dashed lines show loci of constant
  deprojected size, for 10 kpc, 25 kpc, 100 kpc, 1 Mpc and 5 Mpc
  respectively. }
\label{fig:plotsinivsd}
\end{center}
\end{figure}

The sine best-fit disk angles for the Molonglo quasars are plotted against
the projected linear sizes of these sources in Figure
\ref{fig:plotsinivsd}. The source sizes are taken from
\cite{molonglo3}, or in some cases, new sizes were obtained from
$\sim$ 1.4 GHz MERLIN radio maps (Down et al., in preparation): this
data is summarised in Table \ref{tab:deproj}. The sample can be
divided into three source types on the basis of previous studies:
core-dominated sources, Compact Steep Spectrum (CSS) sources, and
non-CSS FRIIs. Two of the 19 sources are core-dominated, which are
sources whose jets are oriented at small angles to the line of sight,
and therefore have core-to-lobe radio flux density ratios of greater
than unity; examination of their radio-frequency spectral energy
distributions (Down et al., in preparation) reveals that these quasars
have been boosted into this sample by virtue of their strong
cores. The remainder of the sample is somewhat arbitrarily divided
into six Compact Steep Spectrum sources, with projected sizes of less
than 25 kpc \citep{odea98} and 11 non-CSS FRII sources, with projected
sizes of greater than 25 kpc.

There is a highly tentative correlation with a Kendall $\tau$
coefficient 0.44, probability of 65\%, and significance of $<1
\sigma$, calculated from \textit{ASURV} \citep{asurv}, that for the
eleven non-CSS FRII sources, the quasars with largest projected size
have greater disk angles to the line of sight, as expected
statistically from foreshortening if the accretion disk is
perpendicular to the radio jet. The probability of a correlation is
strengthened slightly to a Kendall $\tau$ coefficient of 0.40 with
probability of 74\% and $1 \sigma$ significance if the six CSS sources
are included in the sample. The weakness of the correlation is likely
to be due to a large intrinsic scatter in the source size.

There are four CSS sources with fitted disk angles less than
$10\degree$ that appear to fall on the same relation as the non-CSS
sources. It is probable that most or all of these are from the same
population as the non-CSS sources, the only difference being projected
size. The remaining two CSS sources, MRC0222-224 and MRC1114-220, have
strong evidence for disks inclined at angles greater than $30\degree$
to the line of sight.

The radio map for MRC0222-224 (see \citet{molonglo3}) shows two radio
lobes with no hint of a core (a possible weak core is present between
these lobes in a higher resolution MERLIN map, Down et al., in
preparation); this quasar is consistent with an intrinsically small
source viewed at a large angle to the accretion disk axis. The Balmer
decrement for this quasar is estimated as \mbox{\ha $/$ \hb{} $\sim$
  23}, more than a factor of two higher than for any of the other
sources in this Molonglo sub-sample (Janssens et al., in
preparation). A simple interpretation is that this quasar is a young
source, possibly surrounded by a cocoon of dust which reddens the
optical emission \citep{baker02}.

The radio map of MRC1114-220 \citep{silva1} reveals a strong radio
core and single-sided jet, indicating that this source probably lies
at a small angle to the line of sight, so that it is unlikely to be as
intrinsically small as it appears. Possible explanations are that the
accretion disk and radio jet are misaligned in this source following a
merger event, or that the jet is precessing; however, this quasar
merits further investigation. 

The two quasars excluded from the sub-sample, MRC0418-288 and
MRC1256-243, have small projected sizes. MRC0418-288 is a CSS source
with projected size $D < 16.6$ kpc, so is likely to be an
intrinsically small, reddened source similar to
MRC0222-224. MRC1256-243 has a larger size of $D \sim 65$ kpc and a
high core-to-lobe flux density ratio at 10 GHz of $R_{10 \mathrm{GHz}}
\sim 3$, so is a core-dominated source: this is likely to have a small
measured disk angle.


\subsection{Deprojected source sizes}
\label{sec:deprojsizes}

\begin{figure}
\begin{center}
\includegraphics[width = 0.5\textwidth]{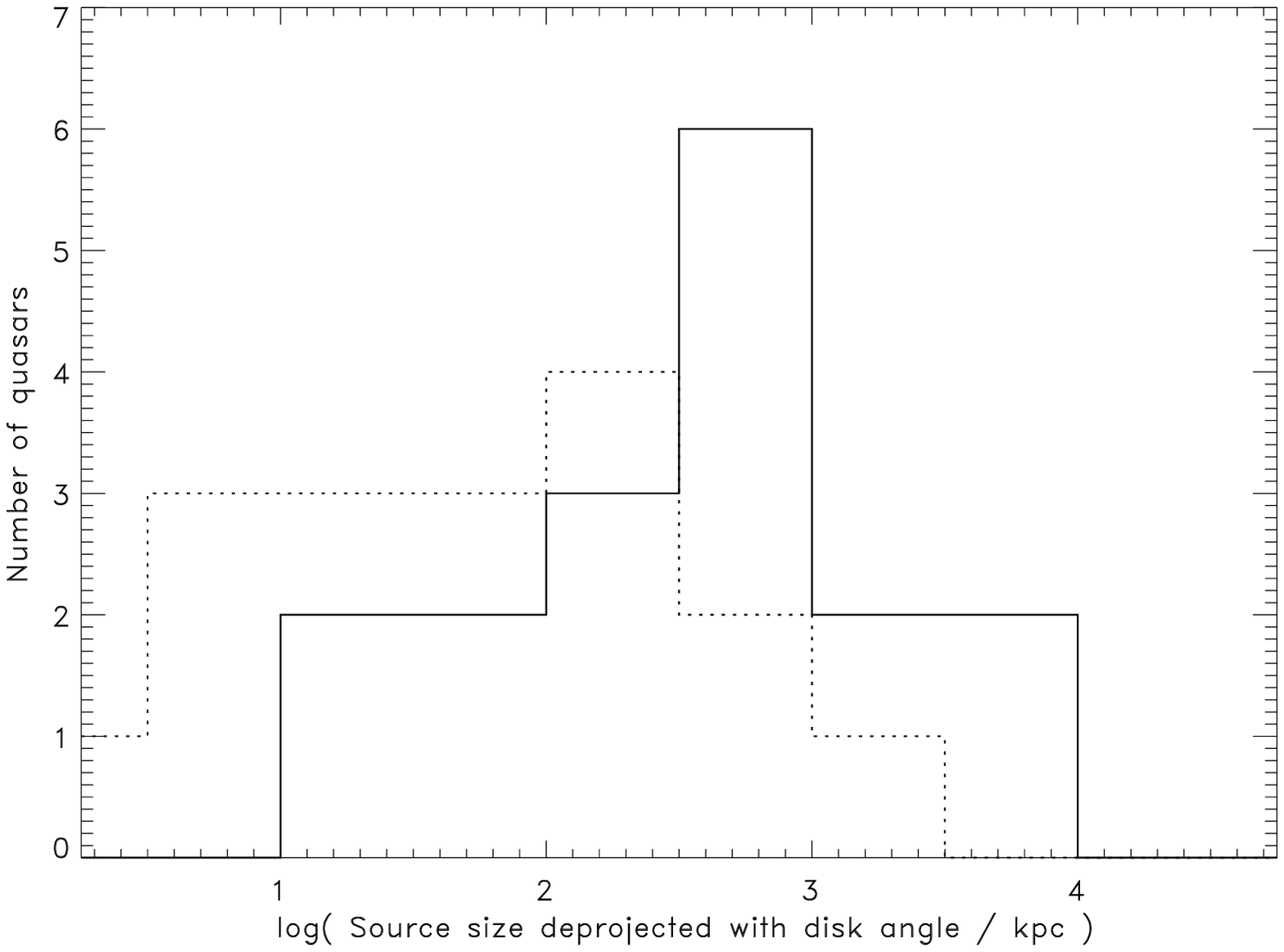}
\caption{The solid line shows source sizes deprojected with the fitted
  disk angles (see Table \ref{tab:deproj}). The projected sizes are
  plotted with a dotted line. Note that the two core-dominated sources
  have no measured size and are therefore excluded.}
\label{fig:histsize}
\end{center}
\end{figure}

The source sizes were deprojected by dividing the apparent source
sizes by the sine of the fitted disk angle, to compensate for simple
geometric projection. This does not take into account the expansion of
the source, but since the hotspots of jets are only expected to
advance at $\sim 0.1$ c \citep{longair79}, this effect is small and is
not considered. The deprojected sizes are given in Table
\ref{tab:deproj}, and are plotted in Figure \ref{fig:histsize}.

\begin{table}
\begin{center}
\begin{tabular}{llllll}
\hline
\multicolumn{2}{c}{}       & {}   & Proj. & Origin & Deproj.     \\
\multicolumn{2}{c}{Quasar} & Type & source size  & of       & source      \\
\multicolumn{2}{c}{}       & {}   & $l$ (kpc) & $l$      & size (kpc)  \\
\hline
\hline
 1 & MRC0222-224 & CSS  & 20.6  & 1 & $ 28\pm^{8}_{1}           $ \\ 
 2 & MRC0327-241 & CD   & --    & {} &  --                       \\
 3 & MRC0346-279 & CD   & --    & {} &  --                       \\ 
 4 & MRC0413-210 & FRII & 41.2  & 2 & $ 550\pm^{70}_{80}        $ \\ 
 5 & MRC0413-296 & FRII & 330.4 & 2 & $ 520\pm^{80}_{30}        $ \\ 
 6 & MRC0430-278 & CSS  & 5.7   & 1 & $ 380\pm^{109000}_{370}   $ \\ 
 7 & MRC0437-244 & FRII & 1045.9& 2 & $ 3430\pm^{100}_{360}      $ \\ 
 8 & MRC0450-221 & FRII & 120.0 & 2 & $ 530\pm^{30}_{40}        $ \\ 
 9 & MRC0549-213 & FRII & 28.7  & 2 & $ 110\pm^{360}_{60}       $ \\ 
10 & MRC1019-227 & CSS  & 21.2  & 1 & $ 140\pm^{90}_{10}         $ \\ 
11 & MRC1114-220 & CSS  & $<$ 7.9& 2 & $ <14\pm^{7}_{1}           $ \\ 
12 & MRC1208-277 & FRII & 359.3 & 2 & $ 1410\pm^{330}_{100}     $ \\ 
13 & MRC1217-209 & FRII & 245.9 & 2 & $ 1700\pm^{480}_{1060}    $ \\ 
14 & MRC1222-293 & FRII & 242.7 & 2 & $ 3240\pm^{660}_{1350}    $ \\ 
15 & MRC1301-251 & FRII & 73.6  & 2 & $ 160\pm^{90}_{50}        $ \\ 
16 & MRC1349-265 & CSS  & 4.5   & 1 & $ 69\pm^{860}_{20}        $ \\ 
17 & MRC1355-215 & FRII & 35.6  & 2 & $ 480\pm^{160}_{160}      $ \\ 
18 & MRC1355-236 & FRII & 132.6 & 2 & $ 760\pm^{220}_{100}      $ \\ 
19 & MRC1359-281 & CSS  & 2.8   & 1 & $ 37\pm^{5}_{7}           $ \\ 
\hline
\end{tabular}
\caption{Summary of projected and deprojected source
  sizes. \textit{Columns 1 and 2}: Quasar index and MRC name;
  \textit{Column 3}: Source type -- CD indicates sources with
  core-to-lobe radio flux ratio at 10 GHz of greater than 1, CSS
  indicates sources with $\alpha_{\mathrm{opt}} > 0.5$ and projected
  source sizes $<$ 25 kpc, FRII indicates non-CSS FRII, which
  encompasses all other sources; \textit{Column 4}: Projected source
  sizes in kpc; \textit{Column 5}: Origin of projected source size
  measurement -- 1 = New measurement (Down et al., in preparation), 2
  = \citet{molonglo3}; \textit{Column 6}: Deprojected source sizes in
  kpc calculated using the fitted disk angles.}
\label{tab:deproj}
\end{center}
\end{table}

The distribution of deprojected source sizes found using the fitted
disk angles is a single-peaked distribution, with the main
concentration of sources in the range 100 kpc -- 1 Mpc. Four of the
CSS sources, MRC0222-224, MRC1114-220, MRC1349-265 and MRC1359-281,
have deprojected sizes of less than 100 kpc. As the projected source
sizes are defined as the distance between the centres of the furthest
separated radio components (following \citet{silva1}), the size may be
underestimated if the whole source is not observed, i.e. in the case
of MRC1114-220, only the core and approaching jet are visible, and so
the size may be underestimated by a factor of $\sim 2$. There are
hints that deprojection slightly tightens the distribution in source
size as expected, but there is still a large scatter.

\begin{figure}
\begin{center}
\includegraphics[width = 0.5\textwidth]{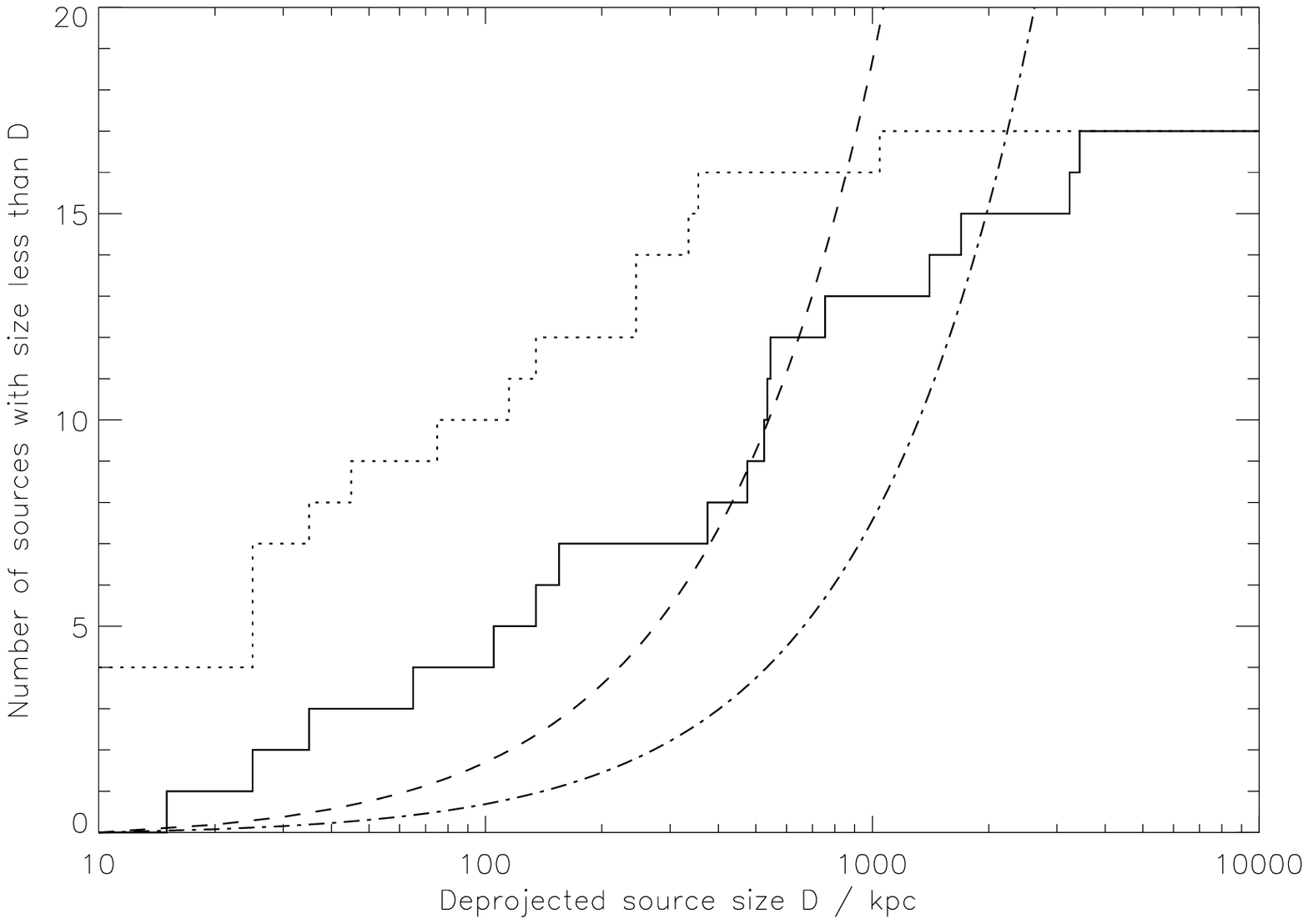}
\caption{Cumulative distribution of source sizes deprojected using the
  best-fit disk angles (solid line), with the projected source sizes
  shown for comparison (dotted line). The curves show the predicted
  distribution if the heads of the sources are expanding at a constant
  rate; the dot-dashed line is normalised at the deprojected size of
  the largest source in the sample, the dashed line is normalised at 1
  Mpc. Note that the two core-dominated sources have no measured
  size, and are boosted into this sample by strong core
  emission, and are therefore excluded from this plot.}
\label{fig:dhistsize}
\end{center}
\end{figure}

The cumulative distribution of the deprojected source sizes found
using the best-fit disk angles is shown in Figure \ref{fig:dhistsize},
with expected distributions of linear sizes, assuming that the
hotspots of the lobes are expanding at a constant rate, shown for
comparison. If the radio jet and accretion disk axes coincide, the
calculated deprojected sizes are broadly consistent with a constant
expansion of the heads of the jets up to $\sim 1$ Mpc, although there
seems to be an excess of small sources.  The distribution drops off at
sizes greater than $\sim 1$ Mpc, which can be explained by the duty
cycle of the quasars: the number of sources larger than a certain
cut-off value will be depleted as they become quiescent
\citep{bird08}. The distribution tails off more gradually than the
highly simplified model, which can be explained by variation in
hotspot advance speeds between quasars, some sources having an
unusually long duty cycle, or by the largest sources being in
especially low-density environments. The excluded sources, MRC0418-288
and MRC1256-243, are predicted to have very small and very large
deprojected sizes respectively, so will not affect the overall
distribution much.

\subsection{Relationships with radio luminosity}
\label{sec:luminosity}

\begin{figure}
\begin{center}
\includegraphics[width = 0.5\textwidth]{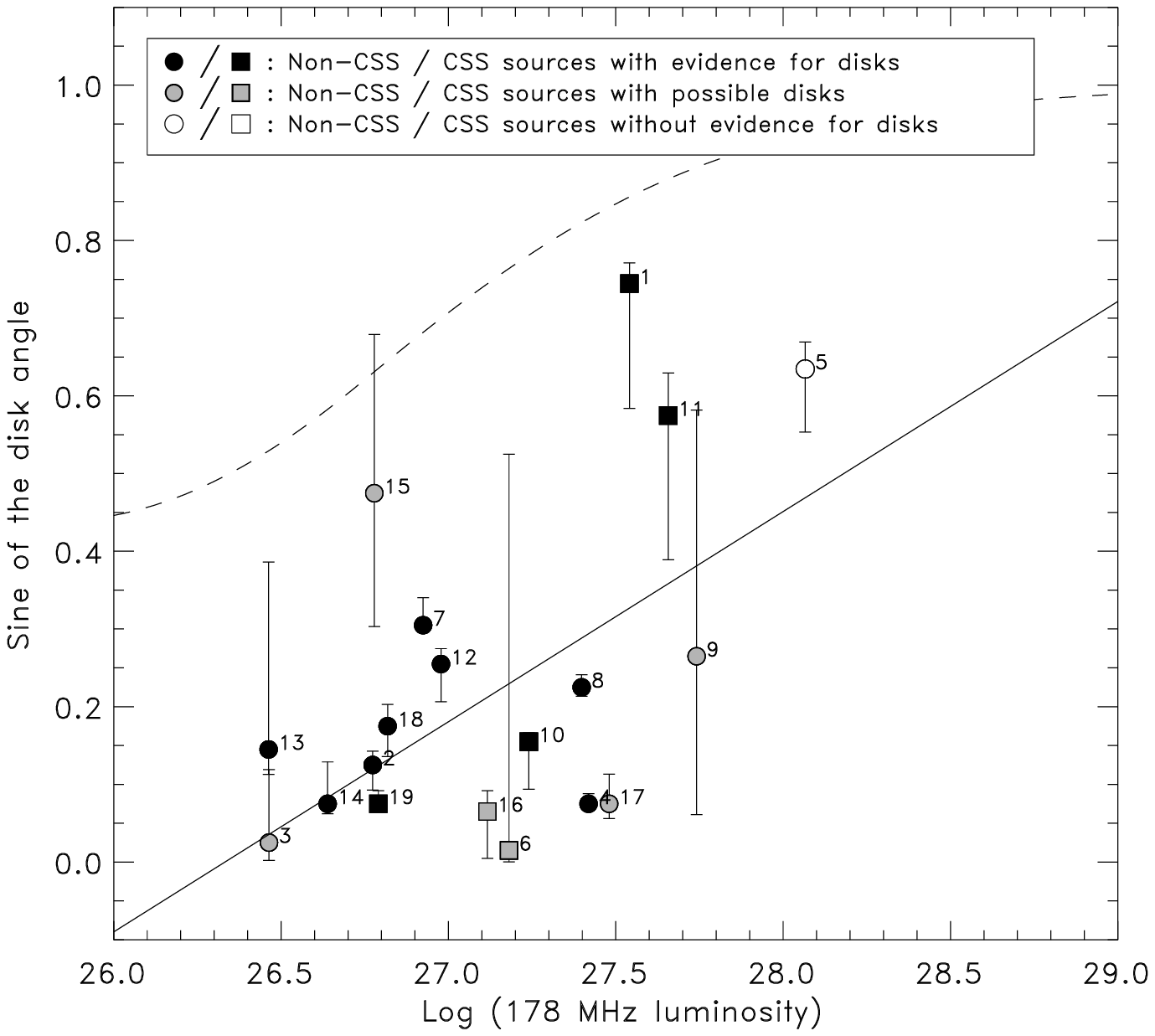}
\caption{Sine best-fit disk angles versus logarithmic 178 MHz
  luminosity. The disk angle error bars include both the errors
  returned from the individual fits and a systematic error calculated
  from the variation in results from using different random seeds to
  initialise the Monte Carlo engines, added in quadrature, but do not
  take into account any error due to model selection. The non-CSS FRII
  sources and core-dominated sources are plotted as circles, and CSS
  sources are plotted as squares. The black points have weak to strong
  evidence for a disk, the grey points have possible disks (there is
  inconclusive evidence, or weak to moderate evidence against a disk)
  and the white points have strong evidence against a disk. Index
  numbers are matched to source names in Table \ref{tab:obsdata}. The
  best-fit line (minimising $\chi^2$) is shown in black. The dashed
  line shows the radio luminosity-dependent critical angle (following
  \citet{willott00}) with fiducial angle $45\degree$ at \loglum = 27,
  and minimum quasar fraction 10\%.}
\label{fig:plotsinivslum}
\end{center}
\end{figure}

The sine of the best-fit disk angles are plotted against the 178 MHz
radio luminosity in Figure \ref{fig:plotsinivslum}. There is a
correlation between these quantities, with a Kendall $\tau$
coefficient of 0.60, probability of 93\% and $1.8 \sigma$
significance. The measurement of the disk angle is entirely
independent of the radio luminosity, so the fact that these parameters
are correlated provides direct, albeit weak, evidence for the receding
torus model \citep{lawrence91}, if the accretion disk axes align with
the radio jets. According to this theory, sources with high radio
luminosity have larger torus opening angles due to dust sublimation
\citep{simpson98}, and therefore higher luminosity sources appear as
quasars, rather than radio galaxies, up to a greater viewing angle
(the critical angle). The prediction from this model is therefore that
more luminous quasars have jet angles ranging up to a higher cut-off
value, and if the disk angle is assumed to match the jet angle, the
calculated disk angles support this prediction.

A calculation of the torus opening angle, following the model of
\citet{willott00}, normalised by a critical angle of $45\degree$ at
\mbox{\loglum = 27}, and modified by a minimum quasar fraction of 10\%
which dominates the opening angle at low radio luminosities
(e.g. \citet{vardoulaki08}), is also marked on Figure
\ref{fig:plotsinivslum}. The measured disk angles all fall within the
calculated envelope of opening angles, and so assuming that the
obscuring tori are aligned with the accretion disks, these results are
in accordance with the receding torus model.

The missing source MRC1256-243 has luminosity $\log L_{178
  \mathrm{MHz}} \sim 27.4$, which is consistent with a low disk
angle. MRC0418-288 has luminosity $\log L_{178 \mathrm{MHz}} \sim
26.8$, which is a relatively low radio luminosity for the predicted
large disk angle; however, an angle of $\sim 30\degree$ would still be
consistent with the receding torus scheme.

\subsection{Distribution of angles}
\label{sec:angdist}

The expected distribution of jet angles ($\theta$) can be modelled for
the Molonglo quasar sample, including a Doppler-boosted core
component. From Bayes' Theorem
\begin{equation}
\pprob(\theta | L_{\mathrm{tot}} > L_{\mathrm{min}}) \propto \pprob(L_{\mathrm{tot}} > L_{\mathrm{min}} | \theta) \times \pprob(\theta)
\label{eq:peq}
\end{equation}
where $L_{\mathrm{tot}}$ is the total luminosity of the quasar and $L_{\mathrm{min}}$ is the limiting luminosity of the survey, and
\begin{equation}
\pprob(\theta) = 2 \pi \sin \theta \ud \theta \,\,.
\label{eq:peven}
\end{equation}
The proportionality is required because there is no solid
information about the overall number of sources for which $L_{\mathrm{tot}} > L_{\mathrm{min}}$.

Marginalising the lobe luminosity, then
\begin{eqnarray}
\pprob(L_{\mathrm{tot}} > L_{\mathrm{min}} | \theta) & \propto &  
\int_{0}^{\infty} \pprob(L_{\mathrm{tot}} > L_{\mathrm{min}} | \theta, L_{\mathrm{lobe}}) \nonumber \\
{} & {} & \times \pprob(L_{\mathrm{lobe}}) \ud L_{\mathrm{lobe}} \,\,.
\label{eq:probeq}
\end{eqnarray}

From \citet{willott01}, the radio luminosity function of high
luminosity AGN is 
\begin{equation}
\rho(L) = \rho_{\mathrm{norm}} \Big(\frac{L}{L_{\mathrm{norm}}}\Big)^{-\alpha} ,
\label{eq:willott01}
\end{equation}
where $\rho$ is the comoving space density of sources in logarithmic
luminosity space and $\alpha \sim 2.3$. Since this relation was found
for luminosities of 151 MHz and 178 MHz, then the total luminosity is
low enough in frequency to be approximated as the lobe
luminosity. Then
\begin{eqnarray}
\pprob(L_{\mathrm{lobe}}) & \propto & L_{\mathrm{lobe}}^{-\alpha} \times \frac{\ud \log L_{\mathrm{lobe}}}{L_{\mathrm{lobe}}} \nonumber \\
{} & \propto & L_{\mathrm{lobe}}^{-\alpha - 1} \,\,.
\label{eq:pllobe}
\end{eqnarray}

\noindent Defining the core-to-lobe flux ratio $R$ in the same way as \citet{jackson99}, then
\begin{equation}
R = R_c \frac{1}{\gamma^2} \left( \frac{1}{(1 + \beta
  \cos\theta)^2} + \frac{1}{(1 - \beta \cos\theta)^2} \right) \,\,,
\label{eq:rlana}
\end{equation}
where $R_c$ is some fiducial value of $R$, and is found by \citet{jackson99} to be $R_c \sim 0.01$ for FRII sources; $\gamma$ is the Lorentz factor and $\beta$ is the velocity of the jet in units of c. The total flux density, $S_{\mathrm{tot}}$, is then defined as 
\begin{equation}
S_{\mathrm{tot}} = S_{\mathrm{lobe}} (1 + R) \,\,,
\label{eq:stot}
\end{equation}
and since flux density is proportional to luminosity, then
\begin{equation}
L_{\mathrm{tot}} = L_{\mathrm{lobe}} (1 + R) \,\,.
\label{eq:ltot}
\end{equation}

Now for each value of $L_{\mathrm{lobe}}$ and $\theta$, the total
luminosity is uniquely defined, and $\pprob(L_{\mathrm{tot}} >
L_{\mathrm{min}} | \theta, L_{\mathrm{lobe}})$ becomes simply 0 or 1.
The outcome of this is that for a given $\theta$, there is one
limiting lobe flux $L_{\mathrm{lim}}(\theta)$ above which the
probability of detecting a source is unity, and below which it is
zero, and this simply changes the limits on the integration so that,
substituting equations (\ref{eq:peven}), (\ref{eq:probeq}) and
(\ref{eq:pllobe}) into equation (\ref{eq:peq}) we find
\begin{equation}
\pprob(\theta | L_{\mathrm{tot}} > L_{\mathrm{min}}) \propto  \int_{L_{\mathrm{lim}}(\theta)}^{\infty} L_{\mathrm{lobe}}^{-\alpha - 1} \ud L_{\mathrm{lobe}} \times \sin \theta \ud \theta \,\,.
\label{eq:fineq}
\end{equation}
This integrates to
\begin{eqnarray}
\pprob(\theta | L_{\mathrm{tot}} > L_{\mathrm{min}}) & \propto & [- L_{\mathrm{lobe}}^{-\alpha}]_{L_{\mathrm{lim}}(\theta)}^{\infty} \times \sin \theta \nonumber \\
& \propto & L_{\mathrm{lim}}(\theta)^{-\alpha} \times \sin \theta \,\,.
\label{eq:fineq2}
\end{eqnarray}

\begin{figure}
\begin{center}
\includegraphics[width = 0.5\textwidth]{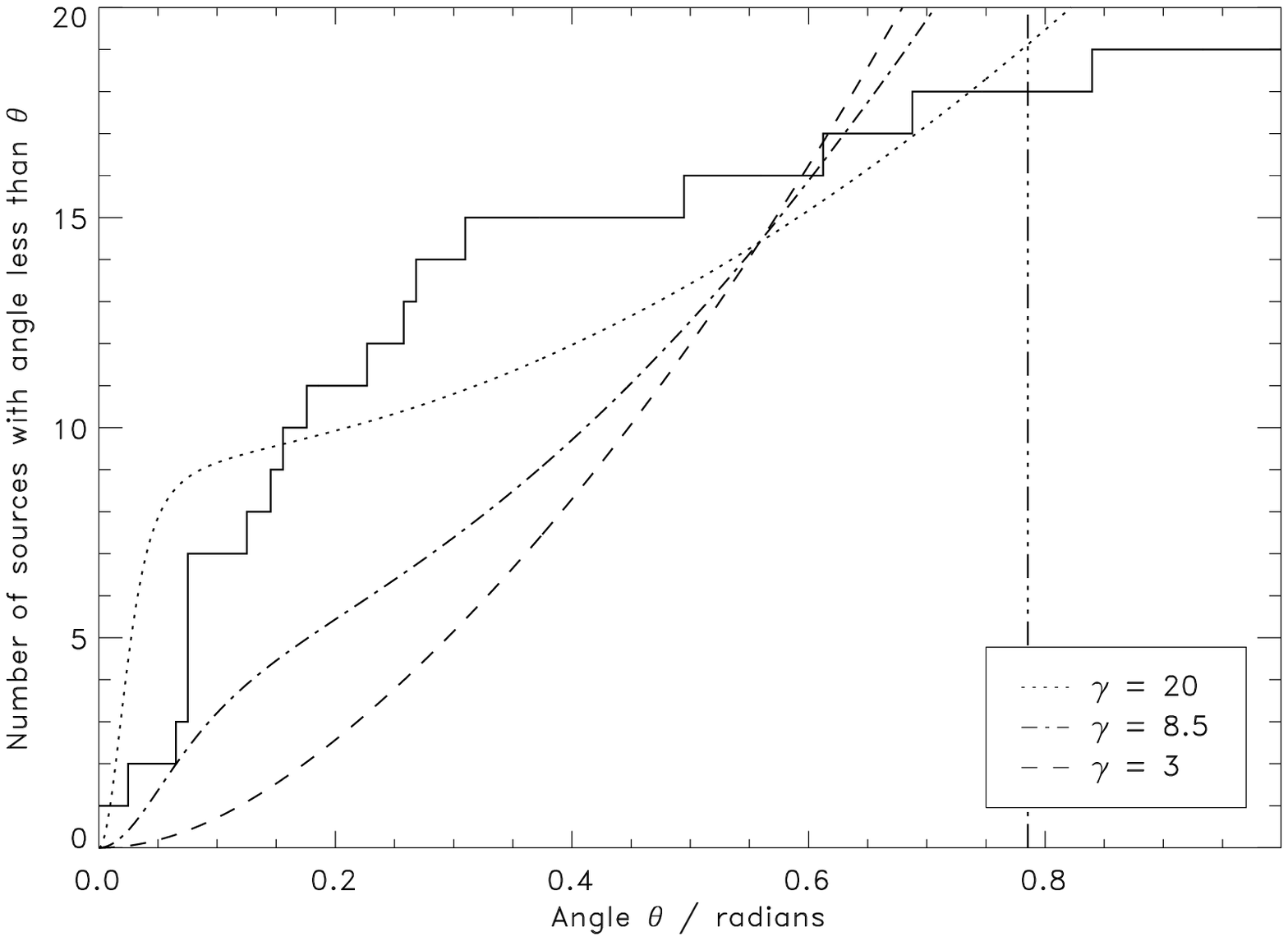}
\caption{The cumulative distribution of the best-fit disk angles
  (solid black line). The curves show the theoretical distributions of
  jet angle, modelled from the \citet{jackson99} relation and
  normalised up to an angle of $45\degree$, for different values of
  the Lorentz factor: the dashed line shows the expected cumulative
  distribution for $\gamma = 3$, the dot-dashed line for $\gamma =
  8.5$, and the dotted line for $\gamma = 20$. The vertical
  dot-dot-dot-dashed line indicates a disk angle of $45\degree$.}
\label{fig:pthetafitdisk}
\end{center}
\end{figure}

Figure \ref{fig:pthetafitdisk} shows the cumulative probability
distribution for the disk angles, together with the theoretical
distributions of jet angle for different values of the Lorentz factor
$\gamma$. If the accretion disk axis and radio jets align (i.e. if the
disk is perpendicular to the jet), then the distribution of disk
angles is most consistent with a value of the jet Lorentz factor
around $\gamma \sim 20$. It should be noted that MRC0418-288 was
excluded from this sub-sample of quasars on the basis of optical
faintness. This source has a high probability of being at a large
angle to the line of sight, since it may well be a reddened source in
which the dusty torus is blocking some sight lines to the optically
bright nucleus (e.g. \citet{baker02}). MRC1256-243 is a core-dominated
source, which is likely to have a very small disk angle. An extra
quasar added to the each end of the disk angle distribution function
would make it more in agreement with the modelled angular distribution
for $\gamma = 20$.

This Lorentz factor differs from the $\gamma \sim 3$ measured from the
3C sample. This is explicable if $\gamma$ is dependent on angle, such
that $\gamma \sim 20$ close to the line of sight, when viewing the
quasars down the axes of their jets, and $\gamma \sim 3$ when viewed
at larger angles (e.g. \citet{hardcastle06}). The angular
distribution of the Molonglo quasars, which are selected at
intermediate radio frequency, is dominated by Doppler boosting,
whereas this is a small effect for the 3CRR sample. The distribution
of the fitted Lorentz factor for the 3CRR sample is centred around
$\gamma \sim 3$, though with a slight asymmetry biased towards higher
values of $\gamma$, whereas the Molonglo sample is predicted to have
more probability of higher $\gamma$ factors. It is not sufficient to
model the jet with a single Lorentz factor. The luminosity -- jet
angle relation is also expected to have some scatter in $\gamma$ due
to intrinsic differences in the sources.

The distribution in fitted disk angle levels off at $48\degree$. This
is consistent with the opening angle which might be expected for a
sample of powerful FRII sources. The fact that no quasars are observed
with disk angles larger than $48\degree${} adds to the evidence for the
unification scheme which suggests radio galaxies and quasars are the
same objects viewed at decreasing angles to the line of sight.


\subsection{Velocity shifts}
\label{sec:shifts}

\subsubsection{Velocity shift measurements}
\label{sec:velshiftmeas}

The prior ranges on the positions of the single-peaked broad line
components allowed shifts of $\Delta z = \pm 0.015$ compared to the
fitted positions of the narrow lines, corresponding to line shifts of
$\sim 4500$ \kms. If no narrow lines were fitted, the shift of the
Gaussian line was measured relative to the Lorentzian line. The
accretion disk was allowed to shift to a similar degree, $\Delta
\lambda = \pm 100$ \AA{} with respect to the laboratory
wavelength. The fitted velocity shifts from the best-fit models are
given in Table \ref{tab:velshifts}.

\begin{table*}
\begin{minipage}{160mm}
\begin{center}
\begin{tabular}{llllllll}
\hline
\multicolumn{2}{|l|}{}       & \multicolumn{3}{|l|}{\textit{For best-fit model including disk:}}  & \multicolumn{3}{|l|}{\textit{For best-fit model without disk:}} \\ 
\multicolumn{2}{|l|}{Quasar} & {} & $\Delta v_{\mathrm{disk}}$ & $\Delta v_{\mathrm{Lorentzian}}$ & {} & $\Delta v_{\mathrm{Lorentzian}}$ & $\Delta v_{\mathrm{Gaussian}}$ \\
\multicolumn{2}{|l|}{}       & Model & (\kms) & (\kms) & Model & (\kms) & (\kms) \\ 
\hline
(1) & (2)  & (3) & (4) & (5) & (6) & (7) & (8) \\
1  & MRC0222-224    & 1 & 750	$\pm {}^{690	}_{ 250	}$ & -4460   $\pm {}^{	40	}_{  740	}$            & {}	& {}		& {}       \\
2  & MRC0327-241    & 3 & -830  $\pm {}^{230	}_{ 740	}$ & -3920   $\pm {}^{   580	}_{  2750	}$    & {}	& {}		& {}       \\
3  & MRC0346-279    & 5 & -1100   $\pm {}^{   850    }_{   290 }$ ${}^{\ast}$   & --                  & 9	& --		& 1570 $\pm {}^{2400}_{220}$  ${}^{\dagger}$  \\	    				   
4  & MRC0413-210    & 1 & -1490 $\pm {}^{380	}_{ 340	}$ & 310     $\pm {}^{   20	}_{  100	}$            & {}	& {}		& {}       \\
5  & MRC0413-296    & 1 & -350  $\pm {}^{140	}_{ 150	}$ & 2750    $\pm {}^{   110	}_{  90    	}$    & 7	& 2930 $\pm {}^{90}_{80}$ & -140 $\pm {}^{110}_{70}$	             \\   
6  & MRC0430-278    & 1 & -1350 $\pm {}^{1130	}_{ 3330}$ & -400    $\pm {}^{   100	}_{  120	}$            & 7	& 400 $\pm {}^{120}_{190}$ & -1490 $\pm {}^{140}_{2250}$    \\   
7  & MRC0437-244    & 1 & -2660 $\pm {}^{90	}_{ 500	}$ & 410     $\pm {}^{   30	}_{  50		}$    & {}	& {}		& {}       \\
8  & MRC0450-221    & 1 & -1330 $\pm {}^{170	}_{ 330	}$ & -490    $\pm {}^{   300	}_{  200	}$            & {}	& {}		& {}       \\
9  & MRC0549-213    & 5 & 2160    $\pm {}^{   2690   }_{   1600 }$ ${}^{\ast}$   & --                  &  {}      & {}		& {}       \\
10 & MRC1019-227    & 1 & 340	$\pm {}^{610	}_{ 170	}$ & 220     $\pm {}^{   100	}_{  130	}$            & {}	& {}		& {}       \\ 
11 & MRC1114-220    & 1 & 1780	$\pm {}^{410	}_{ 280	}$ & 220     $\pm {}^{   90	}_{  40		}$    & {}	& {}		& {}       \\
12 & MRC1208-277    & 1 & -1740 $\pm {}^{520	}_{ 200	}$ & -1220   $\pm {}^{   60	}_{  40		}$    & {}	& {}		& {}       \\
13 & MRC1217-209    & 5 & 900     $\pm {}^{   1050	}_{  1820}$ ${}^{\ast}$  & --               &  {}	& {}	        & {}       \\
14 & MRC1222-293    & 1 & -1650 $\pm {}^{550	}_{ 570	}$ & -230    $\pm {}^{   40	}_{  90		}$    & {}	& {}		& {}       \\
15 & MRC1301-251    & 1 & 400	$\pm {}^{690	}_{ 320	}$ & -4460   $\pm {}^{   40	}_{  3610	}$    & 7	& -4190 $\pm {}^{190}_{110}$ & 4460 $\pm {}^{330}_{40}$     \\ 
16 & MRC1349-265    & 1 & 640	$\pm {}^{630	}_{ 240	}$ & 40      $\pm {}^{   100	}_{  250	}$            & 7	& 40 $\pm {}^{140}_{130}$ & 1040 $\pm {}^{120}_{110}$   \\
17 & MRC1355-215    & 1 & 1620	$\pm {}^{890	}_{ 270	}$ & -310    $\pm {}^{   60	}_{  110	}$            & 7	& -310 $\pm {}^{90}_{40}$ & 1670 $\pm {}^{290}_{190}$    \\ 
18 & MRC1355-236    & 1 & 310	$\pm {}^{60	}_{ 560	}$ & -1940   $\pm {}^{   270	}_{  80		}$    & {}	& {}	  	& {}	   \\     	
19 & MRC1359-281    & 1 & -830  $\pm {}^{280	}_{ 580	}$ & -590    $\pm {}^{   180	}_{  90		}$    & {}	& {}		& {}	   \\	     	                   
\hline                          
\end{tabular}                   
\caption{Velocity shifts of the broad line components. \textit{Columns
    1 and 2}: Quasar index and MRC name;
  \textit{Column 3}: Best-fit disk model; \textit{Column 4}: Disk
  emission shift relative to narrow \ha{} in \kms{} for the best-fit
  disk model; \textit{Column 5}: Lorentzian line shift relative to
  narrow \ha{} in \kms{} for the best-fit disk model; \textit{Column
    6}: Best-fit model in cases where this does not include a disk;
  \textit{Column 7}: Broad Lorentzian shift relative to narrow \ha{}
  in \kms, for the best-fit model in cases where this does not include
  a disk; \textit{Column 8}: Broad Gaussian shift relative to narrow
  \ha{} in \kms, for the best-fit model in cases where this does not
  include a disk.  ${}^{\ast}$ No narrow \ha{} was fitted in this
  model, and so the shift of the disk emission relative to the broad
  Lorentzian is given in \kms.  ${}^{\dagger}$ No narrow \ha{} was
  fitted in this model, and so the shift of the broad Gaussian
  relative to the broad Lorentzian is given in \kms.  }
\label{tab:velshifts}
\end{center}
\end{minipage}
\end{table*}

In seven cases, the measured velocity shifts cannot be trusted. This
occurs in cases where the fitted parameters of either the broad
Lorentzian line or narrow \ha{} are not constrained by the prior
range. There are four potential reasons for this. (1) The velocity
shift of the broad Lorentzian is unconstrained by the fit; this
component is fitting to a broadband bump in the continuum left by
imperfect continuum subtraction, rather than to the emission line
profile, so this fitted value is not correct. This affects the quasars
MRC0327-224 and MRC1301-251. (2) The broad Lorentzian is clearly
fitting to an additional narrow line component, and has unconstrained
width as a result. This affects MRC1208-277. (3) Narrow \ha{} is
fitting to an extra broad component, and therefore the fitted position
of this line is not reliable. The relative shifts of the broad
components to each other will be of the right magnitude so long as the
extra broad component fitted by narrow \ha{} is small. This affects
MRC0413-296, MRC0430-278 and MRC1222-293. (4) In the case of
MRC0222-224, there is some degeneracy in the fit of the broad
Lorentzian, such that the shift of this line is not properly
constrained.

\subsubsection{Discussion of the velocity shifts}

\begin{figure}
\begin{center}
\includegraphics[width = 0.5\textwidth]{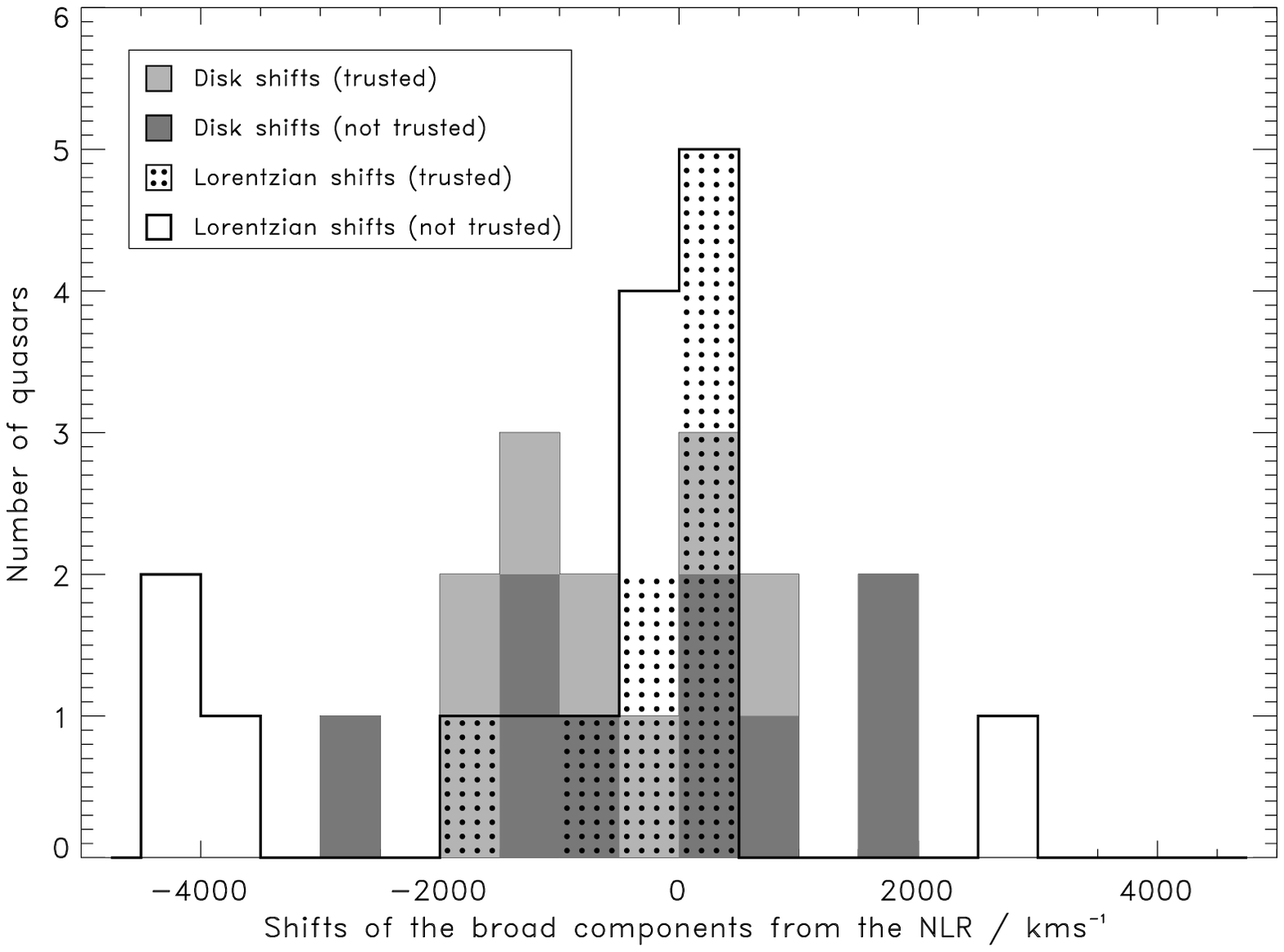}
\caption{Distribution of line shifts for the best-fit models which
  include an accretion disk. The disk shifts with respect to narrow
  \ha{} are shown as a grey-shaded histogram, and the Lorentzian line
  shifts with respect to narrow \ha{} are shown as a solid-line
  histogram. The subsets of these with reliable measurements are
  marked with dark-grey shading and dot shading respectively. Note
  that three sources are excluded as no narrow lines were fit for
  these spectra.}
\label{fig:plotshiftdisklor}
\end{center}
\end{figure}

Figure \ref{fig:plotshiftdisklor} shows the shifts of the broad
emission components for the best-fit models which include an accretion
disk. Note that these models all include a Lorentzian line in addition
to the disk emission. Three of the sources were best fitted with
models which did not include narrow lines, and hence do not have
shifts measured relative to the NLR; none of these quasars have
published redshifts from high-resolution observations of narrow lines,
so these were excluded. The distribution of fitted disk shifts is
relatively evenly distributed between -2000 \kms{} and +2000 \kms,
with the exception of MRC0437-244 with a disk shift of -2660 \kms{}
relative to the narrow lines. This distribution does not change
significantly if only the reliable shifts are considered.

The fitted Lorentzian line shifts are distributed in three groups. The
largest group have relatively modest velocity offsets to the NLR, with
9/16 shifts within $\pm 500$ \kms{} of the narrow line position, and a
small tail out to -2000 \kms. There is one source only with a
Lorentzian line shift relative to the narrow lines of more than +500
\kms: MRC0413-296 has a fitted shift of +2750 \kms, but this is not
reliable, since the narrow lines were fitting to an extra broad
component in this spectrum and hence the \ha{} line centre used for
reference was not reliably fitted. There are also three sources with
blueshifts of -3500 -- -4500 \kms. Of these three quasars, two have
Lorentzian lines which are fitting to a component of the continuum
rather than the broad \ha{} emission, and the other has degeneracy in
the fitted values of the Lorentzian line, so the shifts for these
sources are not trustworthy. Therefore, all dependable Lorentzian line
shifts fall in the central group, with moderate redshifts with respect
to the NLR, or moderate to high blueshifts with respect to the NLR.

Table \ref{tab:avshifts} gives the average shifts of the broad line
components and their standard errors. The spread of shifts is very
wide, but when only the reliable velocity shifts are considered, both
the accretion disk and the Lorentzian line have average blueshifts of
order -200 \kms. Taken overall, there is a trend within this sample of
powerful quasars for the broad emission to be blueshifted, in
agreement with the majority of the literature (see Section
\ref{sec:velshifts}). It should be borne in mind that the outflows and
inflows may be symmetric, but if dust in the plane of the accretion
disk obscures the far-side emission from view, the shifts measured are
for the near-side quasar emission.

Separating the sample into 10 low-$z$ ($z < 1$) and 6 high-$z$ ($z >
1.5$) sources and considering only the trustworthy shifts (6 and 3
sources respectively for low-$z$ and high-$z$), the average Lorentzian
line shift is -380 \kms{} for the low-$z$ sources and +40 \kms{} for
the high-$z$ sources, consistent with this emission originating close
to the systemic velocity. The average accretion disk shift for the
trustworthy sources is -890 \kms{} for the low-$z$ sources and +620
\kms{} for the high-$z$ sources, so the disk emission is blueshifted
with respect to systemic velocity at low-$z$ and redshifted at
high-$z$.

\begin{table*}
\begin{minipage}{160mm}
\begin{center}
\begin{tabular}{lllllll}
\hline
{} & \multicolumn{3}{c}{\textit{All shifts}} & \multicolumn{3}{c}{\textit{Trusted shifts only}} \\
Source of emission & Average shift & Standard error & No. of sources & Average shift & Standard error & No. of sources \\
(and $z$ range) & (\kms) & (\kms) & {} & (\kms) & (\kms) & {} \\
\hline
\hline
Accretion disk (all)        & -400 	& 310	& 16 & -180  & 470 & 9 \\ 
Accretion disk (low-$z$)    & -920 	& 320	& 10 & -890  & 460 & 6 \\ 
Accretion disk (high-$z$)   & 460	& 440	& 6  & 620   & 260 & 3 \\ 
Lorentzian line (all)       & -880 	& 470	& 16 & -240  & 230 & 9 \\ 
Lorentzian line (low-$z$)   & -1210 	& 520	& 10 & -380  & 320 & 6 \\ 
Lorentzian shift (high-$z$) & -330 	& 890	& 6  & 40    & 150 & 3 \\ 
\hline
\end{tabular}
\caption{Summary of the mean averages and standard errors of the
  shifts of the broad emission components, for best-fit models which
  include an accretion disk. Shifts are given for both the entire
  sample, and only those sources whose shifts are reliable; and are
  given for sub-samples divided by redshift: the sample as a whole,
  the low-$z$ fraction ($z < 1$), and the high-$z$ fraction ($z >
  1.5$). The number of sources in each sub-sample is shown.}
\label{tab:avshifts}
\end{center}
\end{minipage}
\end{table*}

\begin{figure*}
\begin{minipage}{160mm}
\begin{center}
\includegraphics[width = 1.0\textwidth]{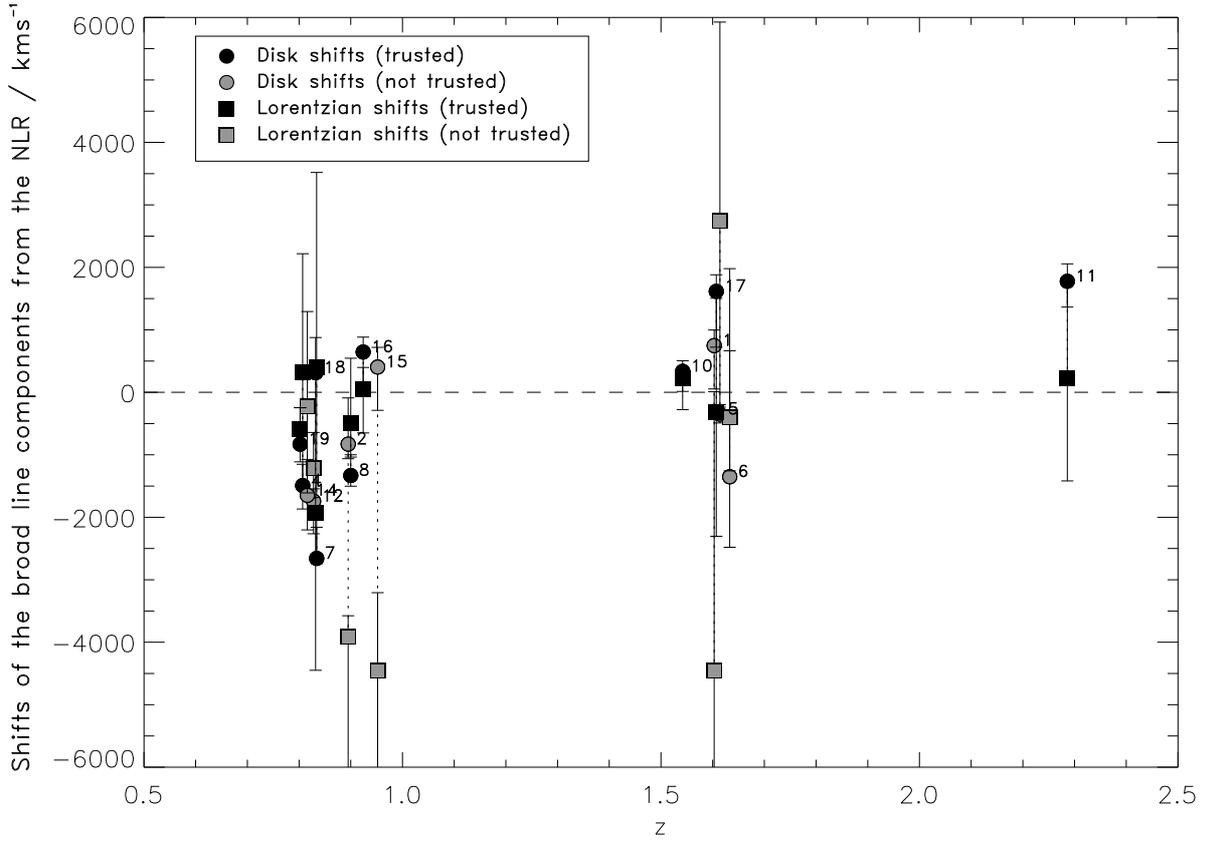}
\caption{Accretion disk shifts (circles) and Lorentzian line shifts (squares) with respect to narrow \ha{} for the best-fit models which include a disk, plotted against the source redshift. Black points indicate sources in which the shifts are reliable, and grey points indicate sources in which the shifts are not trusted (see Section \ref{sec:velshiftmeas}). Dotted lines join the disk and Lorentzian shifts from the same source, where the error bars do not overlap. The lower limits of the error bars for sources 2, 15 and 1 which are not shown on the plot are -9700 \kms, -7600 \kms{} and -9400 \kms{} respectively.}
\label{fig:plotshiftred}
\end{center}
\end{minipage}
\end{figure*}

Figure \ref{fig:plotshiftred} shows the broad-line velocity shifts
from the best-fit disk models plotted against quasar redshift. The
redshift and the shift of the accretion disk emission with respect to
the narrow lines are correlated with a Kendall $\tau$ coefficient of
0.850, significant at the $2 \sigma$ level (98\% probability of a
correlation). This correlation may arise from the dependence of both
these quantities on optical luminosity \citep{wills80} or radio
luminosity. The relation between AGN redshift and luminosity is
well-known, and occurs because sources at greater distances are more
powerful. This effect, the Malmquist Bias, is illustrated in Table
\ref{tab:luminosities}, which shows that the sources with $z > 1.5$
are more luminous than the $z < 1$ sources at radio frequencies, and
in terms of optical continuum and emission lines. There is no
correlation apparent between the quasar redshift and the shift of the
fitted Lorentzian line with respect to the narrow lines, with
a Kendall $\tau$ of 0.133 (significance of $< 0.5
  \sigma$, 28\% probability of a correlation). It appears that the
shifts between the single-peaked broad emission line and the NLR are
modest.

\begin{table}
\begin{center}
\begin{tabular}{lllllll}
\hline
{} & \multicolumn{2}{l}{$z < 1$} & \multicolumn{2}{l}{$z > 1.5$} & \multicolumn{2}{l}{All $z$} \\
{} & \textit{Av.} & \textit{Err.} & \textit{Av.} & \textit{Err.} & \textit{Av.} & \textit{Err.} \\
\hline
\hline
$L_{151 \mathrm{MHz}}$    & 26.9  &  0.1 & 27.8 & 0.1 & 27.3 & 0.1 \\  
$b_{\mathrm{J}}$ & -24.8 & 0.3 & -25.7 & 0.4 & -25.1 & 0.3 \\
$L_{\mathrm{Nar} \, H_{\alpha}}$ &  0.135 & 0.030 & 0.541 & 0.221 & 0.285 & 0.095 \\     
$L_{\mathrm{Brd} \, H_{\alpha}}$  &  3.93 & 1.05 & 19.54 & 6.49 & 9.69 & 3.02 \\            
\hline
\end{tabular}
\caption{Average luminosities for the quasar sub-sample divided by
  redshift into a low-$z$ fraction ($z < 1$), and a high-$z$ fraction
  ($z > 1.5$), and for the combined sample. \textit{Row 1}:
  Logarithmic 151 MHz luminosity; \textit{Row 2}: Optical
  $b_{\mathrm{J}}$ magnitudes from the UK Schmidt IIaJ plates, where
  $b_{\mathrm{J}} = B - 0.23(B-V)$ \citep{bahcall80}; \textit{Row 3}:
  Luminosities of the narrow \ha{} lines in units of 10$^{36}$ W
  sr$^{-1}$; \textit{Row 4}: Luminosities of the broad
  \ha{} lines in units of 10$^{36}$ W sr$^{-1}$.}
\label{tab:luminosities}
\end{center}
\end{table}

\begin{figure}
\begin{center}
\includegraphics[width = 0.5\textwidth]{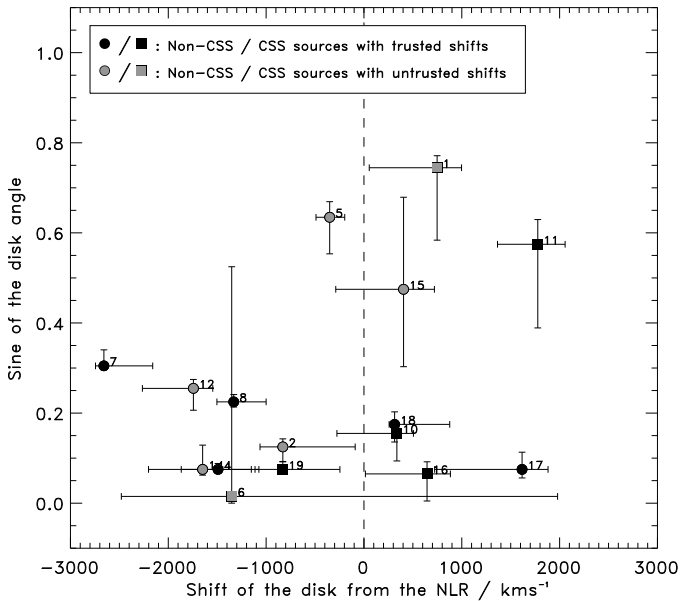}
\caption{Best-fit accretion disk angle plotted against the shift of the disk emission with respect to narrow \ha{} for non-CSS sources (circles) and CSS sources (squares). The points with reliable shifts are plotted in black; those sources whose shifts are not trusted are plotted in grey.}
\label{fig:plotshiftangled}
\end{center}
\end{figure}

Large-scale emission is most commonly seen alongside the radio jets or
within the ionisation cone of the source, e.g. \citet{nesvadba08}. If
it is the case that the material flows alongside the jet, the greatest
velocity shifts should be observed for sources at small angles to the
line of sight. No correlation is found between the magnitude of the
velocity shift of the disk with respect to the NLR and the sine of the
disk angle to the line of sight (Kendall $\tau$ coefficient -0.05,
significance of $< 0.1 \sigma$, probability of correlation 11\%); the
velocity shifts are plotted against the disk angle in Figure
\ref{fig:plotshiftangled}. However, the sources at $\gtrsim 30
\degree$ show velocity shifts of less than $\pm 1000$ \kms{} (note
that these are not reliable shifts), with the exception of
MRC1114-220, which is a young radio source, and hence is more likely
to have a misaligned disk. The narrow line gas may in fact be
entrained along the jet, but moving in a transverse direction:
\citet{nesvadba06} and \citet{nesvadba08} suggest from integral field
studies of a small sample of high-redshift radio galaxies that the
ionised gas is being accelerated away from the nucleus of the galaxy
by expanding cocoons of hot gas around the jets. There are also
indications that the distribution of gas surrounding CSS sources does
not mirror that seen in the larger sources, with more dust and gas
seen in edge-on CSS sources, and so it is probable that the velocity
distribution of the material in these young sources will also differ.

There is no evidence for a correlation between the magnitude of the
disk shift with respect to the NLR and the width of the narrow lines
(deconvolved from instrumental effects), with Kendall $\tau$ of
-0.283, significance $< 1 \sigma$, a probability for anticorrelation
of 56\%. The narrow lines would be expected to have broader profiles
for greater velocity shifts if the shift between the disk emission and
the narrow line emission is due to the narrow lines forming in a
moving mass of gas.

The measured velocity shifts for the broad components relative to the
narrow lines imply that the single-peaked broad lines arise from a
region with a similar velocity to the NLR, while the disk is
redshifted relative to these regions for higher-$z$, and hence
brighter, sources, and blueshifted in relation to these for
lower-$z${} sources. The accretion disk is expected to be fixed at the
systemic redshift, and therefore these results can be interpreted as
the single-peaked broad lines and narrow lines arising from outflowing
material in optically-luminous sources, and infalling material in less
luminous sources. This could be the result of a quasar wind model of
AGN feedback, in which powerful quasars fuel outflows which sweep the
gas out of the galaxy \citep{silk98}; the outflows are expected to
scale with the quasar luminosity.


\section{Effects of Model Assumptions}
\label{sec:assumptions}

\subsection{Models with the NLR and accretion disk fixed at the same redshift}
\label{sec:assumptmodel}

If the narrow lines are formed at the systemic redshift, as much of
the literature indicates, then there should be minimal velocity shift
between the accretion disk and the NLR. In order to test the effect of
this alternative assumption, a second set of models was fitted, with
the disk emission fixed at the same redshift as the narrow
lines. Aside from the exclusion of one parameter describing the shift
between these two emission components, the models are identical, and
in the cases of the models with no narrow lines, the disk is allowed
to shift as usual; this change therefore affects only four models (1,
3, 10 and 11).

\subsection{Notes on individual quasars}
\label{sec:assumptnotes}

\noindent\textbf{MRC0222-224}: Model 1 was selected with
higher evidence by $\Delta \ln(\mathrm{Evidence}) = 6.0$ than Model
10, the next best model.

\noindent\textbf{MRC0327-241}: The best-fit model is Model
1.  Models 3, 5 and 10 are also within the $\Delta
\ln(\mathrm{Evidence}) = 5$ bound of this model; however, all these
models include disk emission, so this quasar has strong evidence for
a disk.

\noindent\textbf{MRC0346-279}: This low signal-to-noise spectrum does not 
differentiate well between the models. Model 9 was selected by the
Bayesian evidence, with higher evidence by $\Delta
\ln(\mathrm{Evidence}) = 0.80$ over the next best fit.

\noindent\textbf{MRC0413-210}: Model 1 was selected, with
$\Delta \ln(\mathrm{Evidence}) = 7.7$ over the next best model.

\noindent\textbf{MRC0413-296}: As before, although the spectrum of 
this quasar clearly contains narrow lines, the narrow line parameters
instead fitted broad components of emission. The best
model is Model 7, with evidence over the next best-fit model of
$\Delta \ln(\mathrm{Evidence}) = 12.1$. This quasar has strong
evidence against the presence of a disk.

\noindent\textbf{MRC0430-278}: The preferred model for this spectrum 
is Model 7, though many models fall within the Jeffreys'
criterion. The margin in logarithmic evidence of Model 7 over Model 3,
the best-fit model which includes disk emission, is only $\Delta
\ln(\mathrm{Evidence}) = 2.3$, so this is a ``possible disk'' quasar.

\noindent\textbf{MRC0437-244}: Model 5, including a disk, was selected
with strong evidence.

\noindent\textbf{MRC0450-221}: Model 1, including a disk, was selected
with strong evidence. The fitted disk angle is not within the errors
of the fitted angle from the first analysis: 20$\degree\pm {}^{ 1}_{
  1}$ from this analysis compared to 13$\degree\pm {}^{ 1}_{ 1}$ for
the case in which the accretion disk is allowed to shift.

\noindent\textbf{MRC0549-213}: Many of the models fall within the 
Jeffreys' criterion of the best-fit model, including three of the disk
models, so this quasar is a ``possible disk'' source. The selected
model was Model 6, but this is preferred over the best-fit disk model,
Model 1, by only $\Delta \ln(\mathrm{Evidence}) = 1.7$.

\noindent\textbf{MRC1019-227}: Model 1 is preferred by only $\Delta
\ln(\mathrm{Evidence}) = 1.5$ over Model 7, so this source has a ``weak
disk'' classification.

\noindent\textbf{MRC1114-220}: Model 1 was selected by $\Delta
\ln(\mathrm{Evidence}) = 6.3$ over Model 7, the best-fit model without
a disk, so there is strong evidence for a disk. The fitted angle, 
15$\degree\pm {}^{ 1}_{ 4}$, is not in agreement with the angle found 
from the analysis in which the disk is allowed to shift with respect
to the narrow lines (35$\degree\pm {}^{ 4}_{ 12}$).

\noindent\textbf{MRC1208-277}: Model 7 was selected with evidence
difference $\Delta \ln(\mathrm{Evidence}) = 86.3$ over Model 3, which
contains a disk, so this has strong evidence against a disk.  The
fitted angle is 46$\degree\pm {}^{ 4}_{ 3}$, which is strongly in
disagreement with the case in which the disk is allowed to shift with
respect to the NLR (15$\degree\pm {}^{1}_{ 3}$).

\noindent\textbf{MRC1217-209}: Model 5 was selected, but many of the 
models fall within the Jeffreys' criterion of this model. There is an
evidence difference between Model 5 and Model 9, the best-fit model
without a disk, of $\Delta \ln(\mathrm{Evidence}) = 1.4$.

\noindent\textbf{MRC1222-293}: Model 7 was selected as the best-fit 
model, with evidence difference of $\Delta \ln(\mathrm{Evidence}) =
50.4$ over Model 1, which is the highest-evidence model including an
accretion disk. The fitted angle is 19$\degree\pm {}^{ 1}_{ 3}$, which
is in disagreement with the angle found from the analysis is which the
disk is allowed a velocity shift with respect to the NLR
(4$\degree\pm {}^{ 3}_{ 1}$).

\noindent\textbf{MRC1301-251}: Model 7 is preferred by a margin of 
$\Delta \ln(\mathrm{Evidence}) = 0.05$ over Model 10, the
highest-evidence model including a disk, and by a margin of $\Delta
\ln(\mathrm{Evidence}) = 0.8$ over Model 1, so any of these models are
plausible.

\noindent\textbf{MRC1349-265}: Model 7 was selected by 
$\Delta \ln(\mathrm{Evidence}) = 1.9$ over Model 1, the
highest-evidence model including a disk.

\noindent\textbf{MRC1355-215}: Model 7 was selected over the next best
model, Model 1, by $\Delta \ln(\mathrm{Evidence}) = 3.7$.

\noindent\textbf{MRC1355-236}: Model 1 was selected with strong
evidence.

\noindent\textbf{MRC1359-281}: There is strong evidence for Model 1.
The fitted angle is 7$\degree\pm {}^{ 1}_{ 1}$, which does not agree
with the angle fitted from the first analysis, 4$\degree\pm {}^{ 1}_{ 1}$,
for the case in which the accretion disk is not fixed at the redshift
of the narrow lines.

\subsection{Fits to the emission spectra}
\label{sec:assumptfits}

Table \ref{tab:diskanglesns} gives the results of the model selection
for the case in which the NLR and the accretion disk are fixed at the
same redshift. For 14 out of 19 sources, the $1 \sigma$ error range
of the best-fit disk angle from the new analysis overlaps with the
$1 \sigma$ error range of the best-fit disk angle for the case in
which the disk is allowed to shift with respect to the NLR. The
remaining five are MRC0450-221, MRC1114-220, MRC1208-277, MRC1222-293
and MRC1359-281.

The narrow line centres of a number of the models were not strongly
constrained in the fitting process, and since the disk and NLR are
fixed at the same redshift in this regime, an incorrectly fitted
narrow line centre leads to an offset in the disk position as well,
giving an erroneous fitted angle. This potentially affects MRC0327-241
(in which the narrow line position differed by 20\AA{} between the fit
from the analysis in which the disk shift is allowed and the one in
which it is not); MRC1114-220, in which there was degeneracy in the
fitted position of the narrow \ha; MRC0413-296, MRC0430-278,
MRC1222-293 and MRC1359-281, in which the narrow lines were fit with
very broad widths ($\gtrsim$ 30 \AA), and hence their positions may
not have been fixed to very good accuracy. This effect might plausibly
be the reason that the disk angles from the two analyses do not agree for
MRC1114-220 and MRC1359-281. However, at least MRC0450-221,
MRC1208-277 and MRC1222-293 have different disk angles as a direct
result of the fixing of the accretion disk at the redshift of the NLR.

In addition to the inconsistency in disk angles for these five sources
between the two analyses, the two sources with the most extreme
changes in fitted disk angle (MRC1208-277 and MRC1222-293) have
changed from having strong Bayesian evidence in favour of a disk, in
the regime where a disk shift is allowed from the NLR, to having
strong Bayesian evidence \textit{against} the presence of disk, when
the disk is fixed at the redshift of the NLR. The volume of prior
parameter space was reduced in this second analysis by the removal of
a variable, which, if this variable was extraneous, would have
increased the Bayesian evidence for the presence of accretion
disks. The fact that the evidence remained similar in the majority of
cases, but declined sharply in these two cases, implies that the
variable describing the velocity shift between the disk and the NLR is
required.

\begin{table*}
\begin{minipage}{180mm}
\begin{center}
\begin{tabular}{lllllllllllll}
\hline
\multicolumn{2}{|l|}{}       & Best-fit & Evidence & \multicolumn{2}{|l|}{For best-fit model} & \multicolumn{2}{|l|}{For best-fit model} & {} & \multicolumn{3}{|l|}{\textit{Analysis with disk shift allowed}} \\ \cline{10-12}
\multicolumn{2}{|l|}{Quasar} & model   & for disk     & \multicolumn{2}{|l|}{with disk}    & \multicolumn{2}{|l|}{without disk} & Disk & Best-fit & Evidence & Disk \\ 
\multicolumn{2}{|l|}{}       & {} & {}     & $\Delta \ln E^{2}_{1\mathrm{d}}$ & $\Delta \ln E^{\mathrm{n}}_{1\mathrm{d}}$ & $\Delta \ln E^{2}_{1\mathrm{n}}$ & $\Delta \ln E^{\mathrm{d}}_{1\mathrm{n}}$ & angle & Model & for disk & angle \\
\hline
\hline
1  & MRC0222-224 & 1 & 	SD & 6.02  & 11.71 &      & 	 & 53$\degree\pm {}^{ 7}_{11}$  & 1 & 	SD & 48 $\degree\pm {}^{2}_{12}$ \\	        
2  & MRC0327-241 & 1 & 	SD & 1.77  & 7.06  &      & 	 &  7$\degree\pm {}^{ 1}_{ 1}$  & 3 & 	SD &  7 $\degree\pm {}^{1}_{2}$ \\	        
3  & MRC0346-279 & 9 (1) & PD &    &       & 0.80 & 1.08 & 10$\degree\pm {}^{ 3}_{ 5}$  & 9 (5) &      PD & 1 $\degree\pm {}^{5}_{1}$ \\ 	
4  & MRC0413-210 & 1 & 	SD & 7.66  & --    &      & 	 &  9$\degree\pm {}^{ 3}_{ 4}$  & 1 & 	SD &  4 $\degree\pm {}^{1}_{1}$ \\	        
5  & MRC0413-296 & 7 (1) & ND &    &       & 12.12& --   & 42$\degree\pm {}^{ 2}_{ 2}$  & 7 (1) &      ND & 39 $\degree\pm {}^{3}_{6}$ \\	
6  & MRC0430-278 & 7 (3) & PD &    &       & 1.40 & 2.05 &  8$\degree\pm {}^{ 3}_{ 4}$  & 7 (1) &      PD & 1 $\degree\pm {}^{31}_{1}$ \\      
7  & MRC0437-244 & 5 & 	SD & 16.87 & 62.52 &      & 	 & 21$\degree\pm {}^{ 1}_{ 1}$  & 1 & 	SD & 18 $\degree\pm {}^{2}_{1}$ \\ 	        
8  & MRC0450-221 & 1 & 	SD & 8.06  & 39.27 &      & 	 & 20$\degree\pm {}^{ 1}_{ 1}$  & 1 & 	SD & 13 $\degree\pm {}^{1}_{1}$ \\ 	        
9  & MRC0549-213 & 6 (1) & PD &    &       & 0.90 & 1.70 &  7$\degree\pm {}^{17}_{ 6}$  & 6 (5) &      PD & 15 $\degree\pm {}^{20}_{12}$ \\    
10 & MRC1019-227 & 1 &  WD & 1.53  & --    &      &      &  6$\degree\pm {}^{ 1}_{ 1}$  & 1 &          WD &  9 $\degree\pm {}^{1}_{4}$ \\	
11 & MRC1114-220 & 1 & 	SD & 4.14  & 6.25  &      & 	 & 15$\degree\pm {}^{ 1}_{ 4}$  & 1 & 	SD & 35 $\degree\pm {}^{4}_{12}$ \\	        
12 & MRC1208-277 & 7 (3) & ND &    &       & 86.26 & --  & 46$\degree\pm {}^{ 4}_{ 3}$  & 1 & 	SD & 15 $\degree\pm {}^{1}_{3}$ \\	        
13 & MRC1217-209 & 5 & 	WD & 1.37  & --    &      & 	 &  8$\degree\pm {}^{14}_{ 2}$  & 5 & 	WD &  8 $\degree\pm {}^{14}_{2}$ \\             
14 & MRC1222-293 & 7 (1) & ND &    &       & 26.92& 50.36& 19$\degree\pm {}^{ 1}_{ 3}$  & 1 & 	SD &  4 $\degree\pm {}^{3}_{1}$ \\	        
15 & MRC1301-251 & 7 (10)& PD &	   &       & 0.05 & --   & 31$\degree\pm {}^{ 2}_{ 2}$  & 7 (1) &      PD & 28 $\degree\pm {}^{14}_{11}$ \\	
16 & MRC1349-265 & 7 (1) & PD &    &       & 1.94 & --	 &  4$\degree\pm {}^{ 1}_{ 3}$  & 7 (1) &      PD & 4 $\degree\pm {}^{2}_{3}$ \\	
17 & MRC1355-215 & 7 (1) & PD &    &       & 3.65 & --   &  4$\degree\pm {}^{ 1}_{ 2}$  & 7 (1) &      PD & 4 $\degree\pm {}^{2}_{1}$ \\	
18 & MRC1355-236 & 1 & 	SD & 7.47  & --    &      & 	 & 14$\degree\pm {}^{ 1}_{ 5}$  & 1 & 	SD & 10 $\degree\pm {}^{2}_{2}$ \\	        
19 & MRC1359-281 & 1 &  SD & 7.53  & --    &      &      &  7$\degree\pm {}^{ 1}_{ 1}$  & 1 &          SD &  4 $\degree\pm {}^{1}_{1}$ \\ 	
\hline
\end{tabular}
\caption{Summary of the model fitting results for the analysis in
  which the accretion disk is fixed at the same redshift as the
  NLR. \textit{Columns 1 and 2}: Quasar index and MRC name;
  \textit{Column 3}: Index of the best-fit model. If the selected
  model does not include an accretion disk, the index of the
  highest-evidence model which includes an accretion disk is shown in
  brackets.  \textit{Column 4}: Indication of whether there is
  evidence for a disk according to the odds ratio, with codes as
  follows: SD = strong evidence for a disk; MD = moderate evidence for
  a disk; WD = weak evidence for a disk; PD = a possible disk,
  i.e. the best fit is a non-disk model, but there is only
  inconclusive, weak or moderate evidence for this; ND = strong
  evidence against the presence of a disk; \textit{Columns 5 and 6}:
  Natural logarithmic evidence difference between the best-fit model,
  in cases where this includes an accretion disk, and the second best
  model (\textit{Column 5}, $\Delta \ln E^{2}_{1\mathrm{d}}$), and in
  cases where the second best fit also includes a disk, between the
  best-fit model and the best-fit model without a disk (\textit{Column
    6}, $\Delta \ln E^{\mathrm{n}}_{1\mathrm{d}}$); \textit{Columns 7
    and 8}: Natural logarithmic evidence difference between the
  best-fit model, in cases where this does not include an accretion
  disk, and the second best model (\textit{Column 7}, $\Delta \ln
  E^{2}_{1\mathrm{n}}$), and in cases where the second best fit also
  does not include a disk, between the best-fit model and the best-fit
  model with a disk (\textit{Column 8}, $\Delta \ln
  E^{\mathrm{d}}_{1\mathrm{n}}$); \textit{Column 9}: Best-fit angle of
  the disk and error in degrees. \textbf{For analysis in which the
    disk is allowed to shift with respect to the NLR --} \textit{Column
    10}: Index of the best-fit model; \textit{Column 11}: Indication
  of whether there is evidence for a disk according to the odds ratio,
  codes as before; \textit{Column 12}: Best-fit angle of the disk and
  error in degrees. }
\label{tab:diskanglesns}
\end{center}
\end{minipage}
\end{table*}

\begin{figure}
\begin{center}
\includegraphics[width = 0.5\textwidth]{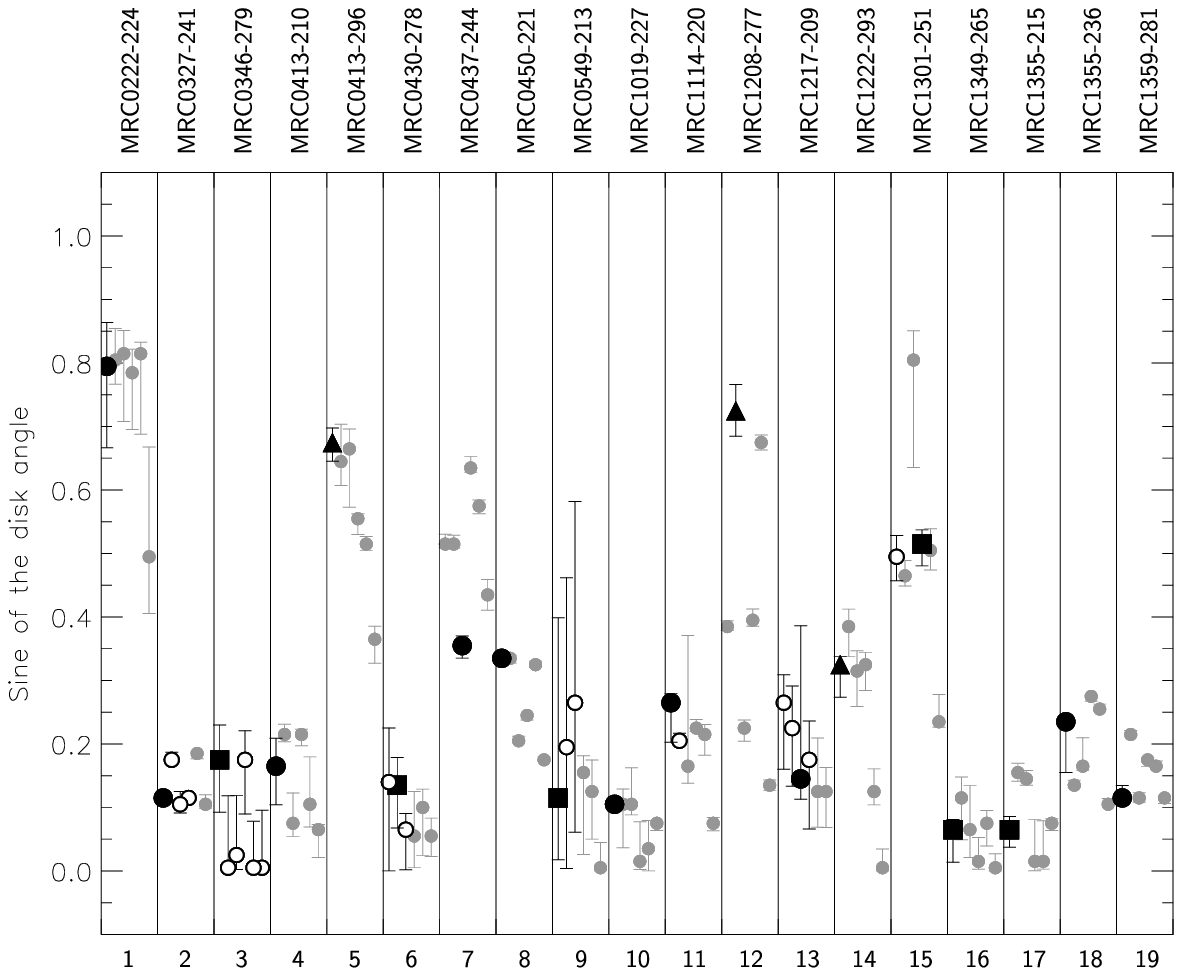}
\caption{Disk angles for model fits with an accretion disk, for the
  analysis where the disk emission is fixed with respect to the narrow
  lines. The black points indicate the best-fit disk model for each
  quasar. Circular symbols indicate evidence for a disk, square
  symbols indicate a possible disk (inconclusive evidence, or weak to
  moderate evidence against a disk), while triangular symbols indicate
  strong evidence against the presence of a disk. White points mark
  fits which fall within one Jeffreys' criterion of the best fit, and
  these can be seen in all cases to be consistent with the best-fit
  disk angle. All angles derived from non-best-fit disk cases are
  plotted in pale grey. }
\label{fig:plotallanglesns}
\end{center}
\end{figure}

Figure \ref{fig:plotallanglesns} shows the fitted disk angles for the
best-fit models including a disk, in the analysis where the disk is
fixed at the same redshift as the NLR. Comparing these fitted angle
distributions with those from the case where a disk -- NLR velocity
shift is allowed (shown in Figure \ref{fig:plotallangles}), it can be
seen that these distributions are reasonably similar; only MRC1208-277
and MRC1222-293 have fitted angles which change dramatically. There is
a general tendency for the disk angle error bars to be smaller in the
new analysis, which is due to the reduced number of variables
tightening the fit.  It should be noted that the error bars plotted
are simply $1 \sigma$ error bars given the quasar spectrum and the
model: no error has been added to account for the choice of model.

\subsection{Relationships with radio source size}
\label{sec:assumpsizes}

\begin{figure}
\begin{center}
  \includegraphics[width = 0.5\textwidth]{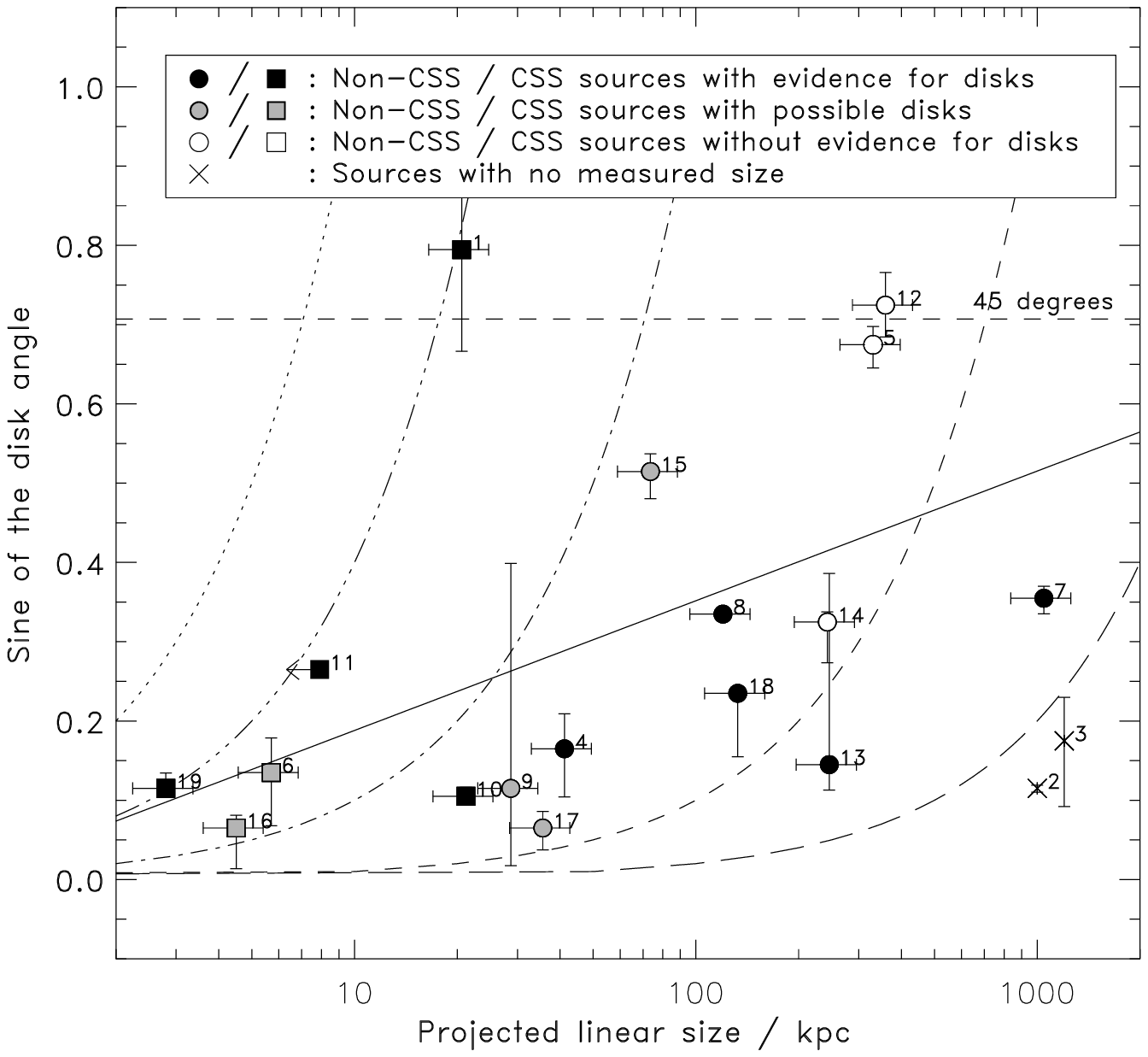}
  \caption{Sine best-fit disk angles versus the projected linear size
    of the quasar in kpc for the analysis in which the disk is fixed
    at the redshift of the NLR. The disk angle error bars include both
    the errors returned from the individual fits and a systematic
    error calculated from the variation in results from using
    different random seeds to initialise the Monte Carlo engines,
    added in quadrature, but do not take into account any error due to
    model selection. The non-CSS FRII sources are plotted with
    circles, and CSS sources are plotted with squares. The two sources
    plotted with crosses are core-dominated sources without measured
    projected sizes. The black points have weak to strong evidence for
    a disk, grey points have possible disks, and the white points have
    Bayesian evidence against the presence of a disk.  Index numbers
    are matched to source names in Table \ref{tab:obsdata}. The black
    line shows the best linear fit to all sources excluding the
    core-dominated sources. There is a $2 \sigma$ correlation by
    Kendall's $\tau$ test between the sine of the disk angle and the
    projected size both for all sources excluding the core-dominated
    sources, and for the 11 non-CSS FRII sources. The dotted,
    dot-dot-dot-dashed, dot-dashed, dashed and long-dashed lines show
    loci of constant deprojected size, for 10 kpc, 25 kpc, 100 kpc, 1
    Mpc and 5 Mpc respectively. }
\label{fig:plotsinivsdns}
\end{center}
\end{figure}

Figure \ref{fig:plotsinivsdns} shows, for the new analysis, sine disk
angle plotted against projected linear source size, which if the radio
jets are perpendicular to the disks should be correlated due to
geometric effects. Fixing the disk at the redshift of the NLR has
strengthened this from a $1 \sigma$ correlation, in the case that the
model allowed a disk velocity shift (see Figure
\ref{fig:plotsinivsd}), to a $2 \sigma$ correlation (Kendall $\tau$
coefficient 0.779, probability that a correlation exists 97\%). This
correlation remains at a significance level of $2 \sigma$ if the CSS
sources are removed from the sample (Kendall $\tau$ coefficient 0.982,
probability that a correlation exists 96\%).

\begin{figure}
\begin{center}
\includegraphics[width = 0.5\textwidth]{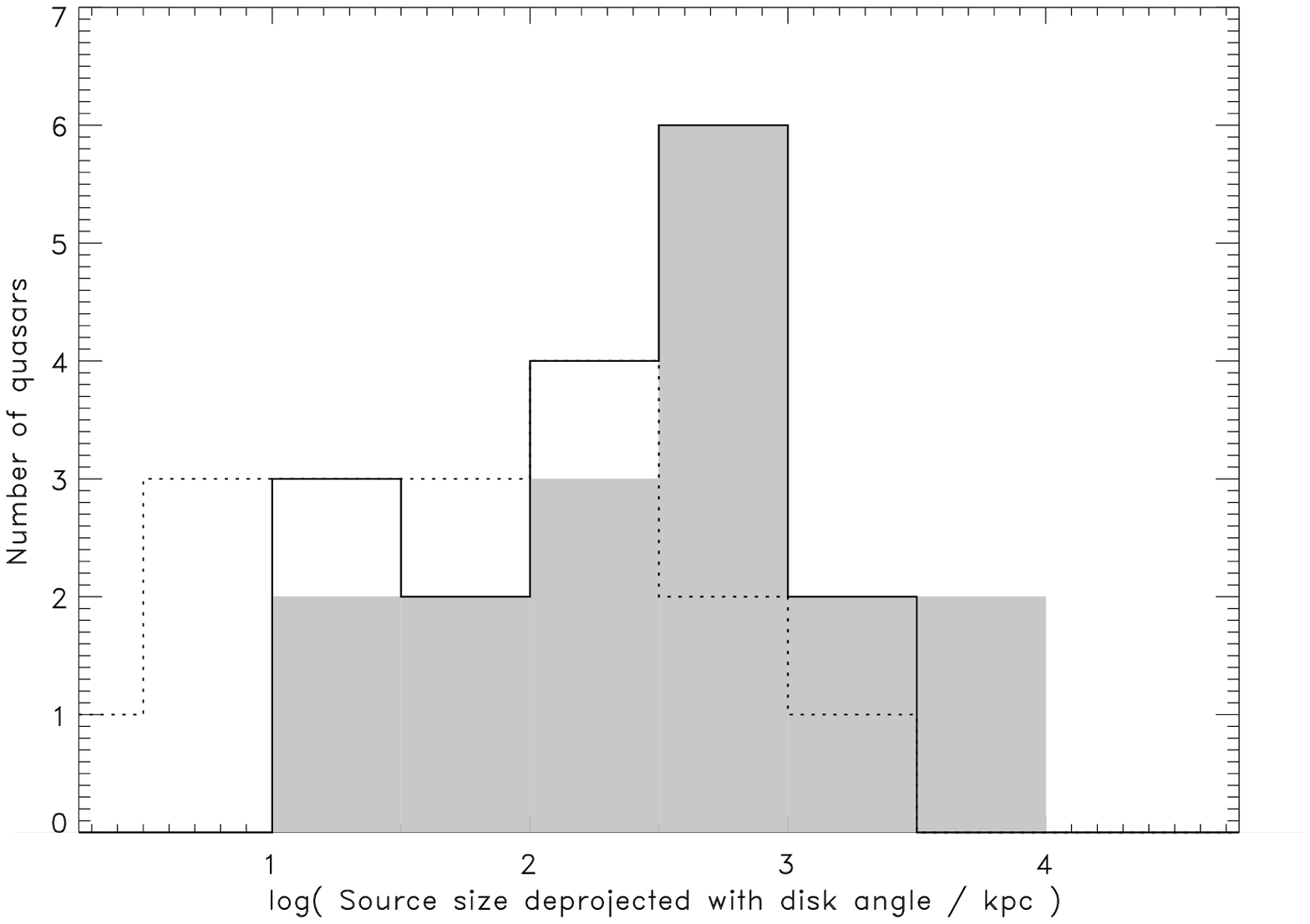}
\caption{The solid black line shows source sizes deprojected with the
  fitted disk angles from the models where the disk emission is fixed
  at the same redshift as the NLR (see Table \ref{tab:deprojns}). The
  source sizes from deprojecting with the fitted angles from the
  analysis in which the disk emission is allowed to shift with respect
  to the NLR are shaded in grey. The projected sizes are plotted with
  a dotted line. Note that the two core-dominated sources have no
  measured size and are therefore excluded.}
\label{fig:histsizens}
\end{center}
\end{figure}

The source sizes were deprojected using the method in Section
\ref{sec:deprojsizes}, and the resultant sizes are given in Table
\ref{tab:deprojns}, with those from the original analysis for
comparison. A histogram of the deprojected sizes is shown in Figure
\ref{fig:histsizens}; the fitted angles from the new analysis
deproject the source sizes to a slightly narrower range than the
original analysis (see Figure \ref{fig:histsize}). This does not imply
a preference for the models with no disk shift, since the range of
deprojected sizes merely gives an indication as to the typical size to
which a source expands before it becomes quiescent.

\begin{table}
\begin{center}
\begin{tabular}{llllll}
\hline
\multicolumn{2}{c}{}       & {}   & Proj.      & Deproj.     & Deproj.     \\
\multicolumn{2}{c}{}       & {}   & source     & source      & source      \\
\multicolumn{2}{c}{Quasar} & Type & size       & size (kpc)  & size (kpc)  \\
\multicolumn{2}{c}{}       & {}   & (kpc)      & (No disk    & (Disk shift \\
\multicolumn{2}{c}{}       & {}   & {}         & shift)      & allowed)    \\
\hline
\hline
 1 & MRC0222-224 & CSS  & 20.6  &  $    26\pm^{   5}_{   2}   $&  $ 28\pm^{8}_{1}           $ \\ 
 2 & MRC0327-241 & CD   & --    &  --                        &   --                        \\
 3 & MRC0346-279 & CD   & --    &  --                        &   --                        \\ 
 4 & MRC0413-210 & FRII & 41.2  &  $   250\pm^{ 150}_{  50}   $&  $ 550\pm^{70}_{80}        $ \\ 
 5 & MRC0413-296 & FRII & 330.4 &  $   490\pm^{  20}_{  20}   $&  $ 520\pm^{80}_{30}        $ \\ 
 6 & MRC0430-278 & CSS  & 5.7   &  $    42\pm^{  42}_{  10}   $&  $ 380\pm^{109000}_{370}    $ \\ 
 7 & MRC0437-244 & FRII & 1045.9&  $  2950\pm^{ 170}_{ 120}   $&  $ 3430\pm^{100}_{360}      $ \\ 
 8 & MRC0450-221 & FRII & 120.0 &  $   360\pm^{  10}_{  10}   $&  $ 530\pm^{30}_{40}         $ \\ 
 9 & MRC0549-213 & FRII & 28.7  &  $   250\pm^{1400}_{ 180}   $&  $ 110\pm^{360}_{60}        $ \\ 
10 & MRC1019-227 & CSS  & 21.2  &  $   200\pm^{  20}_{  20}   $&  $ 140\pm^{90}_{10}         $ \\ 
11 & MRC1114-220 & CSS  & $<$ 7.9&  $    <30\pm^{   9}_{   2}   $&  $ <14\pm^{7}_{1}           $ \\ 
12 & MRC1208-277 & FRII & 359.3 &  $   500\pm^{  30}_{  30}   $&  $ 1410\pm^{330}_{100}     $ \\ 
13 & MRC1217-209 & FRII & 245.9 &  $  1700\pm^{ 480}_{1060}   $&  $ 1700\pm^{480}_{1060}    $ \\ 
14 & MRC1222-293 & FRII & 242.7 &  $   750\pm^{ 140}_{  30}   $&  $ 3240\pm^{660}_{1350}    $ \\ 
15 & MRC1301-251 & FRII & 73.6  &  $   140\pm^{  10}_{  10}   $&  $ 160\pm^{90}_{50}        $ \\ 
16 & MRC1349-265 & CSS  & 4.5   &  $    69\pm^{ 260}_{  14}   $&  $ 69\pm^{860}_{20}        $ \\ 
17 & MRC1355-215 & FRII & 35.6  &  $   550\pm^{ 410}_{ 130}   $&  $ 480\pm^{160}_{160}      $ \\ 
18 & MRC1355-236 & FRII & 132.6 &  $   570\pm^{ 290}_{  10}   $&  $ 760\pm^{220}_{100}      $ \\ 
19 & MRC1359-281 & CSS  & 2.8   &  $    24\pm^{   2}_{   4}   $&  $ 37\pm^{5}_{7}           $ \\ 
\hline
\end{tabular}
\caption{Summary of projected and deprojected source
  sizes. \textit{Columns 1 and 2}: Quasar index and MRC name;
  \textit{Column 3}: Source type -- CD indicates sources with
  core-to-lobe radio flux ratio at 10 GHz of greater than 1, CSS
  indicates sources with $\alpha_{\mathrm{opt}} > 0.5$ and projected
  source sizes $<$ 25 kpc, FRII indicates non-CSS FRII, which
  encompasses all other sources; \textit{Column 4}: Projected source
  sizes in kpc (sources of these values are given in Table
  \ref{tab:deproj}); \textit{Column 5}: Deprojected source sizes in
  kpc, calculated using the fitted disk angles from the
  models in which the accretion disk is fixed at the same redshift as
  the NLR; \textit{Column 6}: Deprojected source sizes in kpc,
  calculated using the fitted disk angles from the models
  in which the accretion disk is allowed to have a different redshift
  from the NLR.}
\label{tab:deprojns}
\end{center}
\end{table}

\begin{figure}
\begin{center}
\includegraphics[width = 0.5\textwidth]{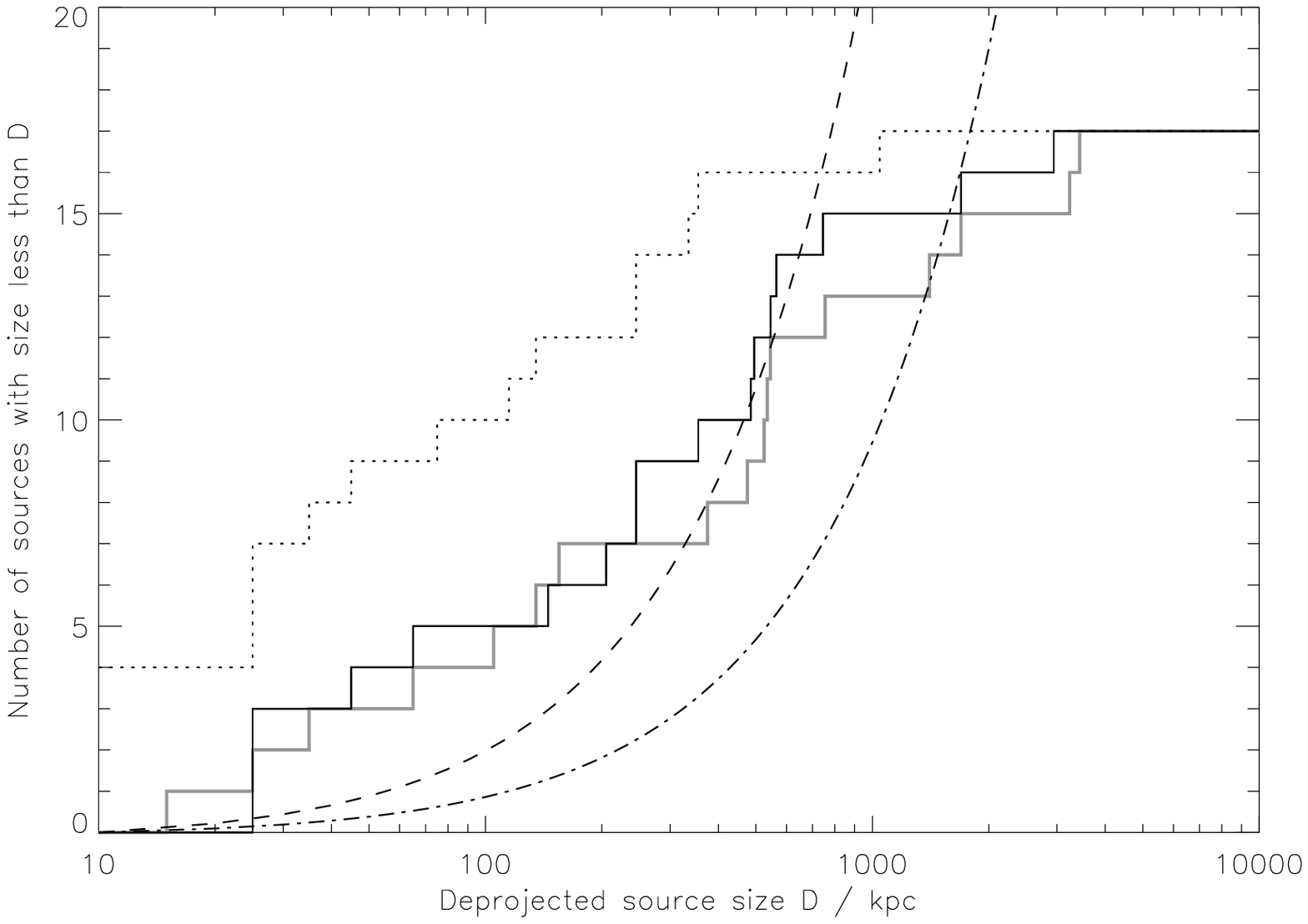}
\caption{Cumulative distribution of source sizes deprojected using the
  best-fit disk angles from the models where the disk emission is
  fixed at the same redshift as the NLR (solid line). The projected
  source sizes (dotted line) and the source sizes deprojected with the
  angles from the analysis where the disk emission is allowed to shift
  with respect to the NLR (grey line) are shown for comparison. The
  curves show the predicted distribution if the heads of the sources
  are expanding at a constant rate; the dot-dashed line is normalised
  at the deprojected size of the largest source in the sample, the
  dashed line is normalised at 1 Mpc. Note that the two core-dominated
  sources have no measured size, and are boosted into this sample by
  strong core emission, and are therefore excluded from this plot.}
\label{fig:dhistsizens}
\end{center}
\end{figure}

Figure \ref{fig:dhistsizens} shows the cumulative distribution of the
deprojected source size for the analysis of the models with the disk
fixed at the redshift of the NLR. This distribution is marginally more
consistent with the comparison models of uniformly expanding sources
than the distribution found in the original analysis (see Figure
\ref{fig:dhistsize}). Both distributions are consistent, within the
limits of small number statistics, with the picture of sources
expanding uniformly up to some cut-off size, though there are
indications in both cases of an excess of small sources.

\subsection{Relationships with radio luminosity}
\label{sec:assumptluminosity}

\begin{figure}
\begin{center}
\includegraphics[width = 0.5\textwidth]{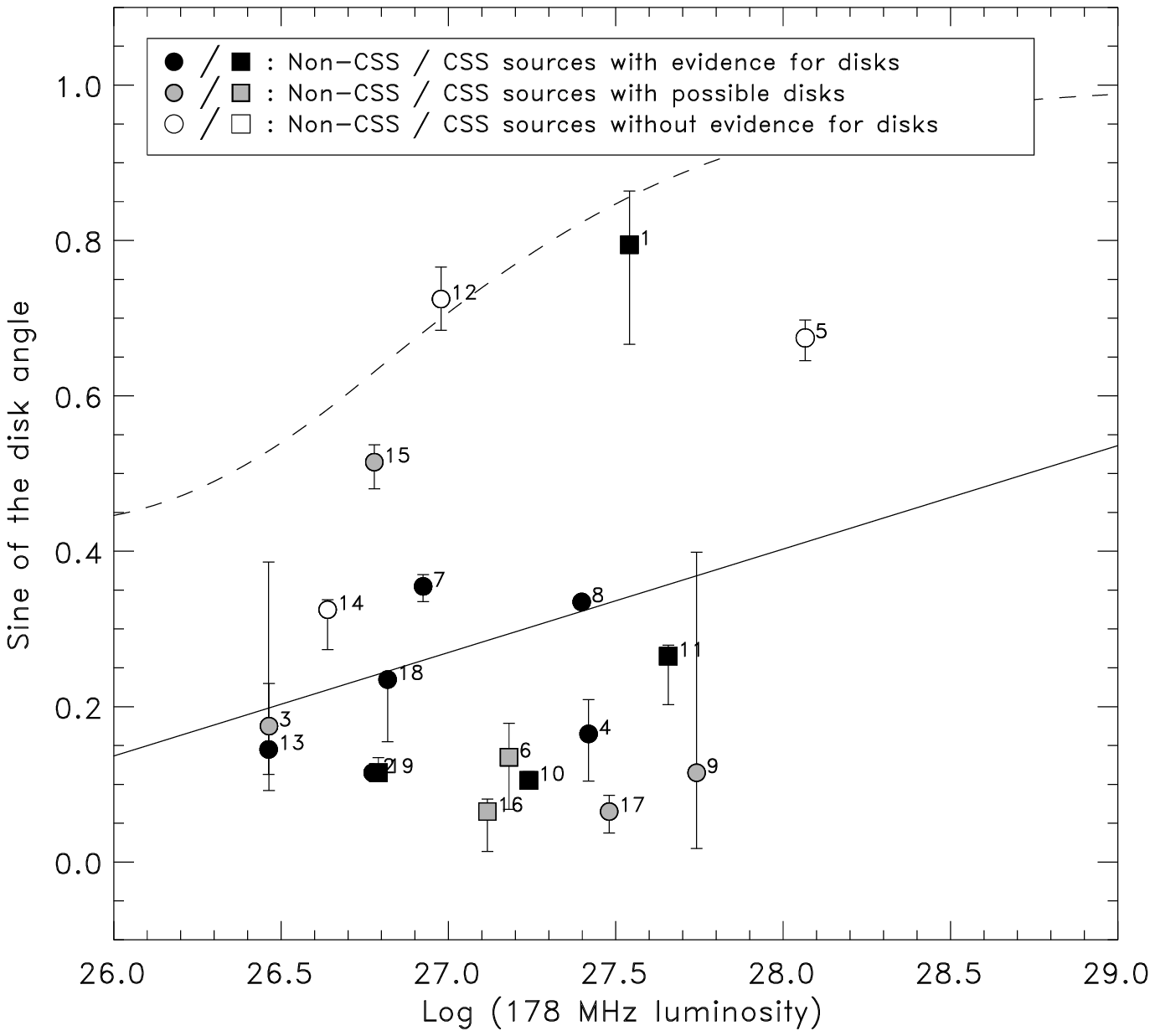}
\caption{Sine best-fit disk angles versus logarithmic 178 MHz
  luminosity. The disk angle error bars include both the errors
  returned from the individual fits and a systematic error calculated
  from the variation in results from using different random seeds to
  initialise the Monte Carlo engines, added in quadrature, but do not
  take into account any error due to model selection. The non-CSS FRII
  sources and core-dominated sources are plotted as circles, and CSS
  sources are plotted as squares. The black points have weak to strong
  evidence for a disk, the grey points have possible disks (there is
  inconclusive evidence, or weak to moderate evidence against a disk)
  and the white points have strong evidence against a disk. Index
  numbers are matched to source names in Table \ref{tab:obsdata}. The
  best-fit line (minimising $\chi^2$) is shown in black. The dashed
  line shows the radio luminosity-dependent critical angle (following
  \citet{willott00}) with fiducial angle $45\degree$ at \loglum = 27,
  and minimum quasar fraction 10\%.}
\label{fig:plotsinivslumns}
\end{center}
\end{figure}

The sine best-fit disk angles for the analysis with the accretion
disks fixed at the NLR redshift are plotted against the 178 MHz radio
luminosity in Figure \ref{fig:plotsinivslumns}. These angles are
consistent with the envelope of angles predicted by the receding torus
model. However, the correlation seen between the sine of the disk
angle and the low-frequency radio luminosity in the original analysis,
at a significance of $>1.5 \sigma$ according to Kendall's $\tau$ test,
has now fallen to $<0.5 \sigma$ (Kendall's $\tau$ coefficient 0.105,
probability of correlation 25\%). This illustrates the strong
dependence of the analysis results on the model assumptions, such that
a large change in fitted angle for only two quasars is enough to
disrupt this correlation.

In order the test the fragility of the luminosity -- disk angle
correlation, the statistical analysis was repeated 24 times, with the
disk angle modified in each case by a Gaussian scatter with standard
deviation of 2.16, the standard deviation of the distribution of
differences between the disk angles with the disk fixed at the
redshift of the NLR, and the disk angles fitted when the accretion
disk was allowed a velocity shift. Only 8 out of 24 test correlations
have higher significance than the original correlation between
luminosity and disk angle of a disk not fixed with respect to the NLR,
which has a significance of $1.8 \sigma$; however, 13 out of 24 test
correlations have a significance of $> 1.5 \sigma$.  A larger sample
of quasars might go some way towards confirming whether this is a real
correlation, but it is clearly necessary to fully test a wider range
of emission models.

\subsection{Distribution of angles}
\label{sec:assumptangdist}

\begin{figure}
\begin{center}
\includegraphics[width = 0.5\textwidth]{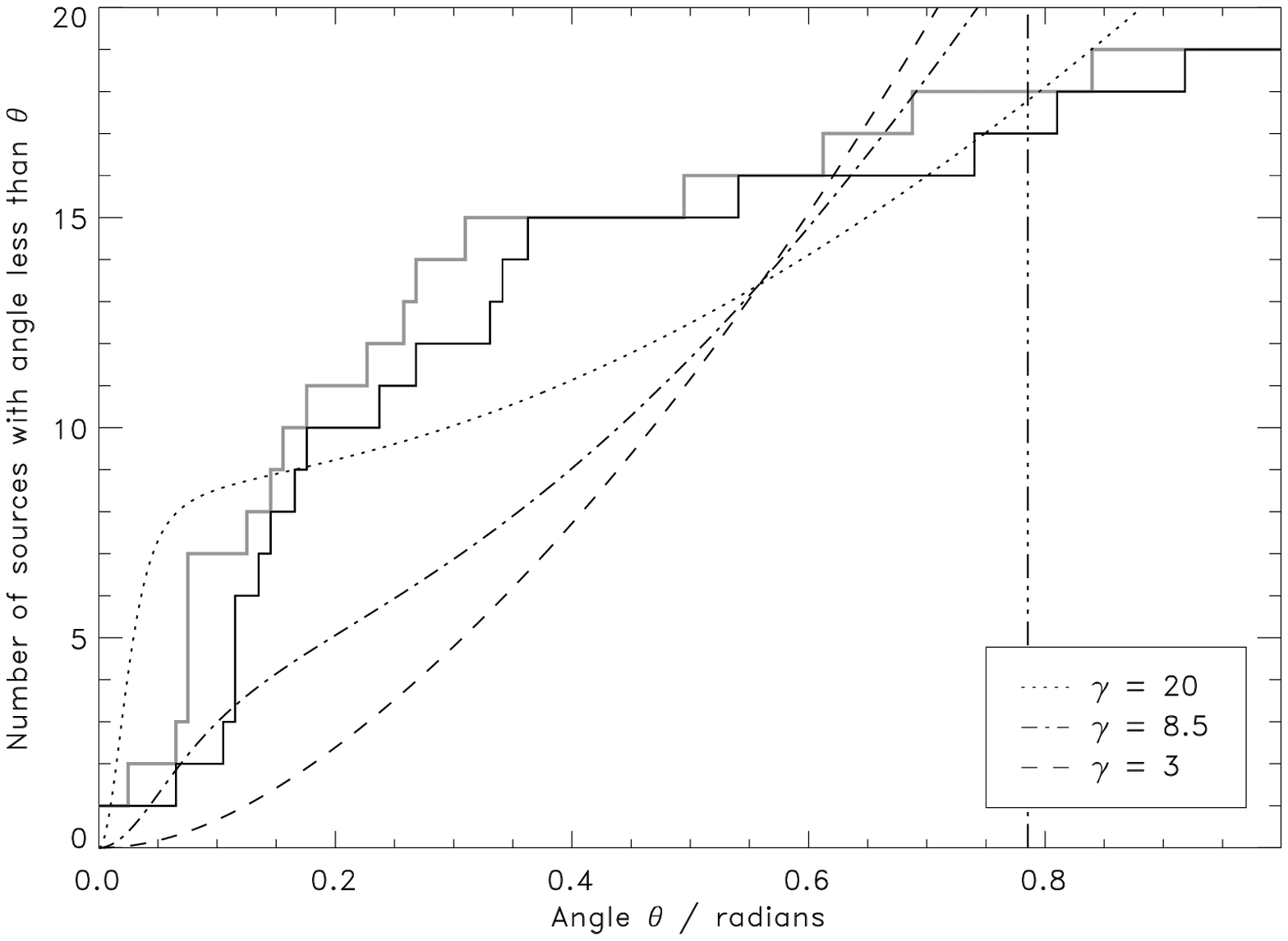}
\caption{The cumulative distribution of the best-fit disk angles from
  the analysis in which the disk redshift is fixed at the redshift of
  the NLR (solid black line). The grey line shows the cumulative
  distribution for the best-fit disk angles for the analysis in which
  the accretion disk is allowed to shift with respect to the NLR. The
  curves show the theoretical distributions of jet angle, modelled
  from the \citet{jackson99} relation and normalised up to an angle of
  $45\degree$, for different values of the Lorentz factor: the dashed
  line shows the expected cumulative distribution for $\gamma = 3$,
  the dot-dashed line for $\gamma = 8.5$, and the dotted line for
  $\gamma = 20$. The vertical dot-dot-dot-dashed line indicates a
  fitted disk angle of $45\degree$.}
\label{fig:pthetafitdiskns}
\end{center}
\end{figure}

The cumulative probability distribution for the disk angles and the
theoretical distribution of jet angle for different values of the
Lorentz factor is shown in Figure \ref{fig:pthetafitdiskns}. With the
assumption that the axis of the accretion disk is coincident with the
radio jets, then the distribution of disk angles is consistent with a
value of the jet Lorentz factor around $\gamma \sim 20$, as for the
previous analysis (see Section \ref{sec:angdist}).

\subsection{Discussion}
\label{sec:assumptdiscussion}

The fitted disk angles have proved to be only partially robust to a
change in the model in which the accretion disk is fixed at the same
redshift as the NLR, although the majority were consistent. The small
number of objects in this sample make it difficult to judge whether
the disappearance of the disk angle -- low-frequency radio luminosity
correlation in this new analysis occurs because the initial results
are a fluke, or is a genuine effect from the alteration of the
models. Since it is not clear that the narrow lines and accretion disk
should be fixed at the same redshift, this new analysis should serve
as a measure of the fragility of the results with respect to a change
in the model.


\section{Conclusions}
\label{sec:conclusions}

All bar one of the 19 quasars are best fitted with an emission model
which includes more than one component of broad-line emission, in both
analyses; there is overwhelming evidence that in this sub-sample of
quasars, a simple one-component broad-line emission model is
insufficient to describe the physical processes at work. There is
strong Bayesian evidence that ten out of nineteen quasars from the
analysis in which a shift is permitted between the BLR and NLR possess
accretion disks, and all but one are consistent with the emission
model of a circular Keplerian accretion disk in addition to a
single-peaked, symmetric emission line arising from a separate
BLR. The one quasar for which this does not apply appears to have very
complex broad emission, consisting of at least three components. If
the model does not allow for a shift between the BLR and NLR, strong
evidence for disks is seen in eight out of the nineteen sources, and
all but three are consistent with the presence of a circular accretion
disk.

\citet{eracleous94}, \citet{eracleous03} and \citet{strateva03} found
that only 3 -- 20$\%$ of AGN have obvious double-peaked lines (see
Section \ref{sec:disks}), in contrast to the $\sim$ 84 -- 95\% of this
sample of quasars which were found to be consistent with the presence
of a thin, optically-emitting accretion disk, despite only a few of
the line profiles appearing obviously double-peaked to the eye. It may
be that accretion disk emission has been seen in so few AGN because
single-peaked broad emission lines, arising from fast-moving clouds
located outside the thin disk or from a wind originating from the disk
itself, have a tendency to swamp the disk emission. In these cases, a
complex model and high signal-to-noise ratio spectra are required in
order to retrieve the accretion disk contribution to the emission. It
is therefore essential to extend this analysis to a larger sample of
low-radio-frequency-selected quasars; optically-selected quasars are
predicted to cover a narrower range in disk angles, as even lightly
reddened quasars (at $\sim 45 \degree$) are much more likely to fall
through the magnitude limit of a survey, and the optical emission may
be angle dependent.

There is a possible correlation of the fitted disk angles with the
projected source size, in agreement with the expected projection
effects if the accretion disk is perpendicular to the jet. There are
two CSS sources which do not appear to lie on the same relation as the
other sources: MRC0222-224 seems to be an intrinsically smaller and
more reddened source than the others, while MRC1114-220 may either be
an outlier, or have a precessing or misaligned disk. When deprojected
using the best-fit disk angles, the source sizes are consistent with a
model in which the accretion disk is perpendicular to the radio jets,
and the heads of the radio jets expand at approximately constant speed
up to a size $\sim 1$ Mpc; the paucity of sources larger than this
limit can be explained by the average duty cycle of these AGN.

The distribution of fitted disk angles is consistent with the
calculation of expected jet angles for Doppler-boosted jets with
Lorentz factors of $\sim$ 20, if the accretion disk rotation axis and
radio jets are assumed to be coincident. The calculated jet angle
distribution may match the disk angle distribution better if a more
physically reasonable model was used for the jet, with a core of
higher Lorentz factor than the surrounding jet sheath, as opposed to
this simple single Lorentz factor model.

There is a weak correlation ($> 1.5 \sigma$ significance by Kendall's
$\tau$ test) between the low-frequency radio luminosity and the sine
of the angle between the accretion disk axis and the line of
sight. This is predicted by the receding torus model of
\citet{lawrence91} if the accretion disks are perpendicular to the
radio jets, such that the disk angles match the jet angles. In this
model, the dust is sublimated by the AGN nucleus, leading to larger
opening angles of the obscuring dusty torus for more radio-luminous
sources, and hence quasars are seen up to larger jet angles. The
largest accretion disk angle measured is 48$\degree$, which is
consistent with the opening angle predicted for powerful FRII radio
sources. A larger sample size is needed to confirm the radio
luminosity -- disk angle correlation, but this is vital, as it would
yield the first direct test of the receding torus model. It should be
noted that when the analysis was performed again with the models in
which the accretion disk was fixed at the redshift of the NLR, this
result disappeared, and so the radio luminosity -- disk angle
correlation is strongly dependent on the model used.

A study of the velocity shifts between the line components revealed
the single-peaked broad lines to be formed at similar redshifts to the
narrow lines. The disk emission tended to be redshifted with respect
to these lines for high-$z${} sources and blueshifted for lower-$z$
sources. Since the accretion disk is expected to be at the systemic
velocity, this would imply that the single-peaked broad lines and the
narrow lines are formed in outflows for more luminous, high-$z${}
sources, and in infalling clouds for closer, less luminous
sources. Such a scenario could arise if powerful winds were causing
outflows of gas from the high-$z$, luminous sources, while these winds
have stalled in the less-luminous quasars seen at $z < 1$, resulting
in infall of the gas back towards the galaxy nucleus.

The results of this paper are all dependent on the validity of the
disk model used in the Bayesian fitting. This model, taken from
\citet{chen89}, describes a circular, Keplerian disk. The illuminating
flux from the central black hole was modelled as falling off with a
radial exponent of 3. In order to investigate the effects of range of
permutations to this basic model, such as a variety of disk
emissivities and warped or elliptical disks, greater spectral
resolution, a larger quasar sample size, and in particular,
multi-epoch data to study the variability of the emission line
profiles are required.

\section*{Acknowledgments}

EJD would like to acknowledge the support of an STFC (formerly
PPARC) studentship, and a Government of Canada Post-Doctoral Research
Fellowship held at the NRC Herzberg Institute of Astrophysics in Victoria,
BC. This paper is based on observations made with the ESO VLT UT1
telescope at Paranal Observatory under program ID 68.B-0409(A). We 
would like to thank the anonymous referee for comments and suggestions
which improved this paper, and C. J. Willott for helpful 
comments.


\appendix

\section{Tests for stability of the fitting}
\label{sec:stabtests}

Detailed tests were run on two of the quasars selected at random
(MRC0437-244 and MRC0450-221), in order to check the Bayesian routines
for stability. In each set of tests, one of the \textit{BayeSys3}
parameters was changed: the Bayesian random seed; the rate at which
the Monte Carlo exploration focuses in on the region of maximum
likelihood; and the ensemble of parallel explorations of the parameter
space during the Monte Carlo optimisation. Plots of the posterior
probabilities of the disk parameters were created, each test being
overlaid in a different shade, to identify fits in which the posterior
probability converged in a different region to the majority of the
fits.

An example of one of the stability tests is shown in Figure
\ref{fig:cosmoseed}: this shows the convergence of some of the disk
parameters for MRC0450-221, as well as the correlations between these
parameters, when the Bayesian fitting is initialised with five
different random seeds. The posterior probability distribution for
each seed is plotted in a different shade of grey. The local velocity
dispersion, $\Delta v$, of the disk material converges in two slightly
different regions of parameter space, depending on the Bayesian seed,
though these areas are within a reasonable error bound of each
other. The local minima do not have vastly different evidence values
in this case ($\Delta \ln(\mathrm{Evidence}) \sim 2.5$ between the
highest and lowest evidence cases); this is due to the fact that the
profile of the disk emission varies only weakly with the local
velocity dispersion \citep{strateva03}. The variation in velocity
dispersion was similar through all of the parameters tested: all other
parameters were shown to be stable in each of the tests for
MRC0450-221. All parameters were stable in the tests for MRC0437-244.

In addition, similar tests were made in which the Bayesian fitting
parameters were unchanged, but the spectra were altered. The first of
these tests added Gaussian random noise to the spectrum, with FWHM of
5\% of the data values, in order to see qualitatively what effect
greater noise has on the fitting process. For MRC0450-221, the
posterior probability for the local velocity dispersion occupied a
similar region in parameter space to the total space occupied by all
fits in the test with different Bayesian random seeds; it appears that the
velocity dispersion parameter is constrained to a fairly large region
of parameter space, and within this, for some tests, local minima are
found. For MRC0437-241, the parameters constrain in similar regions
independent of the noise added.

The second of these tests was the fitting of spectra smoothed with
boxcar widths of 3, 5, 7 and 9 pixels, where each pixel has a scale of
approximately 2 \AA. For MRC0437-244, the smoothed spectrum allowed
the probability to constrain marginally more tightly, but in the same
regions. This was also true of MRC0450-221 for the boxcar smoothing
widths of 3, 5 and 7 pixels. At a boxcar width of 9 pixels, the
posterior probability distributions of the fit alter significantly, as
real structure has been lost from the spectrum.


\label{lastpage}


\begin{thebibliography}{99}
\bibitem[\protect\citeauthoryear{Amico et al.}{2002}]{isaacdrg}Amico, P., Cuby, J.-G., Devillard, N., Jung, Y., Lidman, C., 2002, \textit{Isaac Data Reduction Guide}
\bibitem[\protect\citeauthoryear{Antonucci}{1993}]{antonucci93} Antonucci, R., 1993, ARA\&A, 31, 473
\bibitem[\protect\citeauthoryear{Antonucci \& Miller}{1985}]{antonucci85} Antonucci, R. R. J., Miller, J. S., 1985, ApJ, 297, 621
\bibitem[\protect\citeauthoryear{Bachev}{1999}]{bachev99} Bachev, R., 1999, A\&A, 348, 71 
\bibitem[\protect\citeauthoryear{Bahcall \& Soneira}{1980}]{bahcall80} Bahcall, J. N., Soneira, R. M., 1980, ApJS, 44, 73
\bibitem[\protect\citeauthoryear{Baker et al.}{1999}]{molonglo4} Baker, J. C., Hunstead, R. W., Kapahi, V. K., Subrahmanya, C. R., 1999, ApJS, 122, 29
\bibitem[\protect\citeauthoryear{Baker et al.}{2002}]{baker02} Baker, J. C., Hunstead, R. W., Athreya, R. M., Barthel, P. D., de Silva, E., Lehnert, M. D., 2002, ApJ, 568, 592
\bibitem[\protect\citeauthoryear{Baldwin}{1977}]{baldwin77} Baldwin, J. A., 1977, ApJ, 214, 679
\bibitem[\protect\citeauthoryear{Barthel}{1989}]{barthel89} Barthel, P. D., 1989, ApJ, 336, 606
\bibitem[\protect\citeauthoryear{Bird et al.}{2008}]{bird08} Bird, J., Martini, P., Kaiser, C., 2008, ApJ, 676, 147
\bibitem[\protect\citeauthoryear{Brown et al.}{1989}]{brown89} Brown, L. M. J., Robson, E. I., Gear, W. K., Hughes, D. H., Griffin, M. J., Geldzahler, B. J., Schwartz, P. R., Smith, M. G., Smith, A. G., Shepherd, D. W., Webb, J. R., Valtaoja, E., Terasranta, H., Salonen, E., 1989, ApJ, 340, 129
\bibitem[\protect\citeauthoryear{Capriotti et al.}{1979}]{capriotti79} Capriotta, E., Foltz, C., Byard, P., 1979, ApJ, 230, 681
\bibitem[\protect\citeauthoryear{Chen et al.}{1989}]{chen89b} Chen, K., Halpern, J. P., Filippenko, A. V., 1989, ApJ, 339, 742
\bibitem[\protect\citeauthoryear{Chen \& Halpern}{1989}]{chen89} Chen, K., Halpern, J. P., 1989, ApJ, 344, 115
\bibitem[\protect\citeauthoryear{Collin-Souffrin et al.}{1980}]{collin80} Collin-Souffrin, S., Dumont, S., Heidmann, N., Joly, M., 1980, AA, 83, 190
\bibitem[\protect\citeauthoryear{Collin-Souffrin et al.}{1988}]{collin88} Collin-Souffrin, S., Dyson, J. E., McDowell, J. C., Perry, J. J., 1988, MNRAS, 232, 539
\bibitem[\protect\citeauthoryear{Corbin}{1990}]{corbin90} Corbin, M. R., 1990, ApJ, 357, 346
\bibitem[\protect\citeauthoryear{Devillard}{1997}]{eclipse} Devillard, N., 1997, Messenger, 87, 19
\bibitem[\protect\citeauthoryear{Eracleous \& Halpern}{1994}]{eracleous94} Eracleous, M., Halpern, J. P., 1994, ApJS, 90, 1
\bibitem[\protect\citeauthoryear{Eracleous \& Halpern}{2003}]{eracleous03} Eracleous, M., Halpern, J. P., 2003, ApJ, 599, 886
\bibitem[\protect\citeauthoryear{Espey et al.}{1989}]{espey89} Espey, B. R., Carswell, R. F., Bailey, J. A., Smith, M. G., Ward, M. J., 1989, ApJ, 342, 666
\bibitem[\protect\citeauthoryear{Filippenko}{1988}]{filippenko88} Filippenko, A. V., 1988, Adv. Space Res., 8, 5
\bibitem[\protect\citeauthoryear{Gaskell}{1982}]{gaskell82} Gaskell, C. M., 1982, ApJ, 263, 79
\bibitem[\protect\citeauthoryear{Gurzadyan \& Ozernoy}{1979}]{gurzadyan79} Gurzadyan, V. G., Ozernoy, L. M., 1979, Nature, 280, 214
\bibitem[\protect\citeauthoryear{Halpern \& Chen}{1989}]{halpern89} Halpern, J. P., Chen, K., 1989, Active Galactic Nuclei: Proceedings of the 134th Symposium of the International Astronomical Union, 134, 245
\bibitem[\protect\citeauthoryear{Hardcastle}{2006}]{hardcastle06} Hardcastle, M. H., 2006, MNRAS, 366, 1465
\bibitem[\protect\citeauthoryear{Heckman et al.}{1981}]{heckman81} Heckman, T. M., Miley, G. K., van Breugel, W. J. M., Butcher, H. R., 1981, ApJ, 247, 403
\bibitem[\protect\citeauthoryear{Jackson \& Wall}{1999}]{jackson99} Jackson, C. A., Wall, J. V., 1999, MNRAS, 304, 160
\bibitem[\protect\citeauthoryear{Jarvis \& McLure}{2006}]{jarvis06} Jarvis, M. J., McLure, R. J., 2006, MNRAS, 369, 182
\bibitem[\protect\citeauthoryear{Jeffreys}{1939}]{jeffreys39} Jeffreys, H., 1939, \textit{Theory of probability}, OUP, 3rd ed (1998).
\bibitem[\protect\citeauthoryear{Kapahi et al.}{1998}]{molonglo3} Kapahi, V. K., Athreya, R. M., Subrahmanya, C. R., Baker, J. C., Hunstead, R. W., McCarthy, P. J., van Breugel, W., 1998, ApJS, 122, 29
\bibitem[\protect\citeauthoryear{Koski}{1978}]{koski78} Koski, A. T., 1978, ApJ, 223, 56
\bibitem[\protect\citeauthoryear{Krolik \& Begelman}{1986}]{krolik86} Krolik, J. H., Begelman, M. C., 1986, ApJ, 308, L55
\bibitem[\protect\citeauthoryear{Krolik et al.}{1991}]{krolik91} Krolik, J. H, Horne, K., Kallman, T. R., Malkan, M. A., Edelson, R. A., Kriss, G. A., 1991, ApJ, 371, 541
\bibitem[\protect\citeauthoryear{Large et al.}{1981}]{large81} Large, M. I., Mills, B. Y., Little, A. G., Crawford, D. F, Sutton, J. M., 1981, MNRAS, 194, 693
\bibitem[\protect\citeauthoryear{Lavalley et al.}{1992}]{asurv} Lavalley, M., Isobe, T., Feigelson, E., 1992, Astronomical Data Analysis Software and Systems I, A.S.P. Conference Series, 25, 245
\bibitem[\protect\citeauthoryear{Lawrence}{1991}]{lawrence91} Lawrence, A., 1991, MNRAS, 252, 586
\bibitem[\protect\citeauthoryear{Longair \& Riley}{1979}]{longair79} Longair, M. S.,  Riley, J. M., 1979, MNRAS, 188, 625
\bibitem[\protect\citeauthoryear{McCarthy et al.}{1996}]{molonglo1} McCarthy, P. J., Kapahi, van Breugel, W., Persson, S. E., V. K., Athreya, R. M., Subrahmanya, C. R., 1996, ApJS, 107, 19
\bibitem[\protect\citeauthoryear{McLure \& Jarvis}{2002}]{mclure02} McLure, R. J., Jarvis, M. J., 2002, MNRAS, 337, 109
\bibitem[\protect\citeauthoryear{MacIntosh}{1999}]{mcintosh99} McIntosh, D. H., Rix, H.-W., Rieke, M. J., Foltz, C. B, 1999, ApJ, 517, L73 
\bibitem[\protect\citeauthoryear{Malkan \& Sargent}{1982}]{malkan82} Malkan, M. A., Sargent, W. L. W., 1982, ApJ, 254, 22
\bibitem[\protect\citeauthoryear{Mart\'inez-Sansigre et al.}{2005}]{martinez05} Mart\'inez-Sansigre, A., Rawlings, S., Lacy, M., Fadda, D., Marleau, F. R., Simpson, C., Willott, C. J., Jarvis, J. M., 2005, Nature, 436, 666
\bibitem[\protect\citeauthoryear{Moorwood et al.}{1998}]{moorwood98} Moorwood, A.,  Cuby, J.-G.,  Biereichel, P.,  Brynnel, J.,  Delabre, B.,  Devillard, N.,  van Dijsseldonk, a.,  Gemperlein, H.,  Gilmozzi, R.,  Herlin, T.,  Huster, G.,  Knudstrup, J.,  Lidman, C.,  Lizon, J.-L.,  Mehrgan, H.,  Meyer, M.,  Nicolini, G.,  Petr, M.,  Spyromilio, J.,  Stegmeier, J., 1998, Messenger, 94, 7
\bibitem[\protect\citeauthoryear{Murray \& Chiang}{1997}]{murray97} Murray, N., Chiang, J., 1997, ApJ, 474, 91
\bibitem[\protect\citeauthoryear{Nesvadba et al.}{2006}]{nesvadba06} Nesvadba, N. P. H., Lehnert, M. D., Eisenhauer, F., Gilbert, A., Tecza, M. and Abuter, R., 2006, ApJ, 650, 693
\bibitem[\protect\citeauthoryear{Nesvadba et al.}{2008}]{nesvadba08} Nesvadba, N. P. H., Lehnert, M. D., De Breuck, C., Gilbert, A., van Breugel, W., 2008, A\&A, 491, 407
\bibitem[\protect\citeauthoryear{Netzer}{1985}]{netzer85} Netzer, H., 1985, MNRAS, 216, 63
\bibitem[\protect\citeauthoryear{O'Dea}{1998}]{odea98} O'Dea, C. P., 1998, PASP, 110, 493
\bibitem[\protect\citeauthoryear{Orr \& Browne}{1982}]{orr82} Orr, M. J. L., Browne, I. W. A., 1982, MNRAS, 200, 1067
\bibitem[\protect\citeauthoryear{Osterbrock, Koski \& Phillips}{1976}]{osterbrock76} Osterbrock, D. E., Koski, A. T., Phillips, M. M., 1976, ApJ, 206, 898
\bibitem[\protect\citeauthoryear{Osterbrock \& Cohen}{1979}]{osterbrock79} Osterbrock, Cohen, R., 1979, MNRAS, 187, 61P
\bibitem[\protect\citeauthoryear{P\'erez et al.}{1988}]{perez88} P\'erez, E., Penston, M. V., Tadhunter, C., Mediavilla, E., Moles, M., 1988, MNRAS, 230, 353
\bibitem[\protect\citeauthoryear{Peterson}{1997}]{peterson97} Peterson, B. M., 1997, \textit{Active Galactic Nuclei}, CUP, 1st ed.
\bibitem[\protect\citeauthoryear{Rees et al.}{1982}]{rees82} Rees, M. J., Begelman, M. C., Blandford, R. D., Phinney, E. S., 1982, Nature, 295, 17
\bibitem[\protect\citeauthoryear{Richards et al.}{2002}]{richards02} Richards, G. T., Vanden Berk, D. E., Reichard, T. A., Hall, P. B., Schneider, D. P., SubbaRao, M., Thakar, A. R., York, D. G., 2002, AJ, 124, 1
\bibitem[\protect\citeauthoryear{Scheuer \& Readhead}{1979}]{scheuer79} Scheuer, P. A. G., Readhead, A. C. S., 1979, Nature, 277, 182
\bibitem[\protect\citeauthoryear{Schlegel et al.}{1998}]{schlegel98} Schlegel, D. J., Finkbeiner, D. P., Davis, M., 1998, ApJ, 500, 525
\bibitem[\protect\citeauthoryear{Shakura \& Sunyaev}{1973}]{shakura73} Shakura, N. I., Sunyaev, R. A., 1973, AA, 24, 337
\bibitem[\protect\citeauthoryear{Silk \& Rees}{1998}]{silk98} Silk, J., Rees, M. J., 1998, A\&A, 331, L1

\bibitem[\protect\citeauthoryear{de Silva et al.}{in preparation}]{silva1} de Silva, E., Baker, J. C., Saunders, R. D. E., Hunstead, R. W., in preparation
\bibitem[\protect\citeauthoryear{de Silva et al.}{in preparation}]{silva2} de Silva, E., Baker, J. C., Hunstead, R. W., Saunders, R. D. E., in preparation
\bibitem[\protect\citeauthoryear{Simpson}{1998}]{simpson98} Simpson, C., 1998, MNRAS, 297, L39
\bibitem[\protect\citeauthoryear{Sivia \& Skilling}{2006}]{sivia06} Sivia, D. S., Skilling, J., 2006, \textit{Data Analysis: A Bayesian Tutorial}, OUP, 2nd ed.
\bibitem[\protect\citeauthoryear{Skilling}{2004}]{bayesys} Skilling, J., 2004, \textit{BayeSys and MassInf}, Maximum Entropy Data Consultants Ltd.
\bibitem[\protect\citeauthoryear{Smith et al.}{1981}]{smith81} Smith, M. G., Carswell, R. F., Whelan, J. A. J., Wilkes, B. J., Boksenburg, A., Clowes, R. G., Savage, A., Cannon, R. D., Wall, J. V., 1981, MNRAS, 195, 437 
\bibitem[\protect\citeauthoryear{Strateva et al.}{2003}]{strateva03} Strateva, I. V., Strauss, M. A., Hao, L., Schlegel, D. J., Hall, P. B., Gunn, J. E., Li, L-X., Ivez\'ic, Z., Richards, G. T., Zakamska, N. L., Voges, W., Anderson, S. F., Lupton, R. H., Schneider, D. P., Brinkman, J., Nichol, R. C., 2003, AJ, 126, 1720
\bibitem[\protect\citeauthoryear{Tanaka et al.}{1995}]{tanaka95} Tanaka, Y., Nandra, K., Fabian, A. C., Inoue, H., Otani, C., Dotani, T., Hayashida, K., Iwasawa, K., Kii, T., Kunieda, H., Makino, F., Matsuoka, M., 1995, Nature, 375, 659
\bibitem[\protect\citeauthoryear{Trotta}{2008}]{trotta08} Trotta, R., 2008, Contemporary Physics, 49, 71
\bibitem[\protect\citeauthoryear{Urry \& Padovani}{1995}]{urry95} Urry, C. M., Padovani, P., 1995, PASP, 107, 803
\bibitem[\protect\citeauthoryear{Vanden Berk et al.}{2001}]{vandenberk01} Vanden Berk et al., 2001, AJ, 122, 549
\bibitem[\protect\citeauthoryear{Vardoulaki et al.}{2008}]{vardoulaki08} Vardoulaki, E., Rawlings, S., Simpson, C., Bonfield, D. G., Ivison, R. J.,  Ibar, E., 2008, MNRAS, 387, 505
\bibitem[\protect\citeauthoryear{Veilleux \& Osterbrock}{1987}]{veilleux87} Veilleux, S., Osterbrock, D. E., 1987, ApJS, 63, 295
\bibitem[\protect\citeauthoryear{Vestergaard}{2002}]{vestergaard02} Vestergaard, M., 2002, ApJ, 571, 733
\bibitem[\protect\citeauthoryear{Wilkes}{1984}]{wilkes84} Wilkes, B. J., 1984, MNRAS, 207, 73
\bibitem[\protect\citeauthoryear{Willott et al.}{2000}]{willott00} Willott, C. J., Rawlings, S., Blundell, K. M., Lacy, M., 2000, MNRAS, 316, 449
\bibitem[\protect\citeauthoryear{Willott et al.}{2001}]{willott01} Willott, C. J., Rawlings, S., Blundell, K. M., Lacy, M., Eales, S. A., 2001, MNRAS, 322, 536
\bibitem[\protect\citeauthoryear{Wills}{1980}]{wills80} Wills, D., 1980, ApJ, 240, 721
\bibitem[\protect\citeauthoryear{Wills \& Browne}{1986}]{wills86} Wills, B. J., Browne, I. W. A., 1986, ApJ, 302, 56
\bibitem[\protect\citeauthoryear{}{}]{} 

\end{thebibliography}
\end{document}